\documentclass[usenatbib,useAMS]{mnras}
\usepackage{rotating}
\usepackage{tablefootnote}
\usepackage{multirow}
\usepackage{xcolor}
\usepackage{graphicx}	
\usepackage{amsmath,cancel}	
\usepackage{amssymb}	
\usepackage{multicol}   
\usepackage{bm}		
\usepackage[normalem]{ulem} 
\usepackage{caption} 
\usepackage{url}
\usepackage{longtable}
\usepackage{pdfpages}
\usepackage{multirow}
\usepackage{subcaption}
\usepackage{hyperref}
\usepackage{pdflscape}
\usepackage{wasysym}
\usepackage{supertabular}
\usepackage{float}
\usepackage{orcidlink}

\usepackage{xspace}
\usepackage[export]{adjustbox}
\usepackage{upgreek}

\title[Broad-band outburst of 3C\,138]{A broad-band outburst of the compact steep-spectrum quasar 3C\,138 in 2024--2026}

\author[Mufakharov et al.]{%
T.~V.~Mufakharov,$^{1,2}\orcidlink{0000-0001-9984-127X}$\thanks{E-mail: timur.mufakharov@gmail.com} 
Yu.~V.~Sotnikova,$^{2,3}\orcidlink{0000-0001-9172-7237}$
V.~V.~Vlasyuk,$^{2}\orcidlink{0009-0002-6596-7274}$
S.~Yu.~Sazonov,$^{4}\orcidlink{0009-0000-7620-5086}$
\newauthor
M.~L.~Khabibullina,$^{2}\orcidlink{0000-0001-9515-4552}$
A.~G.~Mikhailov,$^{2}\orcidlink{0000-0002-0279-0777}$
A.~B.~Pushkarev,$^{5,6}\orcidlink{0000-0002-9702-2307}$
T.~An,$^{7}\orcidlink{0000-0003-4341-0029}$
Y.~A.~Kovalev,$^{6,3}\orcidlink{0000-0002-8017-5665}$
\newauthor
Y.~Y.~Kovalev,$^{8}\orcidlink{0000-0001-9303-3263}$
A.~V.~Popkov,$^{9,6}\orcidlink{0000-0002-0739-700X}$
M.~A.~Kharinov,$^{10}\orcidlink{0000-0002-0321-8588}$
G.~S.~Uskov,$^{4}\orcidlink{0000-0002-0274-1350}$
I.~Yu.~Lapshov,$^{4}\orcidlink{0009-0000-4769-452X}$
\newauthor
E.~V.~Filippova,$^{4}\orcidlink{0009-0000-6101-1879}$
A.~Yu.~Tkachenko,$^{4}\orcidlink{0000-0002-7486-1730}$
K.~V.~Iuzhanina,$^{2}\orcidlink{0009-0008-7311-0709}$
A.~K.~Erkenov,$^{2}\orcidlink{0000-0002-6086-9299}$
R.~Yu.~Udovitskiy,$^{2}$
\newauthor
M.~I.~Chazov,$^{2}\orcidlink{0009-0004-4126-8924}$
O.~I.~Spiridonova,$^{2}\orcidlink{0009-0007-7315-3090}$
I.~A.~Rakhimov,$^{10}\orcidlink{0000-0002-9185-6239}$
T.~S.~Andreeva,$^{10}\orcidlink{0000-0003-3613-6252}$
A.~A.~Ogloblin$^{10,11}$
\\
$^{1}$ State Key Laboratory of Radio Astronomy and Technology, Xinjiang Astronomical Observatory, CAS, 150 Science 1-Street, \\ Urumqi 830011, China \\
$^{2}$ Special Astrophysical Observatory of the Russian Academy of Sciences, Nizhny Arkhyz, 369167, Russia\\
$^{3}$ Institute for Nuclear Research, Russian Academy of Sciences, 60th October Anniversary Prospect 7a, Moscow 117312, Russia\\
$^{4}$ Space Research Institute of the Russian Academy of Sciences, Profsoyusnaya street, 84/32, Moscow 117997, Russia \\
$^{5}$ Crimean Astrophysical Observatory of the Russian Academy of Sciences, 298409, Nauchny, Russia \\
$^{6}$ Lebedev Physical Institute of the Russian Academy of Sciences, Leninsky prospekt 53, Moscow 119991, Russia\\
$^{7}$ Department of Astronomy, University of Science and Technology of China, 96 Jinzhai Road, Hefei, Anhui 230026, P. R. China\\
$^{8}$ Max-Planck-Institut f\"ur Radioastronomie, Auf dem H\"ugel 69, Bonn D-53121, Germany\\
$^{9}$ Kourovka Astronomical Observatory, Ural Federal University, Kuibysheva St. 48, Yekaterinburg 620026, Russia\\
$^{10}$ Institute of Applied Astronomy of the Russian Academy of Sciences, Kutuzova Embankment 10, St. Petersburg 191187, Russia\\
$^{11}$ St. Petersburg Academic University of the Russian Academy of Sciences, Khlopina st. 8/3A, St. Petersburg 194021, Russia\\
}

\pubyear{2026}

\begin{document}
\newcommand{\sou}{3C\,138\xspace}
\label{firstpage}
\pagerange{\pageref{firstpage}--\pageref{lastpage}}
\maketitle 

\begin{abstract} 
After several decades of relative quiescence, the compact steep-spectrum quasar began a radio brightening in 2022 and entered an active phase in 2024--2026 pronounced optical-to-$\gamma$-ray activity using dense multifrequency radio monitoring at 1--22\,GHz with \mbox{RATAN-600} and RT-32, optical $R$-band observations with Zeiss-1000 and AS-500/2, X-ray measurements with \textit{Swift}/XRT and \textit{SRG}/ART-XC, and the \textit{Fermi}-LAT $\gamma$-ray light curve. The radio brightening accelerated after 2022 and was strongest at the highest frequencies (38--42 GHz). The radio spectra hardened markedly, with the 11--22\,GHz spectral index evolving from steep to flat or inverted during the active phase. The X-ray flux increased by more than a factor of three during 2025--2026. The photon index hardened from $\Gamma_{\rm X}\simeq 1.6$ to $\Gamma_{\rm X}\simeq 0.9$ and softened back after the peak. Flare decomposition revealed five $\gamma$-ray flares and a sequence of optical subflares during the later stages of the activity. The $\gamma$-ray, X-ray, and optical maxima occur within a $\simeq 12$-day interval, suggesting a common activity episode, whereas the radio brightens more gradually and in a frequency-dependent manner. The sparse two-state spectral energy distributions admit non-unique one-zone synchrotron self-Compton (SSC) solutions, with minor, parameter-dependent broad-line-region external Compton contributions. All four SSC cases yield higher electron-to-magnetic energy-density ratios at the main peak than in the quiet state. These results favour a sustained, core-dominated active phase, with later high-energy and optical flares superposed on the opacity-driven radio evolution of an emerging synchrotron component.
\end{abstract}

\maketitle 

\section{Introduction}
\label{sec:intro}

The quasar \sou at $z = 0.759$ \citep{1966ApJ...144.1244L} is a compact steep-spectrum (CSS) radio source. Such objects are generally interpreted as young or frustrated radio-loud active galactic nuclei (AGN)  \citep{1995A&A...302..317F} in which the compact radio plasma still interacts strongly with the interstellar medium (ISM). Their high-energy detection rate is low: in the largest recent {\it Fermi}-LAT study of young radio sources, 11 out of 162 objects were detected, with the detections concentrated among quasars rather than radio galaxies \citep{2020ApJS..247...33A, 2021MNRAS.507.4564P}. A strong $\gamma$-ray outburst from \sou is therefore unusual among young radio sources and provides a useful probe of its compact jet.

The presence of superluminal motion in the inner jet \citep{2001A&A...370...65S} and a nonuniform Faraday screen near the compact core \citep{2003A&A...406...43C} make \sou a suitable laboratory for testing the interaction of the relativistic jet with the dense environment on parsec scales in AGN.

During 2024--2026, \sou entered a pronounced multiwavelength active phase, exhibiting strong $\gamma$-ray flares \citep{2024ATel16845....1B,2025ATel17107....1M,2025ATel17180....1W,2025ATel17461....1L} accompanied by enhanced X-ray emission \citep{2025ATel17142....1G,2026ATel17645....1U} and significant brightening from optical to radio wavelengths \citep{2025ATel17496....1H,2025ATel17193....1K,2025ATel17077....1L,2025ATel17104....1S,2025ATel17540....1V}. Such broadband activity is unusual for CSS quasars. The radio structure of CSS sources is dominated by optically thin mini-lobes or jets with characteristic extents ranging from hundreds of parsec to several kiloparsecs, seen at relatively large viewing angles \citep{1990A&A...231..333F}. As a result, in contrast to core-dominated and highly Doppler-boosted blazars, CSS sources generally show a low radio variability level \citep{2021AJ....161...88P}. Therefore, the major flare in \sou indicates a substantial change in the physical conditions within the inner jet. In particular, the contemporaneous increase of radio flux density, most pronounced at high frequencies, together with optical brightening, suggests a global reconfiguration of the emitting region rather than a localized or short-lived event. Recent very-long-baseline interferometry (VLBI) observations, complemented by the Very Large Array (VLA) and the Atacama Large Millimetre/submillimetre Array (ALMA) data, instead point to a core-dominated brightening and a compact synchrotron-self-absorbed core whose magnetic field is well below the equipartition value; no separated new knot has yet been resolved \citep{2026A&A...710A..63L}. These results offer an important context for interpreting the broadband variability of the source.

This event also fits into the emerging time-domain view of AGN jets. Recent work has emphasized intermittent jet duty cycles and jet escape as key ingredients of AGN radio activity \citep{2026ApJ..1004L...5A}, while VLBI observations of Mrk~110 have shown that even a historically radio-quiet AGN can launch episodic relativistic ejecta \citep{2025ApJ...987L..26W}.  This  2024--2026 flaring episode provides a rare opportunity to investigate the origin of broadband variability in a CSS quasar whose jet is not viewed at an extreme blazar-like angle: what transient inner-jet conditions allow a young compact radio source to become efficient at GeV production? The jet viewing angle of \sou is not well constrained. The apparent speed of the inner jet components reported by \citet{2001A&A...370...65S}, $\beta_{\rm app}\simeq 3.3$, implies the minimum Lorentz factor $\Gamma_{\rm min}=\sqrt{1+\beta_{\rm app}^2}\simeq3.5$ and the maximum angle to the line of sight $\theta_{\rm max}\simeq34^\circ$ under the standard superluminal-motion interpretation. However, this constraint alone does not uniquely determine the jet orientation. Thus, while some Doppler boosting is expected, the available constraints do not require the extreme beaming typical of classical blazars. The combination of high-energy flaring and frequency-dependent radio brightening therefore provides a probe of intrinsic jet evolution, disturbance propagation, and opacity effects in a stratified relativistic flow.

The physical mechanism responsible for the high-energy emission remains uncertain. Possible scenarios include shock waves that compress the magnetic field and enhance synchrotron and inverse-Compton emission \citep{1985ApJ...298..114M}, magnetic reconnection in a turbulent jet \citep{2016MNRAS.462.3325P}, or external Compton (EC) scattering associated with jet--ISM interaction. The recent flaring activity of \sou offers a valuable case for testing these scenarios in a CSS source. Such events highlight the importance of long-term radio monitoring programmes and multifrequency catalogues of compact AGN, which provide the temporal and spectral context needed to identify emerging jet components and opacity-driven evolution \citep[e.g.][]{2014A&A...572A..59M,2011ApJS..194...29R,2016A&A...596A..45F,2026ApJS..283...69C}.

To characterise this active phase, we conducted coordinated multiwavelength monitoring with an increased cadence in the radio and optical bands, complemented by target-of-opportunity (ToO) X-ray observations \citep{2026ATel17645....1U}. In the following sections, we present the observed data and analyse the variability and spectral evolution of \sou across the radio, optical, X-ray, and $\gamma$-ray bands.

\section{Observations}

\label{sec:obs}

\begin{figure*}
\centerline{\includegraphics[width=1.0\textwidth]{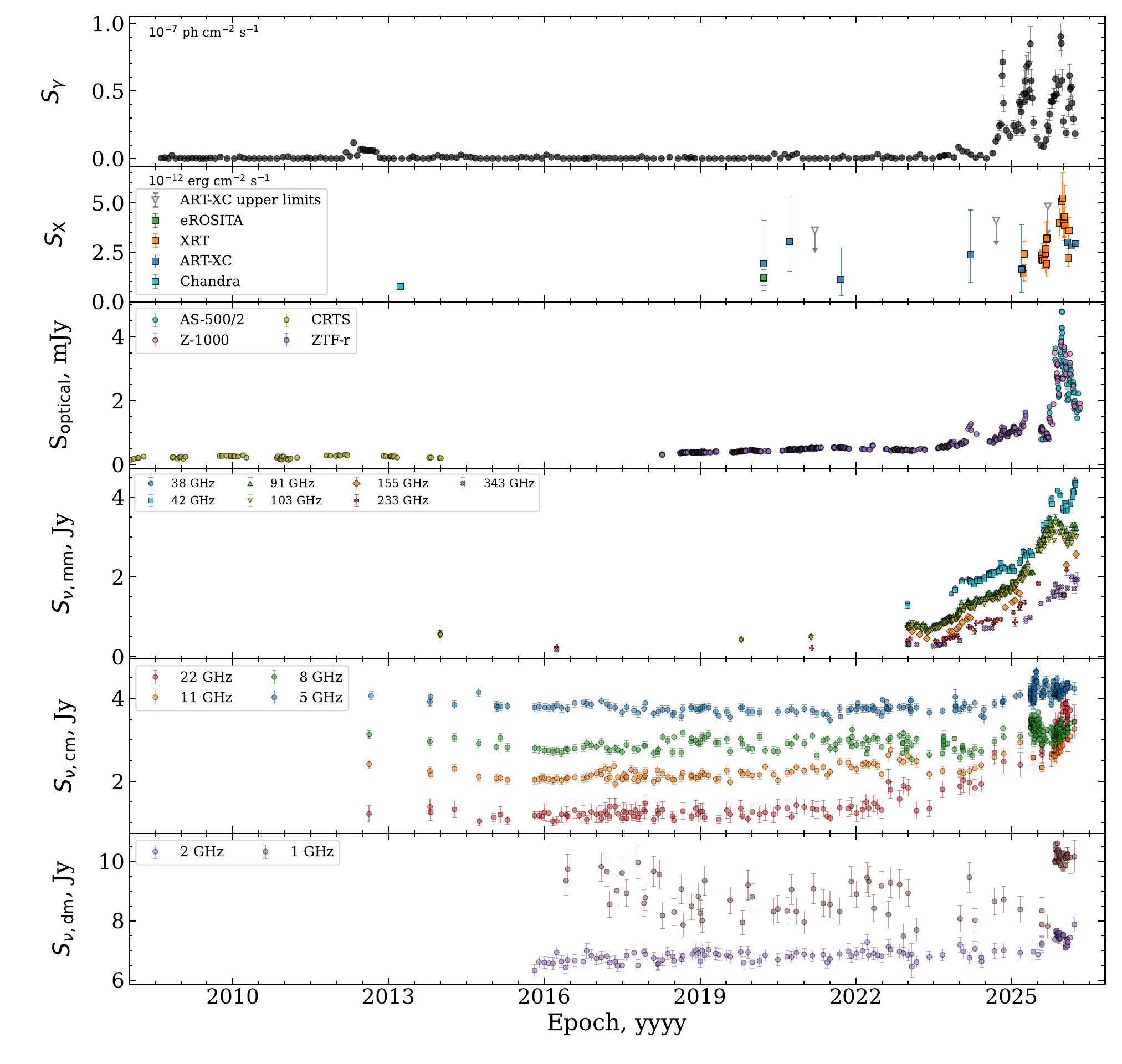}}
\caption{Long-term multiwavelength light curves of \sou over 2008--2026. Top to bottom: {\it Fermi}-LAT $\gamma$-ray photon flux in the \mbox{0.1--300}~GeV band; X-ray flux in the 5--10 keV band from Chandra, eROSITA, \textit{Swift}/XRT, and SRG/ART-XC, including the \mbox{ART-XC} upper limits; optical $R$-band flux density from CRTS, ZTF, and the SAO RAS AS-500/2 and Zeiss-1000 telescopes; millimetre and submillimetre flux densities at 38--343 GHz from ALMA; centimetre-band radio flux densities at 5--22 GHz and 1--2 GHz from \mbox{RATAN-600} and RT-32. The centimetre-band data are separated into two panels to make the frequency-dependent long-term radio evolution more clearly visible.}
\label{fig:light}
\end{figure*}

Figure~\ref{fig:light} shows the multiband light curves for the period over 2008--2026, while Fig.~\ref{fig:light2} presents a more detailed view for the enhanced activity period in 2024--2026. The shaded region in Fig.~\ref{fig:light2} corresponds to the highest observed fluxes during the analysed interval in the $\gamma$-ray, \mbox{X-ray}, and optical bands, which occurred within $\sim 12$ d from 2025.945 to 2025.979.

\subsection{Radio}

\sou has been regularly monitored with the RATAN-600 radio telescope of the Special Astrophysical Observatory of the Russian Academy of Sciences (SAO RAS) as a secondary calibrator. Due to the ring geometry of RATAN-600, the observations were made quasi-simultaneously at six frequencies: 1.2, 2.3, 4.7, 8.2, 11.2, and 22.3 GHz \citep{1993IAPM...35....7P,1979S&T....57..324K,2020gbar.conf...32S}. The data reduction was accomplished using the automated data reduction system \citep{1999A&AS..139..545K,2011AstBu..66..109T,2016AstBu..71..496U,2018AstBu..73..494T} and the Flexible Astronomical Data Processing System (\textsc{FADPS}) standard package modules \citep{1997ASPC..125...46V} for the broadband RATAN-600 continuum radiometers. The RATAN-600 antenna and radiometer parameters are described by \cite{2026MNRAS.547ag333S}. The measurements were calibrated using gain curves constructed from observations of the secondary standard sources PKS\,1151$-$34, 3C\,48, 3C\,147, 3C\,161, 3C\,286, 3C\,295, 3C\,309.1, 3C\,249.1, and NGC\,7027, adopting their standard flux densities; corrections for polarization and source size were taken from \citealt{1977A&A....61...99B,1980A&AS...39..379T, 1994A&A...284..331O,2013ApJS..204...19P,2017ApJS..230....7P}. 
The radio brightening trend has persisted since August 2022 with the flux density growth at higher frequencies (11--22 GHz) preceding that at lower frequencies (5--8 GHz), while the fractional response at 2 GHz remains comparatively weak.

Observations of \sou with the three RT-32 radio telescopes of the Institute of Applied Astronomy of the Russian Academy of Sciences (IAA RAS) have been carried out systematically at the Svetloe, Zelenchukskaya, and Badary radio astronomical observatories \citep{2019..VLBI..Quasar} at two frequencies: 5.05 and 8.63 GHz. Regular experiments at both frequencies started at Svetloe
in August 2021 and have been complemented by dense sessions at Zelenchukskaya (5.05 GHz) and Badary (8.63 GHz) since May 2025. The observations are performed in drift-scan mode and processed with the original software package {\tt CV} \citep{kharinov2012} and the Database of Radiometric Observations. The RT-32 flux density calibration is performed similarly to that of RATAN-600.

In this paper we denote the RATAN-600 and RT-32 frequencies by their rounded values: 1, 2, 5, 8, 11, and 22 GHz. The RATAN-600 and RT-32 flux densities 
$S_{\nu}$ at 1--22 GHz, their errors $\sigma$, and average observing epochs (yyyy.mm.dd, MJD, and yyyy.yyyy) are presented in Table~\ref{tab:TableA1_part} (a fragment). The radio light curves and the evolution of broadband radio spectra are shown in Figs.~\ref{fig:light} and~\ref{fig:radio_spectra}.

\begin{figure*}
\centerline{\includegraphics[width=0.9\textwidth]{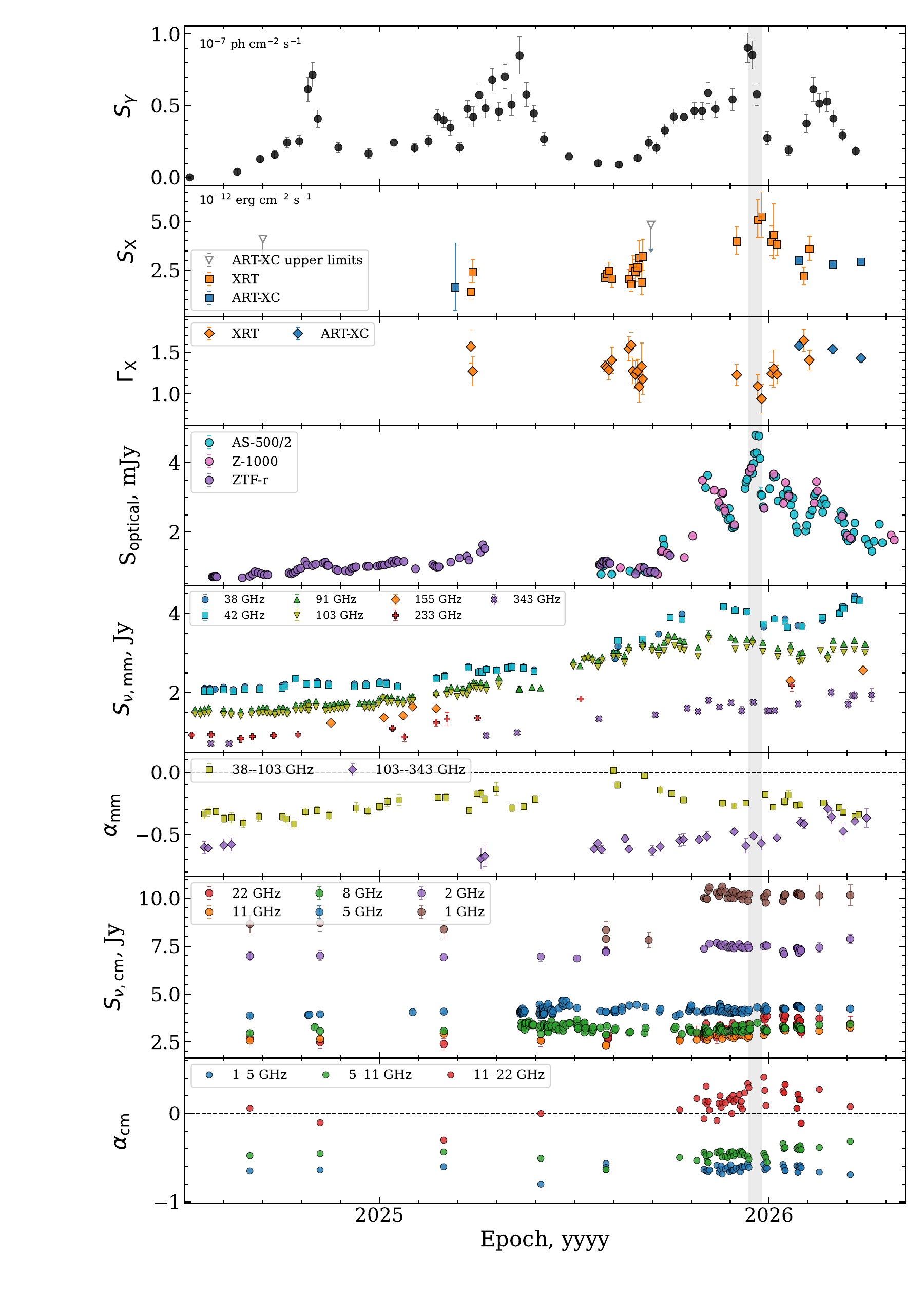}}
\caption{$\gamma$-ray, X-ray, optical, and multifrequency radio light curves of \sou in 2024--2026. The second panel shows the X-ray flux in the 5--10~keV band, while the third panel displays the X-ray photon index $\Gamma_{\rm X}$ derived from power-law spectrum fitting. The additional radio/mm spectral-index panels show the centimetre-band indices at 1--5, 5--11, and 11--22~GHz, alongside the ALMA indices at 38--103 and 103--343~GHz, computed from quasi-simultaneous measurements within a 14-day window. The shaded region marks the $\simeq 12$-day interval spanning the epochs of the flux density maximum in the $\gamma$-ray, X-ray, and optical bands.} 
\label{fig:light2} 
\end{figure*}

\begin{figure}
\centerline{\includegraphics[width=0.5\textwidth]{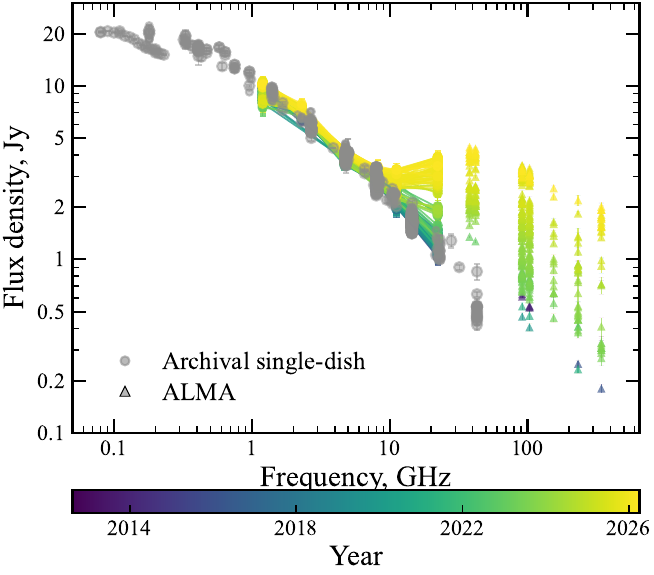}}
\caption{Radio spectra of \sou from RATAN-600 and RT-32. Coloured curves represent individual observing epochs from 2011 to 2026, encoded by the colourbar. Grey symbols show archival literature data from single-dish telescopes compiled in CATS \citep{2009DatSJ...8...34V}. ALMA measurements are shown by triangles and are not connected to the single-dish spectra because the interferometric and single-dish observations have different spatial sensitivities.}
\label{fig:radio_spectra} 
\end{figure}

The millimetre-band flux densities were obtained from the public ALMA Calibrator Source Catalogue\footnote{\url{https://almascience.eso.org/alma-data/calibrator-catalogue}} \citep{2014Msngr.155...19F}. We extracted all available measurements of \sou between 38 and 343 GHz obtained during 2012--2026. Since \sou is a partially resolved source at the ALMA angular resolution, the measured flux densities may depend to some extent on the array configuration and baseline coverage. Nevertheless, the ALMA data provide valuable information on the long-term variability and spectral evolution of the source at millimetre and submillimetre wavelengths. The corresponding light curves are shown in Figs.~\ref{fig:light} and ~\ref{fig:light2}.

\begin{table*}
\caption{A fragment of the RATAN-600 and RT-32 measurements of \sou in 2012--2026: epochs in yyyy.mm.dd (Col.~1), Modified Julian Date (MJD) (Col.~2), epoch in yyyy.yyyy (Col.~3), flux densities at 22, 11, 8, 5, 2, and 1~GHz along with their errors in Jy (Cols.~4--15), and the telescope used (Col.~16). The full version is available as online supplementary material.} 
\begin{tabular}{|l|c|c|c|c|c|c|c|c|c|c|c|c|c|c|c|c|}
\hline
yyyy.mm.dd  & MJD &  yyyy.yyyy & $S_{22}$ & $\sigma$ & $S_{11}$ & $\sigma$ & $S_{8}$ & $\sigma$ & $S_{5}$ & $\sigma$ & $S_{2}$ & $\sigma$ & $S_{1}$ & $\sigma$  & Telescope \\
(1) & (2) & (3) & (4) & (5) & (6) & (7) & (8) & (9) & (10) & (11) & (12) & (13) & (14) & (15) & (16) \\
\hline
2025.08.09 & 60896 & 2025.6033	& --	& --	& --	& --	& --	& --	& 4.06	& 0.07	& -- & --	& -- & -- & RT-32 \\
2025.09.07 & 60925 & 2025.6828	& -- & -- & -- & --	& 3.24	& 0.06	& --	& --	& -- & -- & -- & -- & RT-32\\
2026.01.27 & 61067 & 2026.0712  & 3.45 & 0.23 & 3.31 & 0.06 & 3.27 & 0.02 & 4.37 & 0.07 & 7.39 & 0.07 & 10.24 & 0.09 & R-600 \\
2026.01.28 & 61068 & 2026.0740	& 3.76	& 0.22	& 3.25	& 0.06	& 3.35	& 0.02	& 4.36	& 0.07	& 7.38	& 0.05 & 10.21 & 0.25 & R-600\\
2026.01.29 & 61069 & 2026.0767	& 3.56	& 0.21	& 3.20	& 0.06	& 3.30	& 0.02	& 4.37	& 0.07	& 7.16	& 0.06 & 10.18 & 0.08 & R-600\\
\hline
\end{tabular}
\label{tab:TableA1_part}
\end{table*}

\begin{table}
\caption{\label{TableA2_part} 
A fragment of the SAO RAS $R$-band measurements of \sou in 2025--2026: epoch in yyyy.mm.dd (Col.~1), Modified Julian Date (MJD) (Col.~2), epoch in yyyy.yyyy (Col.~3), flux density and its error in mJy (Cols.~4 and~5), and the telescope designation (Col.~6). The full version is available as online supplementary material.} 
\begin{tabular}{|c|c|c|c|c|c|}
\hline
yyyy.mm.dd & MJD  &  yyyy.yyyy & $R_{\rm flux}$ & $\sigma$ & Telescope \\
(1) & (2) & (3) & (4) & (5) & (6) \\
\hline
2025.08.31 & 60919 & 2025.6645 & 0.895 & 0.033 & AS-500/2\\
2025.09.01 & 60920 & 2025.6673 & 0.966 & 0.018 & Z-1000 \\
2025.09.01 & 60920 & 2025.6673 & 0.947 & 0.028 & AS-500/2\\
2025.09.04 & 60923 & 2025.6753 & 0.988 & 0.015 & AS-500/2\\

\hline
\end{tabular}
\end{table}

\subsection{Optical photometry and spectroscopy}
\label{sec:optical}

The optical $R$-band data for \sou, covering the observing period from 2005 to 2026, have been collected from the following instruments: 
the Catalina Real-time Transient Survey \citep[CRTS;][]{2009ApJ...696..870D} (epochs between 2005 and 2014), the Zwicky Transient Facility \citep[ZTF;][]{2019PASP..131a8002B}, which has been operating since 2018, and the SAO RAS telescopes. The optical monitoring of \sou in the $R$ band has been performed since July 2025 with the 1-m Zeiss-1000 and 0.5-m AS-500/2 reflectors. The description of the SAO RAS facilities is presented in \cite{2024MNRAS.535.2775V}.

All optical data were averaged over individual nights, the CRTS and ZTF measurements were converted into the $R$~band, and the resulting magnitudes were transformed into flux densities using the calibration constant from \citet{1990A&AS...83..183M}. The correction for Galactic extinction was applied, adopting the extinction parameter A$_R$ = 0.632 \citep{2011ApJ...737..103S} according to NED\footnote{\url{https://ned.ipac.caltech.edu}} data. The $R$ band flux densities $R_{\rm flux}$, their errors $\sigma$, and average observing epochs are presented in Table~\ref{TableA2_part} (a fragment).

According to the CRTS and ZTF archive data, the optical brightness of \sou varied slowly between 0.25 and 0.5 mJy from 2005 until late 2023. Unfortunately, we did not find any reliable estimates of \sou optical flux for the epochs between 2014 and 2017. In March and October 2024, the source underwent two optical flares, each reaching approximately 1.0~mJy. After the quasar emerged from solar conjunction in July 2025, its behaviour changed drastically: starting at the same brightness level as observed in October 2024, the source brightened with some variations to $\sim4.7$~mJy by 20 December 2025, as was reported by \cite{2025ATel17540....1V}. After reaching the maximum, the flux density gradually declined to level around 1.5 mJy, remaining above the historical quiescent level, with additional subflares in January and February 2026.

The optical spectroscopy of \sou with prime-focus multimodal SCORPIO-2 \citep{2011BaltA..20..363A} at the \mbox{6-metre} telescope (BTA) was carried out under Director's Discretionary Time (DDT) on November 28/29, 2025. The weather conditions during the observations were good: almost ideal transparency and the seeing (measured as the full width at half maximum of stellar spectra in individual exposures) was about $1\farcs5$. We used a $1\farcs0$--wide slit in combination with the VPHG1026@735 volume phase holographic (VPH) grating from the \mbox{SCORPIO-2} standard set\!\footnote{\url{https://www.sao.ru/hq/lsfvo/devices/scorpio/scorpio.html}}, which provided a spectral resolution of about 6~\AA\ across the 5800--9500~\AA\ wavelength range. 

The slit of SCORPIO-2 was placed across the \sou image in the north-to-south direction. An observation comprised four 5-min exposures. The spectra were processed with the software package described by \cite{1993BSAO...36..107V}. The spectral shape was corrected using the observations of spectrophotometric standards from \cite{1988ApJ...328..315M}.

\subsection{X-rays}
\label{sec:xray}

A few days after the {\it Fermi}-LAT $\gamma$-ray flare on March~23, 2025 \citep{2025ATel17107....1M}, a series of short ToO X-ray observations was conducted with the X-ray telescope (XRT) \citep{2005SSRv..120..165B} aboard the Neil Gehrels {\it Swift} observatory \citep{2004ApJ...611.1005G}. On March 27 and~29, \sou was found at flux levels of $(3.1\pm0.1)\times10^{-12}$ and $(3.9\pm0.2)\times10^{-12}$~erg\,cm$^{-2}$\,s$^{-1}$ (0.4--10 keV energy band, corrected for Galactic absorption), respectively \citep{2025ATel17142....1G}, significantly higher than during the previous {\it Chandra} observation performed in 2013 \citep{2016ApJS..224...40W}. Simultaneously, UV observations were carried out with {\it Swift}/UVOT, which yielded apparent magnitudes of ${\rm UVW2~(1928~\AA)} =18.20\pm0.07$ on March 27 and ${\rm UVW1~(2600~\AA)} =17.52\pm0.05$ on March 29. No previous observations in these bands were found.

Since then, \sou has been frequently monitored in X-rays with {\it Swift}/XRT (as well as with UVOT in the UV band), with a total of 23 short (with a typical exposure of half an hour) observations conducted between March 2025 and February 2026. Following the reports on the increased activity of the quasar in the radio, optical, and $\gamma$-ray bands at the end of 2025, we decided to perform a number of long (nearly 24-hour) X-ray pointed observations, which would also cover a broader energy range from 4 to 30 keV, using the ART-XC telescope \citep{2021A&A...650A..42P} aboard the {\it SRG} observatory \citep{2021A&A...656A.132S}. The first observation took place on January 28--29, 2026, and the source was found \citep{2026ATel17645....1U} at an X-ray flux level that is a factor of $\sim 1.5$ higher than during the XRT March 2025 observations. Two more ART-XC observations were conducted on March 1--2 and March 27--28, 2026.
 
In order to explore the X-ray flux and spectral evolution of \sou during its flaring activity in 2025--2026, we have analysed in a uniform way all available XRT and \mbox{ART-XC} observations, excluding one XRT observation with an exposure shorter than 250 s. Details of the spectral analysis are presented in Section~\ref{sec:xrayspec} below. In short, individual spectra can be described by power laws modified by the Galactic interstellar absorption in the direction of the quasar \mbox{$N_{\rm H}^{\rm MW} = 2.0 \times 10^{21}~\mathrm{cm}^{-2}$} \citep{2016A&A...594A.116H}, assuming solar abundances from \citet{1989GeCoA..53..197A}. The photon index $\Gamma_{\rm X}$ demonstrates significant variations from 0.9 to 1.6 between observations (see the third panel in Fig.~\ref{fig:light2}). From the spectral analysis we also infer X-ray fluxes in the 5--10~keV energy band, which is common to XRT and ART-XC, corrected for Galactic absorption (the effect is less than 0.4 per cent in this energy band).

Prior to the series of ART-XC pointed observations in early 2026, \sou had crossed the ART-XC field of view eight times during the all-sky survey performed by the {\it SRG} observatory in 2019--2022 and 2023--2025. The first four of these snapshot (20--30 seconds, vignetting corrected) observations took place between March 22, 2020, and September 18, 2021, at six-month intervals, while the subsequent four occurred between March 16, 2024, and September 12, 2025, also with a six-month cadence. The quasar was not detected with a significance greater than $2\sigma$ during any of these visits, nor was it confidently detected in the summed maps of either the first two or the first five ART-XC all-sky surveys \citep{2022A&A...661A..38P,2024A&A...687A.183S}. Nevertheless, these data allow us to obtain (following the method described in \citealt{2025MNRAS.540.3170U}) rough estimates of upper limits on the X-ray flux in eight epochs between early 2020 and early 2025. The \mbox{X-ray} flux was converted from the 4--12 keV band used in the ART-XC all-sky survey to the 5--10 keV band assuming a power law with $\Gamma_{\rm X}=1.5$ (the typical value for the ART-XC and XRT pointed observations) and correcting for Galactic absorption.

Another X-ray flux measurement was made by {\it SRG}/eROSITA \citep{2021A&A...647A...1P} on March 22, 2020, yielding $(9\pm3)\times10^{-13}$~erg\,cm$^{-2}$\,s$^{-1}$ in the 2.3--5 keV energy band \citep{2024A&A...682A..34M}. We converted this value to the 5--10~keV band assuming a power-law spectrum with $\Gamma_{\rm X}=1.5$. Finally, {\it Chandra}/ACIS earlier measured a flux of $2.4\times10^{-12}$~erg\,cm$^{-2}$\,s$^{-1}$ in the 0.3--8~keV energy band on March 22, 2013 \citep{2016ApJS..224...40W}. We converted this flux to 5--10~keV assuming $\Gamma_{\rm X}=1.7$ (as in the original paper) and estimated its uncertainty by propagating the relative error into the net count rate and including a typical 15 per cent systematic uncertainty associated with the energy-to-flux conversion factor. 

All information on the X-ray fluxes and spectral slopes is collected in Table~\ref{tab:xray}, while Figs.~\ref{fig:light} and \ref{fig:light2} show how these quantities evolved with time. 




\begin{table*}
    \centering
    \caption{X-ray fluxes and spectral measurements.}
    \renewcommand{\arraystretch}{1.2}
    \begin{tabular}{|r|c|c|c|c|c|}
    \hline
    Date & Exposure & Flux (5--10 keV) & $\Gamma_{\rm X}$ & Wstat(dof) & Telescope\\
    & s & 10$^{-12}$ erg cm$^{-2}$ s$^{-1}$ & \\
    \hline
22.03.2013 & 2002 & $0.77_{-0.12}^{+0.12}$ & & & Chandra \\
    22.03.2020 & 81 & $1.2_{-0.4}^{+0.4}$ & & & eROSITA \\
    22.03.2020 & 24 & $1.9_{-1.4}^{+2.2}$ &  &  & ART-XC\\
22.09.2020 & 29 & $3.0_{-1.5}^{+2.2}$ &  &  & ART-XC\\
19.03.2021 & 26 & $<3.6$ &  &  & ART-XC\\
18.09.2021 & 29 & $1.1_{-0.8}^{+1.6}$ &  &  & ART-XC\\
16.03.2024 & 23 & $2.4_{-1.4}^{+2.3}$ &  &  & ART-XC\\
13.09.2024 & 22 & $<4.1$ &  &  & ART-XC\\
13.03.2025 & 20 & $1.6_{-1.2}^{+2.3}$ &  &  & ART-XC\\
27.03.2025 & 1441 & $1.4_{-0.4}^{+0.5}$ & $1.6_{-0.2}^{+0.2}$ & 59.3(73) & XRT\\
29.03.2025 & 1379 & $2.4_{-0.5}^{+0.7}$ & $1.27_{-0.18}^{+0.18}$ & 61.1(84) & XRT\\
31.07--1.08.2025 & 9784 & $2.14_{-0.19}^{+0.21}$ & $1.34_{-0.07}^{+0.07}$ & 276.2(354) & XRT\\
1--2.08.2025 & 10586 & $2.3_{-0.2}^{+0.2}$ & $1.31_{-0.06}^{+0.06}$ & 292.2(367) & XRT\\
3--4.08.2025 & 3074 & $2.5_{-0.4}^{+0.4}$ & $1.29_{-0.12}^{+0.12}$ & 130.9(156) & XRT\\
6.08.2025 & 1673 & $2.1_{-0.4}^{+0.5}$ & $1.41_{-0.16}^{+0.16}$ & 101.4(92) & XRT\\
22.08.2025 & 1633 & $2.1_{-0.4}^{+0.5}$ & $1.55_{-0.15}^{+0.15}$ & 96.9(112) & XRT\\
24.08.2025 & 1813 & $1.8_{-0.4}^{+0.4}$ & $1.59_{-0.15}^{+0.15}$ & 86(109) & XRT\\
26.08.2025 & 1698 & $2.6_{-0.5}^{+0.6}$ & $1.28_{-0.15}^{+0.15}$ & 81.9(100) & XRT\\
28.08.2025 & 1696 & $2.5_{-0.5}^{+0.6}$ & $1.23_{-0.16}^{+0.16}$ & 91(85) & XRT\\
30--31.08.2025 & 1905 & $2.7_{-0.5}^{+0.6}$ & $1.27_{-0.14}^{+0.14}$ & 86.9(115) & XRT\\
1.09.2025 & 1661 & $3.1_{-0.7}^{+0.9}$ & $1.09_{-0.19}^{+0.19}$ & 55.2(75) & XRT\\
3.09.2025 & 694 & $1.9_{-0.6}^{+0.9}$ & $1.3_{-0.3}^{+0.3}$ & 53.9(34) & XRT\\
4.09.2025 & 1041 & $3.2_{-0.7}^{+0.9}$ & $1.18_{-0.19}^{+0.18}$ & 52.3(77) & XRT\\
12.09.2025 & 19 & $<4.82$ &  &  & ART-XC\\
1.12.2025 & 1478 & $4.0_{-0.7}^{+0.8}$ & $1.23_{-0.13}^{+0.13}$ & 112(134) & XRT\\
21.12.2025 & 1194 & $5.1_{-0.9}^{+1.0}$ & $1.09_{-0.14}^{+0.14}$ & 105.2(116) & XRT\\
24.12.2025 & 941 & $5.2_{-1.1}^{+1.3}$ & $0.94_{-0.17}^{+0.17}$ & 85.4(80) & XRT\\
3.01.2026 & 1311 & $4.0_{-0.7}^{+0.8}$ & $1.24_{-0.14}^{+0.14}$ & 100.9(117) & XRT\\
5.01.2026 & 489 & $4.3_{-1.2}^{+1.6}$ & $1.3_{-0.2}^{+0.2}$ & 67.2(56) & XRT\\
8.01.2026 & 2035 & $3.8_{-0.6}^{+0.6}$ & $1.23_{-0.11}^{+0.11}$ & 138.3(159) & XRT\\
28--29.01.2026 & 83980 & $3.00_{-0.04}^{+0.04}$ & $1.58_{-0.05}^{+0.05}$ & 128.1(123) & ART-XC\\
2.02.2026 & 1648 & $2.2_{-0.4}^{+0.5}$ & $1.65_{-0.13}^{+0.14}$ & 120.8(120) & XRT\\
7.02.2026 & 1631 & $3.6_{-0.6}^{+0.7}$ & $1.41_{-0.12}^{+0.12}$ & 126.3(152) & XRT\\
1--2.03.2026 & 82974 & $2.81_{-0.04}^{+0.04}$ & $1.54_{-0.05}^{+0.05}$ & 94.4(123) & ART-XC\\
27--28.03.2026 & 85485 & $2.94_{-0.05}^{+0.05}$ & $1.43_{-0.05}^{+0.05}$ & 125.1(123) & ART-XC\\
    \hline
    \end{tabular}  
    \label{tab:xray}
\begin{flushleft}
Fluxes are corrected for Galactic absorption; model parameters are provided for pointed observations only. Uncertainties and upper limits are given at the 1$\sigma$ and 2$\sigma$ confidence levels,
respectively. 
\end{flushleft}
\end{table*}

Based on all the available, albeit sparse, X-ray flux measurements, we can tentatively conclude that \sou experienced an X-ray outburst that started in the first half of 2025 and began to decay by late 2025, with the peak flux exceeding the pre-flare levels by a factor of $> 3$.

\subsection{$\gamma$-rays}
\label{sec:gamma}

The Large Area Telescope (LAT), one of the two instruments onboard the {\it Fermi} Gamma-ray Space Telescope, 
observed enhanced $\gamma$-ray activity from a source positionally consistent with the compact steep-spectrum quasar \sou (also known as 4FGL\,J0521.2$+$1637) in late 2024, throughout 2025, and at the beginning of 2026 \citep{2024ATel16845....1B,2025ATel17107....1M,2025ATel17180....1W,2025ATel17461....1L,2026ATel17681....1R}.

For quantitative variability analysis, we constructed the \mbox{$\gamma$-ray} light curve (Fig.~\ref{fig:light}) using the Pass 8 raw data \citep{2009ApJ...697.1071A} from August 4, 2008, to March~29, 2026, and the FermiTools v.2.2.0 software package. Following the standard LAT data reduction procedure, we set the recommended event class (${\rm evclass}=128$), conversion type (${\rm evtype}=3$), a zenith angle cut of $90^\circ$ to eliminate Earth limb contamination, and the energy range of \mbox{0.1--300}~GeV.

We applied the adaptive binning approach \citep{Lott12}, setting the target constant relative flux uncertainty to 25 per cent in each bin. The radio source \sou is positionally associated with the $\gamma$-ray source 4FGL\,J0521.2$+$1637 detected by {\it Fermi}-LAT, with a 1.39$^\prime$ angular separation between the VLBI and $\gamma$-ray coordinates, which is within the {\it Fermi} 95 per cent position confidence ellipse of 1.84$^\prime$$\times$1.54$^\prime$ \citep{4FGL-DR4,RFC}. The source energy spectrum is described by a single power law with a fixed photon index of 2.23. 

\begin{figure*}
\centerline{\includegraphics[width=0.95\textwidth]{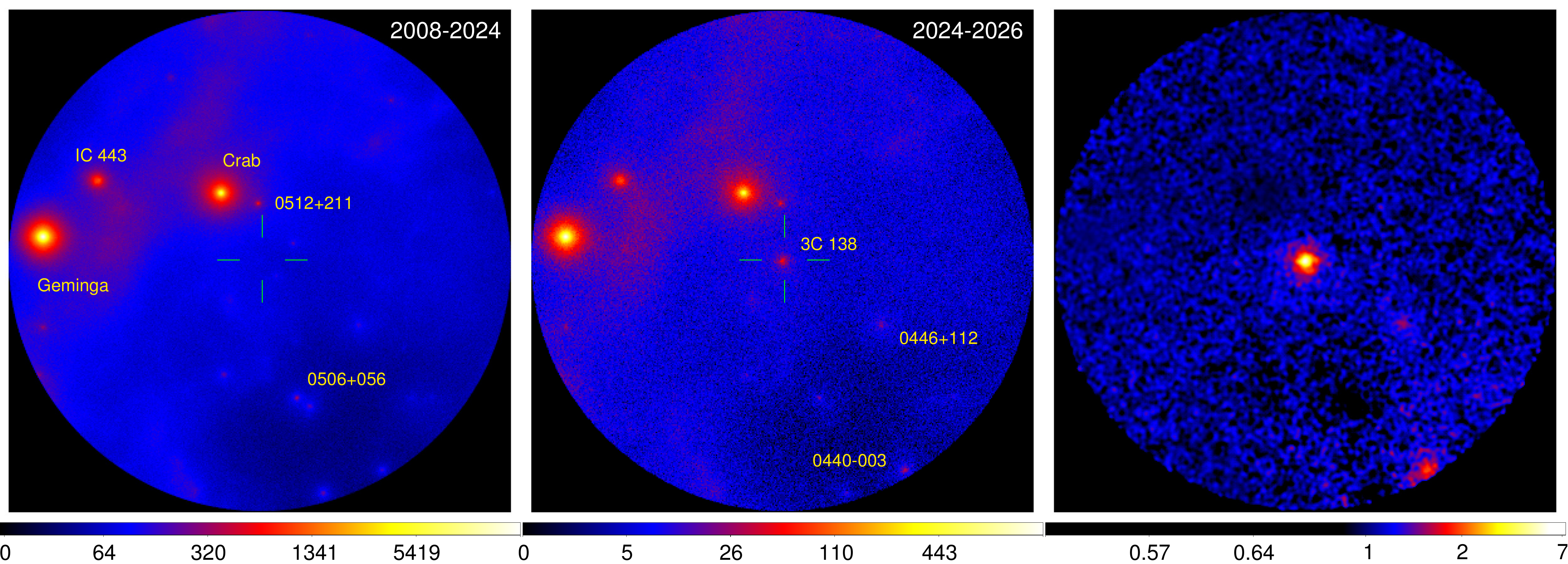}}
\caption{{\it Fermi}-LAT $\gamma$-ray counts maps (0.1--300~GeV) centred on \sou, shown in celestial coordinates using the Aitoff projection. Each map covers a $20^\circ$-radius region with a pixel size of $0\fdg1\times0\fdg1$. Left: the low state of \sou from 2008-08-04 to 2024-09-14, the presumed source position is indicated by cyan crosshairs. Middle: flaring activity period from 2024-09-14 to 2026-03-29. Right: an exposure-weighted count rate ratio (high state / low state) which suppresses steady sources and highlights variable objects. Gaussian smoothing with a kernel $\sigma=1\fdg5$ was applied.}
\label{fig:fermi_maps}
\end{figure*}

The region of interest (ROI) was set as a circle with a radius of $20^\circ$ centred on the target source. The ROI includes 135 catalogued point sources and 2 diffuse sources (supernova remnants S\,147 and IC\,443) along with Galactic and isotropic background diffuse components. In particular, the ROI encompasses very bright $\gamma$-ray emitters, the Crab and Geminga pulsars, and a well-known neutrino source, the quasar TXS\,0506$+$056. The emission from the Crab Pulsar and its nebula, located at an angular distance of $6\fdg2$ from the target, was taken into account to derive more accurate fluxes. Fig.~\ref{fig:fermi_maps} shows the counts maps for the low state and the flaring activity period, along with an image  highlighting the transient behaviour of the source, computed as the ratio of the high-state to low-state count rate maps, each derived by dividing the respective counts map by the corresponding exposure distribution.

\section{Radio spectra}
\subsection{Spectral indices}
\label{sec:indices}

Radio spectral indices were calculated assuming a power-law spectrum:
\begin{equation}
S_{\nu} \propto \nu^{\alpha},
\end{equation}
and were fitted for each epoch with sufficient multifrequency coverage using a linear relation in the log--log space:
\begin{equation}
\ln S_{\nu} = a + \alpha \ln \nu .
\end{equation}
For a given frequency range (low: 1--5 GHz, mid: 5--11 GHz, and high: 11--22 GHz), the slope $\alpha$ was estimated via weighted least squares in $(x,y)=(\ln\nu,\ln S_{\nu})$ with the weights
\begin{equation}
w_i = \frac{1}{\sigma_{\ln S,i}^{2}},
\qquad
\sigma_{\ln S,i} \simeq \frac{\sigma_{S,i}}{S_i},
\end{equation}
where $\sigma_{S,i}$ is the measurement uncertainty of the flux density at a frequency $\nu_i$.

In the summary presented in Table~\ref{tab:indices}, we report the statistics of annual distributions: median spectral indices and 16th--84th percentile ranges, which characterise the scatter within a given year.

The ALMA spectral indices were calculated from measurements grouped within a 14-day window. For the 38--103 GHz index, we used the median fluxes of the low-frequency group at 38/42 GHz and the high-frequency group at 91/103 GHz. For the 103--343 GHz index, we used quasi-simultaneous measurements at 103 and 343 GHz within the same 14-day window.

The radio spectrum has hardened significantly in \mbox{2024--2026} (Fig.~\ref{fig:radio_spectra}): the high-frequency spectral index (\mbox{11--22}~GHz) has evolved from steep ($\alpha \sim -0.8$) to flat or inverted ($\alpha \gtrsim 0.2$), while the low- and mid-frequency indices exhibit a more gradual evolution (Table~\ref{tab:indices} and Fig.~\ref{fig:light2}, bottom panel). The ALMA spectral indices reveal a long-term trend similar to that observed at centimetre wavelengths. Both the 38--103 GHz and 103--343 GHz spectra gradually flatten after 2022, indicating an increasing contribution from a newly emerging compact synchrotron component. The flattening is strongest between 38 and 103 GHz, while the 103--343 GHz spectrum remains optically thin throughout the observed period, suggesting that the turnover frequency of the new component lies below the highest ALMA bands. Unlike the \mbox{11--22}~GHz spectrum, however, the millimetre spectrum remains optically thin, with spectral indices remaining negative even during the peak of the activity.

\begin{table*}
\centering
\caption{Annual median radio spectral indices of \sou. 
The uncertainties correspond to the 16th--84th percentile ranges of the spectral index distributions within each year.}
\begin{tabular}{ccllll}
\hline
Year & $\tilde{\alpha}_{\rm 1-5}$ & $\tilde{\alpha}_{\rm 5-11}$ & $\tilde{\alpha}_{\rm 11-22}$ & $\tilde{\alpha}_{\rm 38-103}$ & $\tilde{\alpha}_{\rm 103-343}$ \\
\hline
2012 &  & $-0.84$ & $-1.02$ & & \\
2013 &  & $-0.66_{-0.04}^{+0.04}$ & $-0.81$ & &  \\
2014 &  & $-0.63_{-0.06}^{+0.06}$ & $-0.82$ & & \\
2015 & $-0.75_{-0.02}^{+0.02}$ & $-0.65_{-0.01}^{+0.03}$ & $-0.83_{-0.07}^{+0.01}$ & &  \\
2016 & $-0.78_{-0.04}^{+0.04}$ & $-0.67_{-0.13}^{+0.04}$ & $-0.80_{-0.20}^{+0.04}$ & & \\
2017 & $-0.69_{-0.08}^{+0.04}$ & $-0.63_{-0.08}^{+0.02}$ & $-0.84_{-0.08}^{+0.14}$ & & \\
2018 & $-0.76_{-0.11}^{+0.09}$ & $-0.58_{-0.03}^{+0.04}$ & $-0.82_{-0.10}^{+0.06}$ & & \\
2019 & $-0.75_{-0.13}^{+0.11}$ & $-0.56_{-0.02}^{+0.02}$ & $-0.82_{-0.15}^{+0.06}$ & & \\
2020 & $-0.69_{-0.04}^{+0.01}$ & $-0.55_{-0.09}^{+0.05}$ & $-0.82_{-0.10}^{+0.09}$ & & \\
2021 & $-0.73_{-0.12}^{+0.05}$ & $-0.48_{-0.06}^{+0.02}$ & $-0.83_{-0.06}^{+0.06}$ & & \\
2022 & $-0.71_{-0.11}^{+0.04}$ & $-0.47_{-0.04}^{+0.06}$ & $-0.74_{-0.13}^{+0.19}$ & $-0.59$ & $-0.71$ \\
2023 & $-0.77_{-0.09}^{+0.13}$ & $-0.51_{-0.08}^{+0.08}$ & $-0.62_{-0.23}^{+0.20}$ & $-0.57_{-0.59}^{+0.55}$ & $-0.77_{-0.79}^{+0.71}$\\
2024 & $-0.67_{-0.18}^{+0.03}$ & $-0.59_{-0.02}^{+0.12}$ & $-0.20_{-0.12}^{+0.10}$ & $-0.37_{-0.41}^{+0.32}$ & $-0.60_{-0.79}^{+0.58}$ \\
2025 & $-0.62_{-0.04}^{+0.03}$ & $-0.48_{-0.05}^{+0.04}$ & $+0.14_{-0.10}^{+0.10}$ & $-0.22_{-0.27}^{+0.14}$ & $-0.57_{-0.62}^{+0.52}$ \\
2026 & $-0.61_{-0.06}^{+0.02}$ & $-0.39_{-0.02}^{+0.02}$ & $+0.16_{-0.10}^{+0.11}$ & $-0.28_{-0.33}^{+0.24}$ & $-0.40_{-0.47}^{+0.36}$ \\
\hline
\end{tabular}
\label{tab:indices}
\end{table*}

\subsection{Evolution of the excess compact synchrotron emission}

\begin{figure}
\centerline{\includegraphics[width=0.5\textwidth]{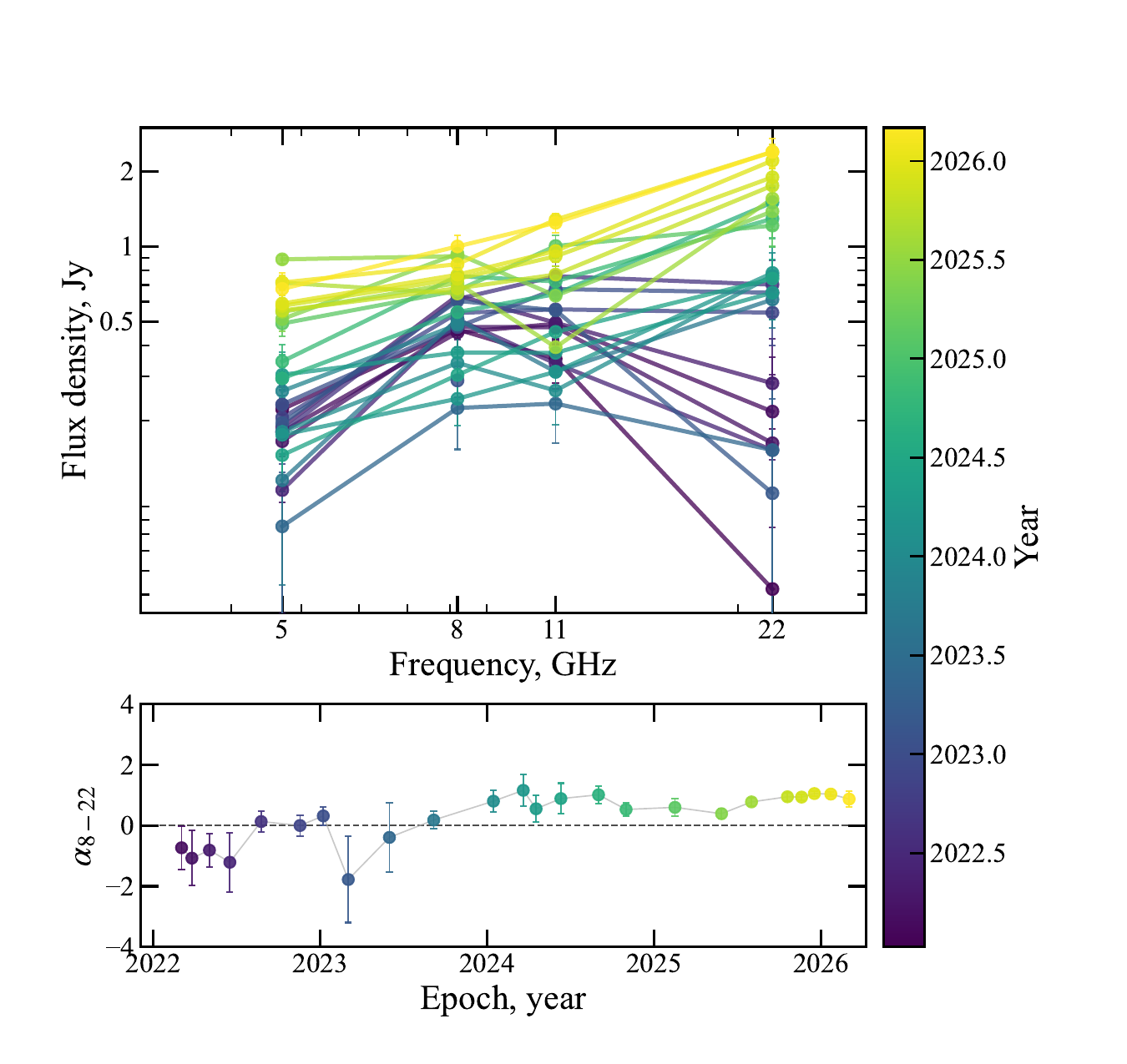}}
\caption{Quasi-simultaneous radio spectra of the excess component in \sou (top) and the corresponding evolution of its 8--22 GHz spectral index (bottom).}
\label{fig:active_spec}
\end{figure}

We extracted the radio spectrum of the newborn compact component as follows: $S_{\rm comp}=S_{\rm total}-\tilde S_{\rm quiet}$, where $\tilde S_{\rm quiet}$ is the averaged flux density in the pre-2022 epoch. The continuum radio spectrum of \sou at 1--22 GHz in the quiet state (pre-2022 epoch) is well fitted by a power law with a spectral index $\alpha\approx-0.7$. The 32 quasi-simultaneous radio spectra of the flaring component, averaged over 30-d windows, are shown in Fig~\ref{fig:active_spec}. Their shape evolves from a peaked spectrum before September 2023 (epoch 2023.680) to a flatter or inverted spectrum afterwards. 

We estimated the magnetic field in the synchrotron self-absorption (SSA) region for the peaked radio spectra as $B_{\rm SSA}\approx 10^{-5}\,b(\alpha_{\rm thin})\,\nu_{\rm m}^{5}\,\theta_{\rm m}^{4}\,S_{\rm m}^{-2}\,\delta\,(1+z)^{-1}$ (e.g., \citealt{1983ApJ...264..296M}), where $b(\alpha_{\rm thin})$ is a dimensionless parameter depending on the optically thin spectral index $\alpha_{\rm thin}$ \citep{2019MNRAS.482.2336P}, $\nu_{\rm m}$ is the peak frequency, $\theta_{\rm m}$ is the angular source size at this frequency, and $S_{\rm m}$ is the flux density at $\nu_{\rm m}$.  
The parameters $\alpha_{\rm thin}$, $\nu_{\rm m}$, and $S_{\rm m}$ were estimated by fitting the SSA model of \citet{1999A&A...349...45T} to the peaked radio spectra. We used the \mbox{\textsc{UltraNest}}\footnote{\url{https://johannesbuchner.github.io/UltraNest/}} package \citep{2021JOSS....6.3001B} for spectral modelling and obtaining uncertainties of these parameters from their posterior distributions.
The angular size cannot be derived from single-dish spectra. The flaring component was identified as the radio core from VLBI observations \citep{2026A&A...710A..63L}. The dependence $\theta \sim \nu^{-1}$ is common for radio cores in AGNs (e.~g. \citealt{1979ApJ...232...34B,2015MNRAS.452.4274P,2022MNRAS.515.1736K}). We therefore scaled $\theta_{\rm m}$ from the 33~GHz SSA-region diameter, $\theta_{33}=0.07$~mas reported by \citet{2026A&A...710A..63L}, as \mbox{$\theta_{\rm m}=\theta_{33}(33\,{\rm GHz}/\nu_{\rm m})$}. We adopted the Doppler factor \mbox{$\delta\approx2.4$} from \citet{2026A&A...710A..63L} as a lower limit.

As a result, we obtained a magnetic field $B_{\rm SSA}=67\pm29$~mG and a minimum-energy field in an equipartition state \mbox{$B_{\rm me}=38\pm3$}~mG \citep{1980ARA&A..18..165M,2005ApJ...622..797K} for the epochs before September~2023 (mean values and standard deviations over the corresponding epochs are quoted here). The two numbers are of the same order of magnitude, but this does not demonstrate equipartition. The uncertainty budget is dominated by the assumed source size, the unresolved nature of the excess spectrum, and the poorly constrained Doppler factor. If our estimates of angular size of the SSA region are correct, then the ratio $B_{\rm SSA} / B_{\rm me} \approx 1.8$ means magnetic field energy density dominating over particle energy density before September 2023. This conclusion is strengthened by the fact that $B_{\rm SSA} / B_{\rm me} \sim \delta^{12/7}$ (this follows from \citealt{1980ARA&A..18..165M,1983ApJ...264..296M,2005ApJ...622..797K}) and we used the lower limit for $\delta$. Moreover, the source-specific VLBI/VLA/ALMA analysis of \citet{2026A&A...710A..63L} finds \mbox{$B_{\rm SSA}=10.1^{+9.6}_{-5.1}$}~mG, \mbox{$B_{\rm me}=200.8^{+44.1}_{-33.8}$}~mG, and \mbox{$B_{\rm SSA}/B_{\rm me}\simeq0.05$} for December 12, 2024, implying a particle-dominated core (i.~e. particle energy density prevails over magnetic field energy density). Such an evolution (a drop in $B_{\rm SSA}$ and an increase in $B_{\rm me}$) is consistent with conversion of magnetic field energy into particle energy, for example through magnetic reconnection processes. We emphasize that this result is based on model-dependent magnetic field estimates from different studies and this interpretation requires confirmation with additional high-resolution observations. We therefore use the single-dish excess spectra primarily as a tracer of opacity evolution rather than a precise measurement of the core energy partition.
 
\section{Optical spectrum}

\begin{figure}
\includegraphics[width=\linewidth]{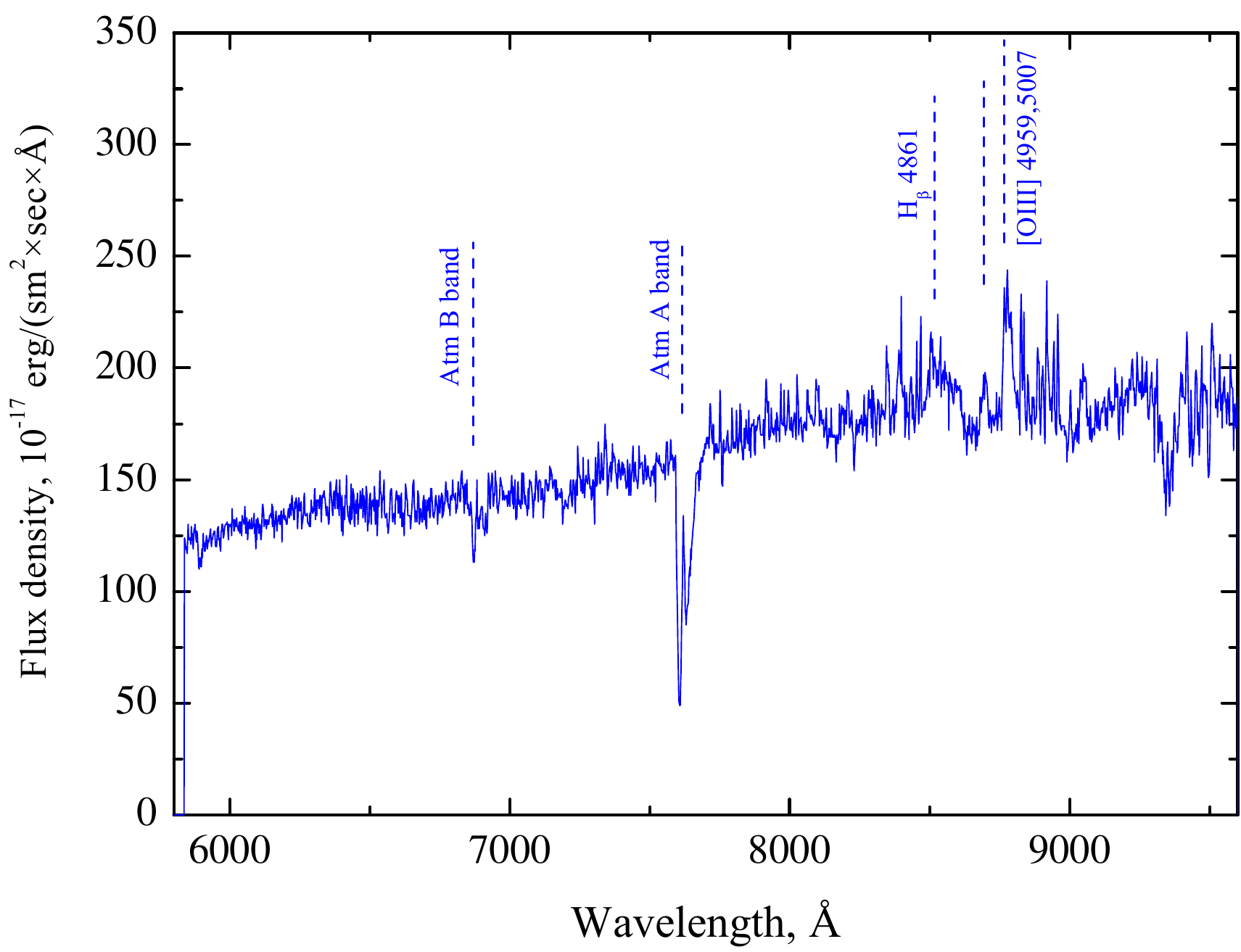}
\caption{
The optical spectrum of \sou in the red~-- near infrared range of 5800--9500~\AA. The vertical lines mark the most prominent features: the [O\,III] and H$_{\beta}$ emission lines along with the atmospheric A and B bands.}
\label{fig:fig4}
\end{figure}

The optical spectrum of \sou (Fig.~\ref{fig:fig4}), taken during the phase of maximum activity, reveals only a few spectral features, namely the prominent emission lines typical of an AGN: [O\,III] $\lambda\lambda$\,4959, 5007~\AA\ and H$_\beta$. All other features are strong telluric absorptions. The main component of the \sou spectrum is the featureless non-thermal power-law continuum. The spectrum is very similar to that with a better signal-to-noise ratio obtained in \citet{2023A&A...671...A32} with MUSE at the VLT when the object was in a quiescent state. The VLT spectrum demonstrates the presence of prominent AGN emission lines and a non-thermal continuum without any signatures of the host galaxy's stellar population.

\begin{figure}
\includegraphics[width=\linewidth]{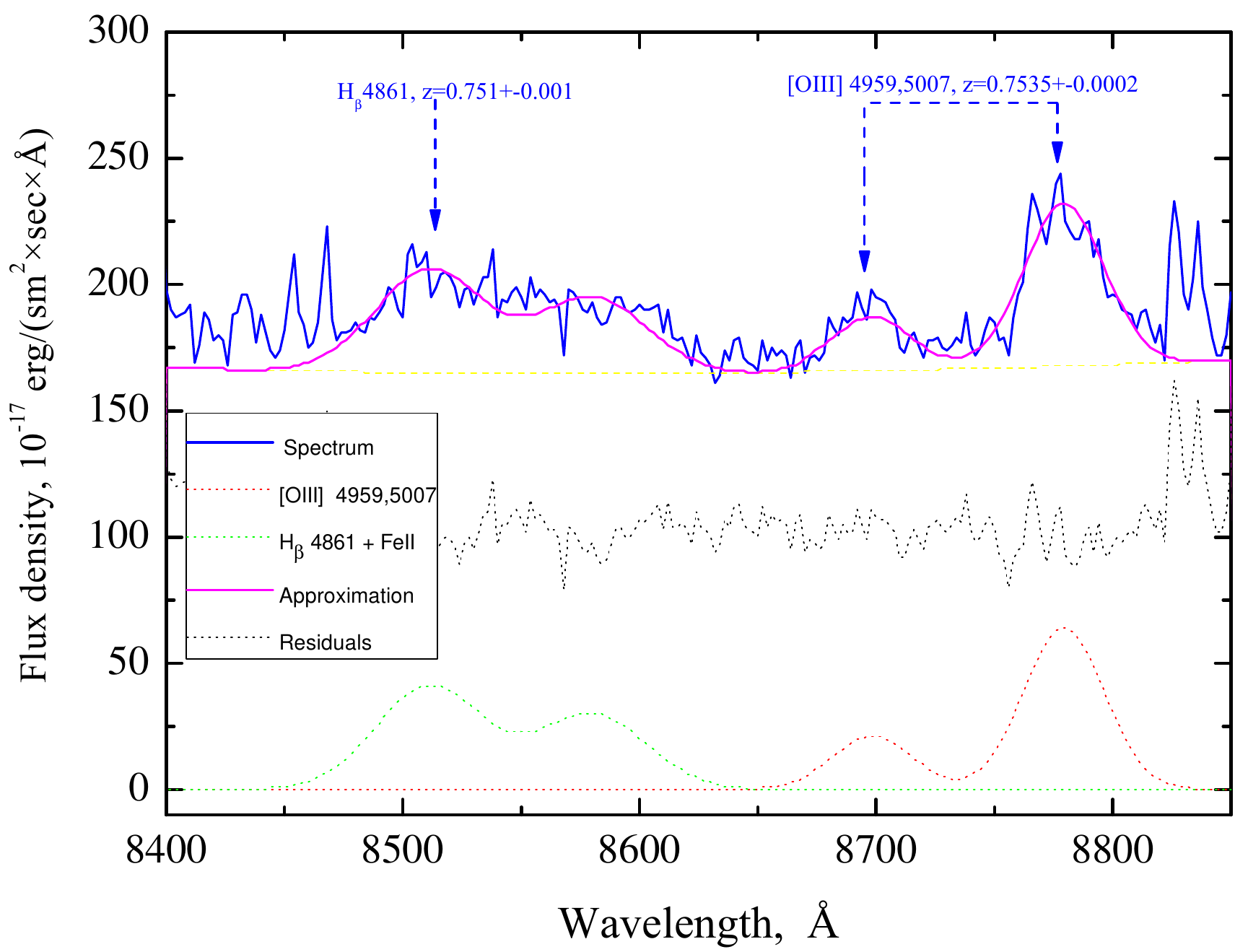}
\caption{A fragment of the \sou spectrum between 8400 and 8900~\AA, showing the fitting of the [O\,III] and H$_{\beta}$ emission lines. The observed spectrum, fitted spectrum, continuum level, residuals, and individual components are presented.}
\label{fig:fig5}
\end{figure}

We approximated the emission lines using Gaussian profiles in the continuum-subtracted spectra, as shown in Fig.~\ref{fig:fig5}.The [O\,III] $\lambda\lambda$\,4959, 5007 lines were fitted using ordinary Gaussian profiles with equal widths and an amplitude ratio of 3 to 1. The H$_{\beta}$ line profile was presented by a combination of Gaussian profiles for the main line and the Fe\,II blend, which was noted in \citet{2023A&A...671...A32}. After correction for the instrumental profile width, the analysis yields full widths at half maximum of 2500~km\,s$^{-1}$ for the broad H$_{\beta}$ component and about 900 km\,s$^{-1}$ for the [O\,III] and Fe\,II blend. The equivalent widths of the [O\,III] and H$_{\beta}$ emission lines slightly exceed 10~\AA, in the observed frame, corresponding to approximately 5.7~\AA~in the source rest frame. This value is not far from the classic limit of 5~\AA, separating the subpopulations of blazars -  BL\,Lac type and flat-spectrum radio quasars (FSRQ) \citep{1991ApJ...374..431S, 1995PASP..107..803U}.

The relatively small equivalent widths are consistent with substantial dilution of the emission lines by the enhanced non-thermal continuum during the active state.

Using data from the Sloan Digital Sky Survey (SDSS; \citealt{2000AJ....120.1579P}), \citet{2022MNRAS.511.214P} provided a detailed study of the optical spectra for a sample of 84 CSS and GPS (Gigahertz-Peaked Spectrum) sources along with 21 Megahertz-Peaked Spectrum (MPS) objects, all located at redshifts $<1$. A careful inspection of these spectra allowed them to estimate the fractions of the stellar and non-thermal components. Most of the studied sources exhibited a significant fraction of the stellar population, but 21 CSS/GPS sources showed a dominating featureless continuum with barely detectable stellar absorption features. Moreover, strong emission lines were observed in their spectra, some of them displaying broad components indicating the presence of an AGN.  Consequently, we conclude that \sou also belongs to this extensive class of CSS/GPS sources with AGN features.

\section{X-ray spectra}
\label{sec:xrayspec}

The SRG/ART-XC and Swift/XRT spectra of the source from pointed observations were extracted following the procedures described in Sections~2.2 and 2.3 of \citet{2025MNRAS.540.3170U}, respectively. The Swift/XRT spectra were filtered below 0.3~keV and grouped to ensure at least one count per energy bin in the source spectrum. The resulting upper energy limit varies between 6.38 and 9.96~keV, depending on the number of counts in the highest-energy channels. The SRG/ART-XC data were grouped with a minimum of five counts per bin in the source spectrum and analysed in the 5--30\,keV energy range. 

\begin{figure}
    \centerline{
    \includegraphics[width=0.5\textwidth]{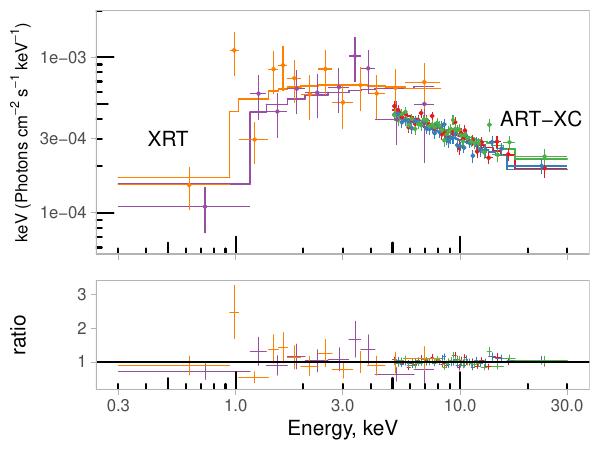}
    }
\caption{
The upper panel shows the unfolded ($F_E$) X-ray spectra of \sou measured with Swift/XRT at the peak of the outburst on 21 and 24 December 2025 (orange and purple, respectively) and afterwards with SRG/ART-XC on 28--29 January, 1--2 March, and 27--28 March 2026 (red, blue, and green, respectively), all fitted by power laws modified by Galactic absorption. The spectra have been rebinned in the plot for better clarity. The bottom panel shows the data-to-model ratio. Uncertainties are shown at the 1$\sigma$ confidence level.
}
\label{fig:xray-spec}
\end{figure}

We analysed each XRT or ART-XC pointed observation individually using \textsc{XSPEC} (version 12.15.1, \citealt{1996ASPC..101...17A}). The quality of the fit was evaluated using the W-statistic (\texttt{cstat} option in \textsc{XSPEC}). We fitted the spectra with a power law modified by photoabsorption in the Galaxy:
\begin{equation}
\text{\textsc{tbabs}} \times \text{\textsc{cflux}} \times \text{\textsc{zpowerlaw}}.
\label{eq:pl}
\end{equation}
Here, \textsc{tbabs} is the Galactic interstellar absorption model \citep{2000ApJ...542..914W},
and \textsc{zpowerlaw} characterises the power-law continuum with two free parameters: photon index $\Gamma_{\rm X}$ and absorption-corrected flux in the observed 5--10\,keV energy band (\textsc{cflux}), with the quasar's redshift being fixed.

All the X-ray spectra are adequately described by this simple model. Adding absorption in excess of the Galactic value does not improve the fits ($p>0.05$), i.e., no intrinsic absorption is required. We also tested the spectra for the presence of a high-energy exponential cutoff in the power-law spectrum of the pointed ART-XC observations and found no evidence for it. We obtained 2$\sigma$ lower limits on the rest-frame cutoff energy of $E_\mathrm{cut} > 29$~keV, $> 49$~keV, and $> 40$~keV for the observations on 28--29 January, 1--2 March, and 27--28 March 2026, respectively.

Figure~\ref{fig:xray-spec} shows the spectra measured by XRT at the peak of the X-ray outburst and by ART-XC at three later epochs, fitted by power laws. Although the different energy ranges of XRT and ART-XC complicate the comparison, it is clear that the spectrum steepened after the 5--10~keV flux reached its maximum. The best-fitting parameters for all studied XRT and ART-XC spectra are presented in Table~\ref{tab:xray}. The third panel in Fig.~\ref{fig:light2} shows the evolution of the X-ray photon index.

\section{Analysis of the light curves}
\subsection{Radio band trends}
\label{sec:trends}

\begin{figure}
\centerline{\includegraphics[width=0.49\textwidth]{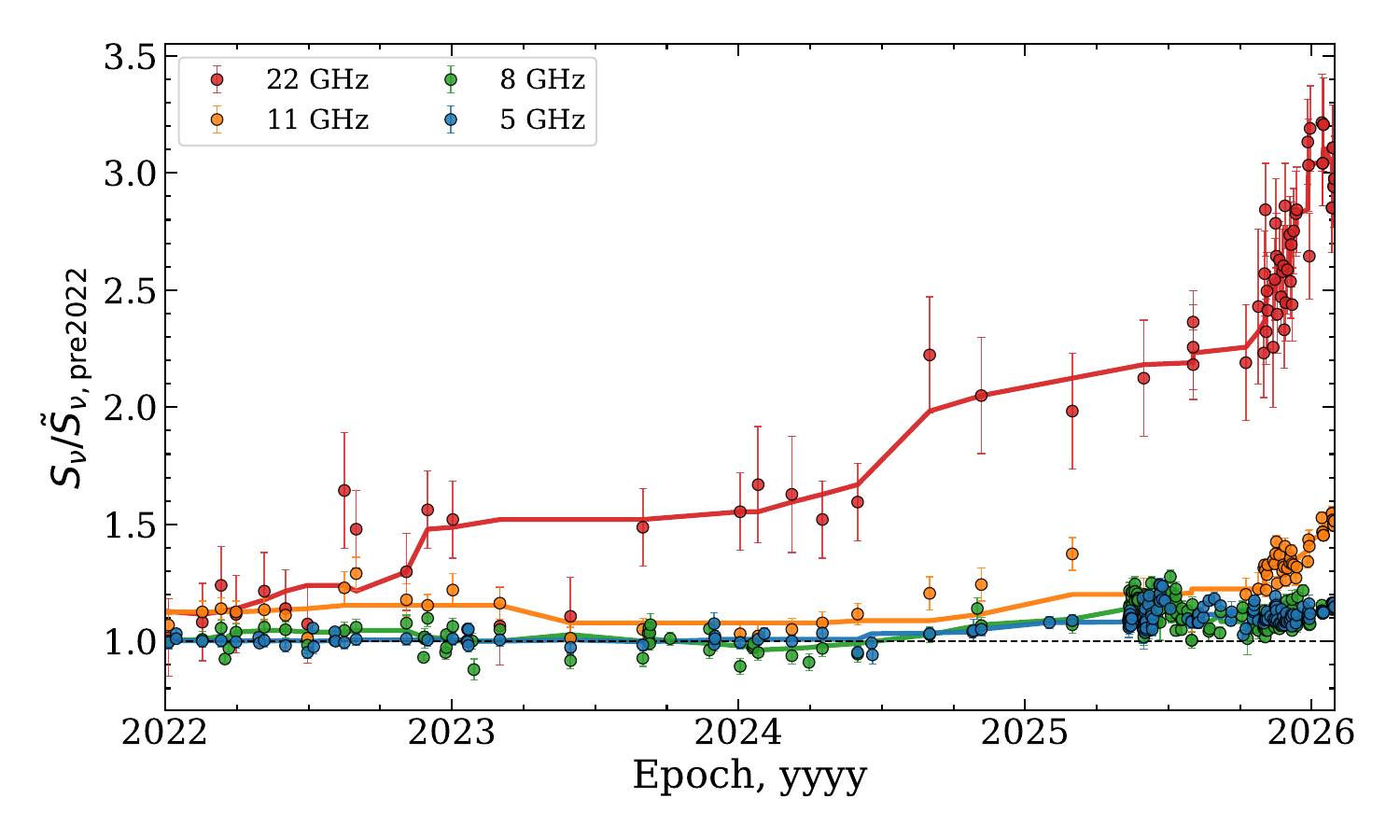}}
\caption{Multifrequency radio light curves of \sou normalized to the pre-2022 flux level. A clear frequency-dependent brightening is observed after 2022, with the strongest increase at 22 GHz.}
\label{fig:trends}
\end{figure}

Multifrequency radio observations show that among the centimetre-band light curves, the strongest and still ongoing flux increase occurs at high frequencies (11--22 GHz), while lower-frequency bands respond more slowly (Fig.~\ref{fig:light}). The radio flux density exhibits a clear transition after 2022, with post-2022 slopes increasing from \mbox{$\sim0.05$--0.2}~Jy\,yr$^{-1}$ across most bands and reaching $\sim0.5$~Jy\,yr$^{-1}$ at 22 GHz (Fig.~\ref{fig:trends}). 

To quantify long-term trends, we fitted a linear model to each light curve,
\begin{equation}
S(t) = mt + b ,
\end{equation}
where $t$ is the epoch in years. The ordinary least-squares (OLS) estimate of the slope is
\begin{equation}
m = \frac{\sum_{i} (t_i-\bar{t})(S_i-\bar{S})}{\sum_{i} (t_i-\bar{t})^{2}},
\end{equation}
where $\bar{t}$ and $\bar{S}$ are sample means. We also computed piecewise OLS slopes separately for $t<2022$ and $t\ge 2022$ to isolate the active-phase evolution.

Long-term trends of the radio emission are listed in Table~\ref{tab:slope} and shown in Fig.~\ref{fig:trends}.

The OLS trends show that the long-term radio brightening is dominated by the post-2022 interval. In the pre-2022 epoch, the slopes are close to zero (i.e., nearly flat), whereas after 2022 they increase significantly, with the strongest rise observed at 22 GHz (post-2022 $m \approx 0.55$~Jy\,yr$^{-1}$) and progressively smaller slopes towards lower frequencies. 

An apparent exception is the 1 GHz band, where a comparatively large slope is obtained. The diagnostic analysis in Appendix~\ref{app:1GHz} shows that this behaviour is affected by uneven sampling and by the change in cadence between the historical and active-state observations. Thus, the exact onset time and the fitted slope at 1 GHz should not be interpreted as a robust inter-band delay. However, the recent rise is unlikely to be purely a sampling artefact and may represent a delayed low-frequency synchrotron response of the same evolving disturbance, becoming visible after the opacity at this frequency decreased.

The ALMA light curves indicate that the millimetre band brightening was already well developed by 2023--2024 and continued throughout 2025--2026. The simultaneous flux increase across all ALMA bands suggests that the newly emerging synchrotron component extends from centimetre to submillimetre wavelengths.

The long-term brightening rate increases strongly from centimetre to millimetre wavelengths. While the 22 GHz flux density increased at a rate of approximately 0.55~Jy\,yr$^{-1}$ after 2022, the strongest growth is observed at 38--42 GHz, reaching $\sim1.1$~Jy\,yr$^{-1}$. This behaviour indicates that the flare is most pronounced at millimetre wavelengths and is consistent with opacity-driven evolution of a compact synchrotron component.

\begin{table}
\centering
\caption{Linear trends of the radio light curves derived using ordinary least-squares fitting. The first column lists observing frequency, while subsequent columns give the long-term slope for two temporal intervals separated at 2022, corresponding to the onset of the recent activity phase. The reported uncertainties represent formal $1\sigma$ errors of the fitted slopes.}
\begin{tabular}{|l|l|l|}
\hline
Band  & Pre 2022 & Post 2022 \\
(GHz) & (Jy yr$^{-1}$) & (Jy yr$^{-1}$) \\
\hline
343  &  & $+$0.53 (0.02) \\
233  &  & $+$0.53 (0.04) \\
155  &  & $+$0.64 (0.04) \\
103  &  & $+$0.85 (0.02) \\
91  &  & $+$0.88 (0.02) \\
42  &  & $+$1.12 (0.05) \\
38  &  & $+$1.07 (0.05) \\
22  & $+$0.01 (0.01) & $+$0.55 (0.03) \\
11  & $+$0.02 (0.01) & $+$0.18 (0.02) \\
8   & $+$0.01 (0.01) & $+$0.11 (0.01) \\
5   & $-$0.04 (0.01) & $+$0.14 (0.01) \\
2   & $+$0.04 (0.01) & $+$0.16 (0.02) \\
1   & $-$0.17 (0.06) & $+$0.42 (0.06)$^{*}$ \\
\hline
\multicolumn{3}{c}{* -- affected by the dense cadence after 2025} \\
\end{tabular}
\label{tab:slope}
\end{table}

\subsection{$\gamma$-rays and optical flares}

To characterise the temporal evolution of flaring activity, we modelled major $\gamma$-ray and optical flares using a parametric exponential profile. For each flare, we fitted a piecewise function of the form
\begin{equation}
S(t) =
\begin{cases}
S_{0} + A \exp\left(\dfrac{t - t_{\rm pk}}{\tau_{\rm r}}\right), & t < t_{\rm pk}, \\
S_{0} + A \exp\left(-\dfrac{t - t_{\rm pk}}{\tau_{\rm d}}\right), & t \ge t_{\rm pk},
\end{cases}
\end{equation}
where $S_{0}$ denotes the baseline flux density, $A$ is the flare amplitude above the baseline, $t_{\rm pk}$ is the epoch of the flare peak, and $\tau_{\rm r}$ and $\tau_{\rm d}$ are the characteristic rise and decay time-scales, respectively.

The fitting was performed using a weighted non-linear least-squares method, with observational uncertainties used as weights when available. Initial parameter values were estimated directly from the data, with the peak epoch set to the time of the maximum observed flux density within the fitting interval, and characteristic time-scales initialized based on the temporal extent of the selected window. 

\begin{figure}
\centerline{\includegraphics[width=0.5\textwidth]{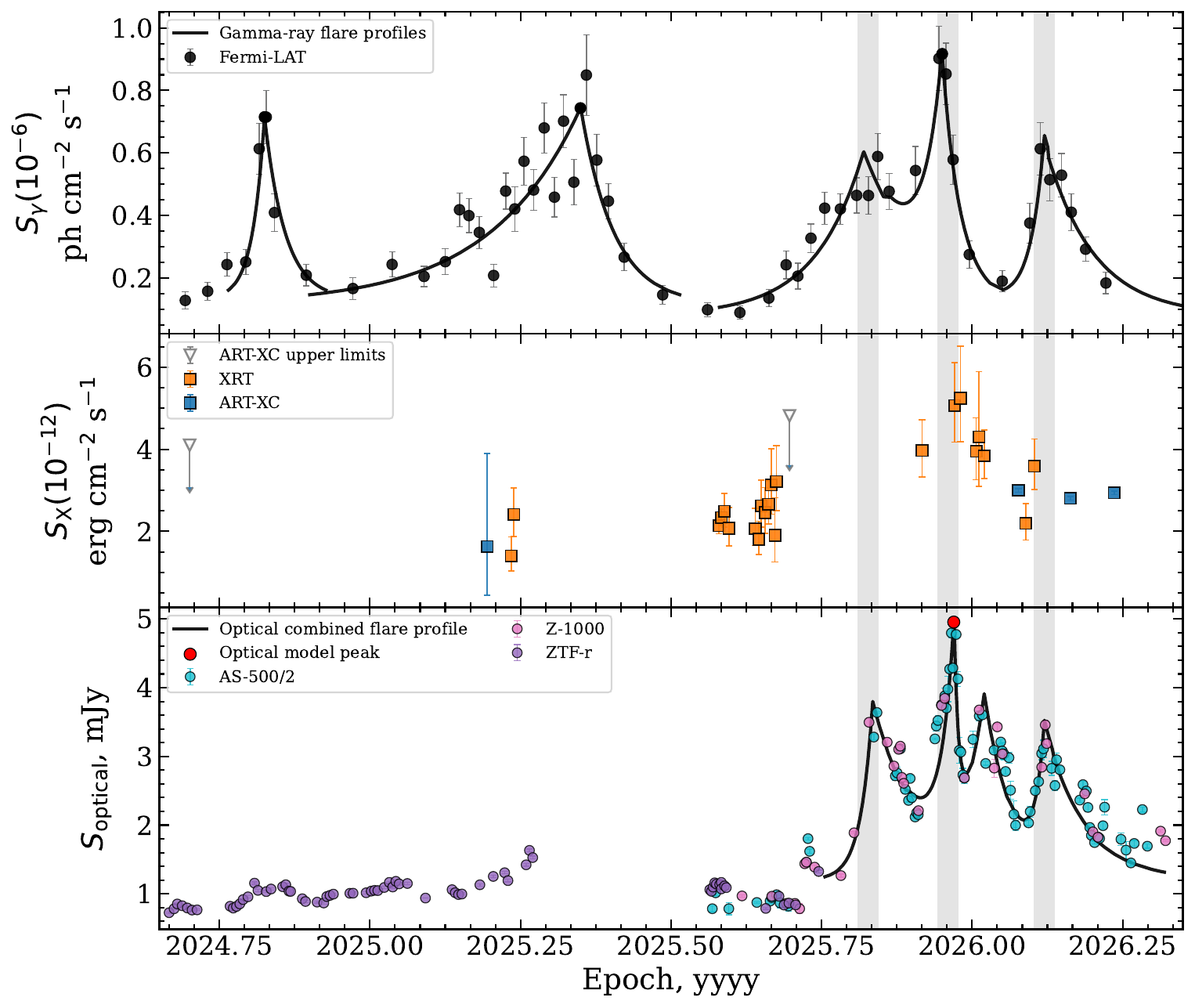}}
\caption{$\gamma$-ray (top), X-ray (middle), and optical $R$-band (bottom) light curves of \sou during the 2024--2026 active phase. The solid black curves show the exponential profiles used to characterise the variability. In the $\gamma$-ray panel, G1 and G2 are shown separately, while the late complex flare is shown as the combined G3--G5 profile. In the optical panel, the solid curve represents the combined R1--R4 model profile. The grey shaded bands mark three $\simeq12$-day intervals corresponding to the temporally matched $\gamma$-ray/optical pairs G3/R1, G4/R2, and G5/R4. The X-ray measurements are shown for comparison and indicate enhanced X-ray activity during the same broadband flaring episode.} 
\label{fig:flares}
\end{figure}

The baseline flux density $S_{0}$ was estimated from the median during the pre-activity epoch ($t < 2024$) and used as the initial value in the fitting procedure. For both the $\gamma$-ray and optical bands, the baseline was allowed to vary within a limited range during the fitting to account for possible long-term trends. For the $\gamma$-ray light curve, five major flaring components were modelled. The first two events are relatively isolated, whereas the later activity is more complex and consists of three partially blended components (G3--G5). In Fig.~\ref{fig:flares}, we show the first two $\gamma$-ray profiles separately and the G3--G5 blend as a combined profile. The parameters of the individual components are listed in Table~\ref{tab:exp}. The fitting intervals were chosen manually to include the dominant rise and decay phases of each event and to minimize contamination from neighbouring variability.

The optical variability during the 2025--2026 active phase exhibits a complex structure consisting of a broad brightening episode with superimposed shorter-time-scale subflares (Fig.~\ref{fig:flares}). We phenomenologically decomposed the optical light curve into four exponential components, R1--R4, using a standard minimization procedure over the 2025.55--2026.35 interval. In Fig.~\ref{fig:flares}, we show the combined optical model profile, i.e., the sum of the individual components. The parameters of the individual subflares are given in Table~\ref{tab:exp}.

\begin{table*}
\centering
\caption{Parameters of the exponential components used to describe the $\gamma$-ray and optical flares of \sou during 2024--2026. The columns give the component peak amplitude or peak flux density, peak epoch $t_{\rm peak}$, and characteristic rise ($\tau_{\rm r}$) and decay ($\tau_{\rm d}$) time-scales. For the $\gamma$-ray components, the flux 
density scale is given in units of $10^{-7}$~ph\,cm$^{-2}$\,s$^{-1}$; for the optical components, the values are given in mJy and refer to the individual fitted components above the adopted baseline/slowly varying level. Time-scales are given in days. The uncertainties represent formal $1\sigma$ statistical errors from the covariance matrix of the weighted non-linear least-squares fit and do not include systematic uncertainties associated with the adopted decomposition.}
\begin{tabular}{lccccc}
\hline
\multirow{2}{*}{Band} & \multirow{2}{*}{Flare} & Component amplitude & $t_{\rm peak}$ & $\tau_{\rm r}$ & $\tau_{\rm d}$ \\
 & & ($\times10^{-7}$ ph cm$^{-2}$ s$^{-1}$ / mJy) & (yyyy.yyy) & (d) & (d) \\
\hline
$\gamma$-ray & G1 & $5.9\pm0.7$ & 2024.825 & $7\pm1$ & $12\pm2$ \\
$\gamma$-ray & G2 & $7.0\pm1.0$ & 2025.350 & $44\pm5$ & $24\pm4$ \\
$\gamma$-ray & G3 & $5.3\pm0.5$ & 2025.820 & $29\pm4$ & $36\pm5$ \\
$\gamma$-ray & G4 & $7.0\pm0.7$ & 2025.950 & $11\pm2$ & $10\pm2$ \\
$\gamma$-ray & G5 & $5.5\pm0.5$ & 2026.120 & $8\pm1$ & $29\pm5$ \\
$R$ band & R1 & $2.6\pm0.1$ & 2025.835 & $7\pm1$ & $26\pm3$ \\
$R$ band & R2 & $2.8\pm0.1$ & 2025.970 & $7\pm1$ & $4\pm1$ \\
$R$ band & R3 & $2.5\pm0.1$ & 2026.020 & $13\pm2$ & $16\pm2$ \\
$R$ band & R4 & $2.0\pm0.1$ & 2026.120 & $6\pm1$ & $23\pm3$ \\
\hline
\end{tabular}
\label{tab:exp}
\end{table*}

\begin{table*}
\centering
\caption{Epochs and values of the highest observed fluxes in different energy bands for \sou during the analysed activity period. The time offsets $\Delta t$ are measured relative to the $\gamma$-ray maximum. The $\gamma$-ray, X-ray, and optical maxima occur within a $\simeq 12$-day interval, indicating a common compact activity episode.}
\begin{tabular}{lcrc}
\hline
Band & $t_{\rm peak}$ (yyyy.yyy) & $\Delta t_{\gamma}$ (d) & Peak flux \\
\hline
$\gamma$-ray (0.1--300 GeV)& 2025.945& 0& $(9\pm1)\times10^{-7}$ ph cm$^{-2}$ s$^{-1}$ \\
Optical ($R$)& 2025.966& $+$8& $4.80\pm0.07$ mJy \\
X-ray (5--10 keV)& 2025.979& $+$12& $5.2^{+1.3}_{-1.1}\times10^{-12}$ erg cm$^{-2}$ s$^{-1}$ \\
\hline
\end{tabular}
\label{tab:maximum}
\end{table*}

This approach allows us to characterise the temporal evolution of the optical emission in terms of a sequence of apparent energy-release events superposed on a longer-term brightening trend. We note that the subflares are partially blended, and therefore the derived parameters become less robust when the blending is strong.

The fitted flare time-scales do not show a strict separation between the $\gamma$-ray and optical bands. The $\gamma$-ray events span a wider range of rise durations, from about 7 to 44 days, mainly because of the more extended G2 and G3 components. In contrast, the optical subflares have more uniform rise durations of about 6--13 days. The decay durations are comparable in the two bands, although individual events differ substantially. The main difference is therefore morphological, the $\gamma$-ray light curve can be represented as a sequence of more distinct events, whereas the optical emission forms a smoother and more strongly blended sequence of subflares. The lack of sufficient optical coverage during the broad G2 event prevents us from assessing whether this $\gamma$-ray flare had an optical counterpart.

The fitted parameters in Table~\ref{tab:exp} provide the estimates of the peak flux, peak epoch, and characteristic rise and decay time-scales for each flare, which are used to compare the temporal behaviour across different energy bands. The uncertainties correspond to formal $1\sigma$ statistical errors derived from the covariance matrix of the weighted non-linear least-squares fit. They do not include systematic uncertainties associated with the adopted flare intervals, partial overlap of neighbouring events, or possible deviations of the actual flare profiles from the assumed exponential model. The fitted peak separations are of order a few days: approximately 5.4~days for G3/R1, 7.3 days for G4/R2, and close to zero for G5/R4, with positive values indicating that the optical peak occurs after the $\gamma$-ray peak. These associations are marked by the grey shaded bands in Fig.~\ref{fig:flares}. The remaining optical subflare, R3, has no obvious isolated $\gamma$-ray counterpart in the adopted decomposition.

Using the characteristic rise and decay time-scales, we estimated upper limits on the comoving size of the emitting region,
\begin{equation}
R' \lesssim \frac{c\,t_{\rm var,obs}\,\delta}{1+z},
\end{equation}
where $t_{\rm var,obs}$ is the observed characteristic rise or decay time-scale, $c$ is the speed of light, $\delta$ is the Doppler factor, and $z=0.759$ is the redshift. For the tabulated time-scales, $t_{\rm var,obs}\simeq4$--44 days, this gives
\begin{equation}
R' \lesssim 0.002-0.063~{\rm pc}
\end{equation}
for $\delta=1$--3. This corresponds to approximately $6\times10^{15}$--$1.9\times10^{17}$~cm. The range $\delta=1$--3 is used here only for an illustrative mildly beamed size estimate; $R'$ scales linearly with $\delta$, and a broader range of jet-parameter combinations is explored separately in Section~\ref{sec:sed}.

These values are physically plausible for compact emission zones in a mildly beamed jet. We note, however, that some of the fitted rise and decay durations, particularly for the later $\gamma$-ray flares and the optical subflares, are sensitive to the adopted fitting intervals and the partial overlap of neighbouring events. Therefore, the fitted parameters should be interpreted as a phenomenological characterisation of the variability rather than as uniquely determined physical cooling times. Although the X-ray sampling is too sparse for an independent flare decomposition, the available measurements provide important evidence for the broadband nature of the active state. The highest X-ray flux is observed during the same period as the strongest $\gamma$-ray and optical activity, coinciding with one of the grey-shaded intervals in Fig.~\ref{fig:flares}, as seen in Table~\ref{tab:maximum}. This near-coincidence supports an association between the X-ray emission and the same compact activity episode responsible for the high-energy and optical flaring, although the cadence does not allow us to measure a reliable X-ray lag.

\section{Variability}
\label{sec:var}

To compare relative variability amplitudes between different bands, we use the fractional variability amplitude $F_{\rm var}$, which is normalized by the mean flux density and, therefore, allows direct comparison between light curves measured in different physical units. We computed $F_{\rm var}$ separately for the pre-2024 and post-2024 intervals. The year 2024 is adopted as a common reference epoch for the onset of the pronounced broadband high-energy and optical activity; for the radio band, the long-term brightening starts earlier, around 2022, as discussed in Section~\ref{sec:trends}.  

The fractional variability amplitude $F_{\rm var}$ was computed following \citet{2003MNRAS.345.1271V}:
\begin{equation}
F_{\rm var} = \frac{\sqrt{S^{2} - \overline{\sigma_{\rm err}^{2}}}}{\bar{S}},
\end{equation}
where $S^{2}$ is the sample variance of the light curve
flux densities,
$\bar{S}$ is the mean flux density, and
$\overline{\sigma_{\rm err}^{2}}$ is the mean squared measurement uncertainty.
The uncertainty on $F_{\rm var}$ was computed as
\begin{equation}
\Delta F_{\rm var}=
\sqrt{
\left(
\sqrt{\frac{1}{2N}}
\frac{\overline{\sigma^{2}_{\rm err}}}{F_{\rm var}\,\bar{S}^{2}}
\right)^{2}
+
\left(
\sqrt{\frac{\overline{\sigma^{2}_{\rm err}}}{N}}
\frac{1}{\bar{S}}
\right)^{2}
},
\end{equation}
where $N$ is the number of measurements. These quantities (\(F_{\rm var}\) and \(\Delta F_{\rm var}\)) were evaluated using only the measurements belonging to the corresponding interval, i.e. either \(t<2024.0\) or \(t\geq2024.0\), rather than from the full light curve.

The resulting values are listed in Table~\ref{tab:fvar}. In the post-2024 interval, the radio $F_{\rm var}$ increases towards higher frequencies, from non-detected or weak variability at 1--2 GHz to $F_{\rm var}=0.15\pm0.01$ at 22 GHz. The variability amplitude continues to increase towards millimetre wavelengths, extending the trend observed in the centimetre RATAN-600 data. This behaviour is consistent with the high-frequency radio emission being dominated by a more compact and rapidly evolving component, while the lower-frequency emission is diluted by larger-scale, slowly varying regions. The optical band shows substantially stronger variability, $F_{\rm var}=0.58$, and the $\gamma$-ray band remains the most variable among the well-sampled bands. The X-ray value should be interpreted cautiously because of sparse sampling, heterogeneous instruments, and the presence of upper limits; it is therefore used only as an indicative measure of the X-ray variability amplitude (X-ray upper limits were excluded from the calculation of $F_{\rm var}$; only flux measurements were used).

\begin{table}
\centering
\caption{Fractional variability amplitudes ($F_{\rm var}$) of the optical, $\gamma$-ray, X-ray, and radio light curves for the pre-2024 and post-2024 intervals. Column $N$ gives the number of measurements used in each band in each time range. Dashes indicate the intervals where intrinsic variability is not detected at a significant level above the measurement noise.}
\begin{tabular}{|l|r|l|r|l|}
\hline
\multirow{2}{*}{Band} & \multicolumn{4}{c}{$F_{\rm var}$ ($\Delta F_{\rm var}$)} \\
 & $N$ & pre-2024 & $N$ & post-2024 \\
\hline
$\gamma$-ray & 82 & 0.93 (0.09) & 61 & 0.60 (0.02) \\
X-ray &  5 & -- & 27 & 0.21 (0.07) \\
$R$ band & 439 & 0.17 ($<0.01$) & 197 & 0.58 ($<0.01$) \\
343 GHz & 11 & 0.11 (0.03) & 25 & 0.30 (0.01) \\
233 GHz & 15 & 0.20 (0.02) & 18 & 0.37 (0.02) \\
155 GHz & 7 & 0.14 (0.02) & 11 & 0.38 (0.01) \\
103 GHz & 66 & 0.17 ($<0.01$) & 107 & 0.35 ($<0.01$) \\
91 GHz & 70 & 0.17 ($<0.01$) & 114 & 0.34 ($<0.01$) \\
42 GHz & 2 & 0.19 (0.02) & 59 & 0.30 ($<0.01$) \\
38 GHz & 3 & 0.12 (0.01) & 62 & 0.30 ($<0.01$) \\
22 GHz & 95 & -- & 61 & 0.15 (0.01) \\
11 GHz & 100 & -- & 58 & 0.09 ($<0.01$) \\
8 GHz & 112 & -- & 170 & 0.05 ($<0.01$) \\
5 GHz &  103 & -- & 158 & 0.03 ($<0.01$) \\
2 GHz & 80 & -- & 51 & -- \\
1 GHz &  48 & -- & 55 & 0.06 ($<0.01$) \\
\hline
\end{tabular}
\label{tab:fvar}
\end{table}

While $F_{\rm var}$ quantifies the fractional scatter within a given interval, the radio active phase is more appropriately characterised by the post-2022 brightening rates and by the normalized light curves shown in Fig.~\ref{fig:trends}. Because the radio evolution is gradual and remains ongoing, the epochs of the largest observed radio flux densities are strongly affected by sampling and should not be interpreted as inter-band flare delays.

\section{Two-state compact-zone SED modelling}
\label{sec:sed}

The different variability time-scales of the optical-to-$\gamma$-ray and radio emission motivate treating the compact high-energy emitting region separately from the radio-emitting structures. We therefore constructed two compact-zone spectral energy distributions (SEDs): a historical quiet-state composite and the main-peak state associated with the approximately 12-day interval containing the closely spaced $\gamma$-ray, optical, and
X-ray maxima.

The quiet-state SED covers 2013 January 1 to 2019 December 25 (MJD 56293--58842) and combines the dereddened mean $R$-band flux density, the 2013 Chandra 5--10~keV measurement, and the LAT spectrum accumulated over the same low-activity interval. The LAT spectrum provides one detection in the 3--10~GeV bin and reliable upper limits in the 1--3, 10--30, 30--100, and 100--300~GeV bins. The main-peak SED corresponds to the interval from 2025 December 9 to 2025 December 23 (MJD 61018--61032.0156) and includes the dereddened $R$-band point, the 5--10~keV X-ray point, and the LAT detections in the 0.1--0.3, 0.3--1, and 1--3~GeV bins. The LAT SED points were calculated from the local power-law normalization at the geometric-mean energy of each bin. The main-peak modelling is therefore restricted to the LAT spectrum below 3~GeV.

\begin{figure}
\centerline{\includegraphics[width=0.5\textwidth]{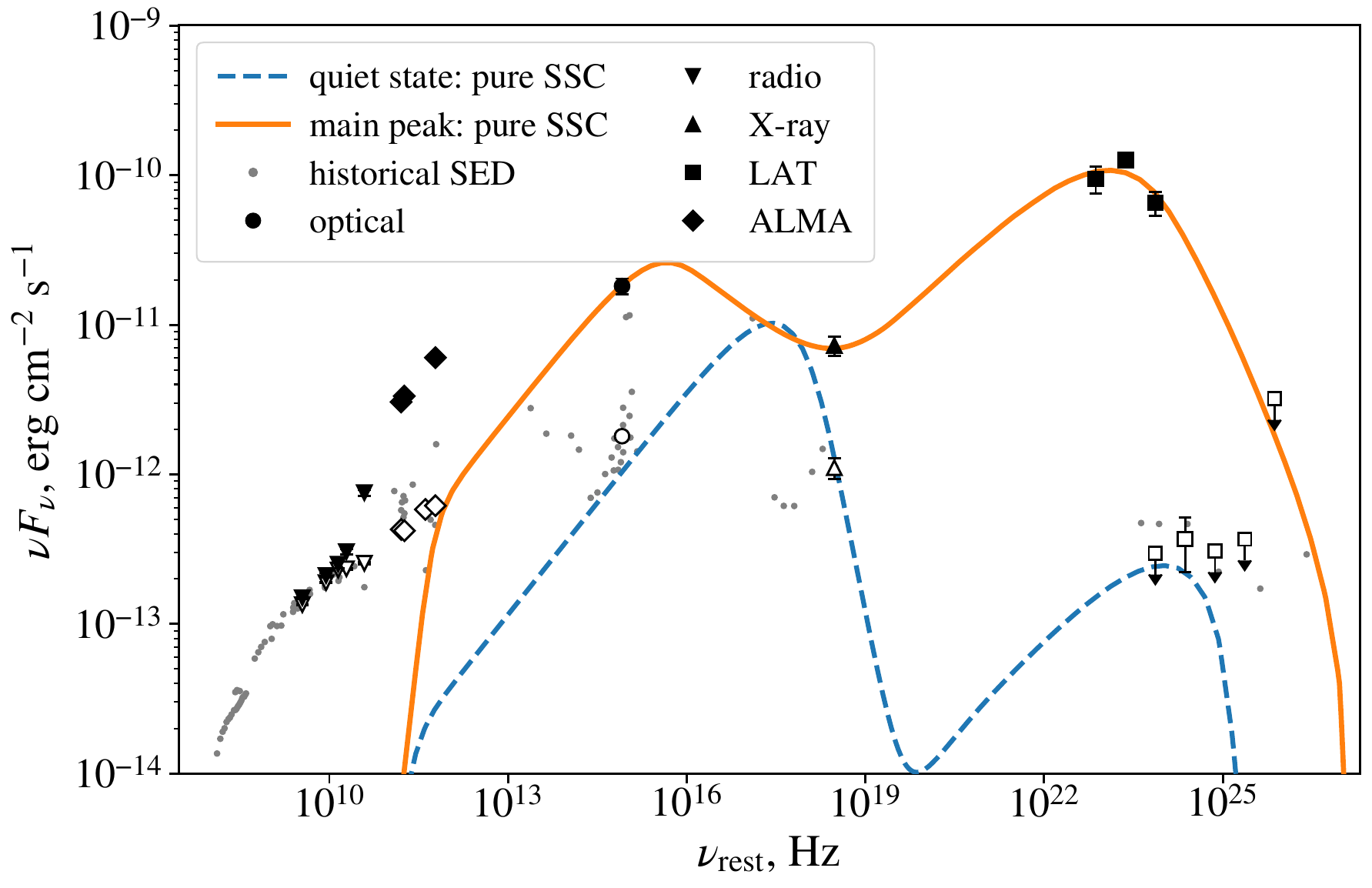}}
\caption{Representative two-state SEDs of \sou for the critical-angle case associated with $\beta_{\rm app}=3.3$: $\Gamma=\delta=3.45$ and $\theta=16.86^\circ$. This parameter combination corresponds to the minimum Lorentz factor compatible with the adopted apparent speed. Dashed and solid curves represent the total synchrotron plus SSC emission for the historical quiet state and the main peak, respectively. Filled and open symbols denote the main-peak and quiet-state measurements, respectively. Downward arrows indicate reliable quiet-state LAT upper limits. Centimetre-band radio and ALMA measurements, together with the grey historical SED points, are shown for context and were not included in the compact-zone fit. Frequencies are given in the source rest frame.}
\label{fig:sed_two_state}
\end{figure}

The centimetre-band radio and ALMA measurements were not included in the compact-zone fit. A compact optical-to-$\gamma$-ray emitting region is expected to be synchrotron self-absorbed at low radio frequencies, whereas the measured radio and millimetre flux densities contain contributions from the VLBI core, the parsec-scale jet, and the more extended CSS structure. These measurements, together with archival non-simultaneous SED points, are shown in
Fig.~\ref{fig:sed_two_state} only as observational context.

The modelling was performed with the \textsc{JetSeT} software
package \citep{2009A&A...501..879T,2011ApJ...739...66T,
2020ascl.soft09001T}. Because the available SED does not constrain the source size independently, we fixed $R=3.0\times10^{16}$~cm in all calculations as a common fiducial size of the compact emitting region. {For $\delta=3.45$ and $2.36$, this radius implies observed light-crossing times of approximately 5.9 and 8.6 d, respectively, compared with the shortest fitted optical decay time of $4\pm1$ d (Table~\ref{tab:exp}). If this decay originates in the same homogeneous region, a smaller radius would be required.}

The emitting region was represented as a homogeneous sphere containing a tangled magnetic field and a broken-power-law electron distribution. In the \textsc{JetSeT} bkn parametrization, the electron distribution can be written as
\begin{equation*}
n_{\rm e}(\gamma)=
N\,
\frac{f(\gamma)}
{\displaystyle
 \int_{\gamma_{\min}}^{\gamma_{\max}}
 f(\gamma')\,{\rm d}\gamma'}
 \end{equation*}
 \begin{equation*} 
\qquad
f(\gamma)=
\begin{cases}
\gamma^{-p}, &
\gamma_{\min}\leq\gamma<\gamma_{\rm br},\\
\gamma_{\rm br}^{\,p_1-p}\gamma^{-p_1}, &
\gamma_{\rm br}\leq\gamma\leq\gamma_{\max}.
\end{cases}
\end{equation*}
Here, $N$ is the total number density of relativistic electrons; $\gamma_{\min}$, $\gamma_{\rm br}$, and $\gamma_{\max}$ are the
minimum, break, and maximum electron Lorentz factors, respectively;
and $p$ and $p_1$ are the spectral indices below and above the break. The low-energy component is produced by synchrotron emission and the
high-energy component by synchrotron self-Compton (SSC) emission. The low-energy electron index was fixed at $p=2$ in all calculations. For the main-peak models, $\gamma_{\min}=10$ was also fixed, whereas $B$, $N$, $\gamma_{\rm br}$, $\gamma_{\max}$, and $p_1$ were varied. For the quiet-state models, $\gamma_{\min}$ was allowed to vary together with these parameters.

A broad random search followed by local
resampling was carried out over the magnetic-field and electron-distribution parameters. Models were ranked with the same descriptive logarithmic-residual score in all calculations, with the
reliable quiet-state LAT upper limits treated as one-sided constraints. This score provides a descriptive comparison of the fitted models and is not a likelihood-based model-selection statistic.

The available SED does not independently constrain the bulk Lorentz factor $\Gamma_{\rm j}$, viewing angle $\theta$, or Doppler factor $\delta$ of the compact emitting region.

We therefore examined several fixed combinations of these parameters, fitting the quiet-state and main-peak SEDs separately while keeping $R$, $\Gamma_{\rm j}$, $\theta$, $\delta$, and $p$ the same in both states.

We considered four parameter combinations based on the published apparent pattern speed $\beta_{\rm app}=3.3$ of a VLBI jet component \citep{2001A&A...370...65S}. For a specified Doppler factor $\delta$, the corresponding bulk Lorentz factor and viewing angle were calculated as
\begin{equation}
\begin{aligned}
\Gamma &=
\frac{\beta_{\rm app}^{2}+\delta^{2}+1}{2\delta}, \\
\theta &=
\arctan\left(
\frac{2\beta_{\rm app}}
{\beta_{\rm app}^{2}+\delta^{2}-1}
\right).
\end{aligned}
\label{eq:gamma_theta}
\end{equation}
\citep{2009A&A...494..527H}. The four cases were:
(i) the variability Doppler factor $\delta=2.36$ reported by
\citet{2026A&A...710A..63L}; (ii) the minimum-Lorentz-factor, or
critical-angle, solution
\[
\Gamma=\delta=\sqrt{1+\beta_{\rm app}^{2}}=3.45,
\qquad
\theta=\arctan(1/\beta_{\rm app})\simeq17^{\circ};
\]

(iii) an illustrative case with $\delta=10$; and (iv) a high-Doppler-factor case with $\delta\simeq27$, for which $\Gamma$ and $\theta$ were recalculated under the $\beta_{\rm app}=3.3$ constraint. The published value $\beta_{\rm app}=3.3$ is adopted here as reported by \citet{2001A&A...370...65S}; it is used as a kinematic reference rather than being recomputed from the measured proper motion using a different cosmology.

These four cases assume that the measured radio-component pattern speed can be used as a reference for the bulk speed of the compact optical-to-$\gamma$-ray emitting region. This identification is a modelling assumption rather than an observationally established property of the source.

\begin{table*}
\centering
\caption{Dependence of the compact-zone SED solutions on the assumed bulk Lorentz factor, viewing angle, and Doppler factor. All four cases adopt $\beta_{\rm app}=3.3$ for a VLBI jet component, under the assumptions described in Section~8. For all four cases, the selected pure-SSC solutions reproduce the three quiet-state detections and the five main-peak detections within the discrepancies quoted in Section~8, while satisfying the reliable quiet-state LAT upper limits. None of the varied parameters in the selected solutions lies on a search boundary. $\mathcal{S}_{\rm SSC}$ and $\mathcal{S}_{\rm EC}$ are descriptive residual scores for the main-peak pure-SSC and SSC+EC-BLR calculations, respectively. Lower scores indicate smaller residuals but do not establish a statistical preference between models. $f_{\rm EC,max}$ is the maximum EC-BLR fraction of the total model flux in the three fitted LAT bins.
$B_{\rm q}$ and $B_{\rm p}$ denote the magnetic field strengths in the quiet-state and main-peak pure-SSC realizations, respectively. $(U_{\rm e}/U_B)_{\rm q}$ and $(U_{\rm e}/U_B)_{\rm p}$ are the corresponding ratios of relativistic-electron to magnetic-field energy densities.}
\label{tab:sed_geometry}
\scriptsize
\setlength{\tabcolsep}{7pt}
\begin{tabular}{lcccccccccc}
\hline
Case &
$\Gamma$ &
$\theta$ (deg) &
$\delta$ &
$\mathcal{S}_{\rm SSC}$ &
$\mathcal{S}_{\rm EC}$ &
$f_{\rm EC,max}$ &
$B_{\rm q}$ (G) &
$B_{\rm p}$ (G) &
$(U_{\rm e}/U_B){\rm q}$ &
$(U{\rm e}/U_B){\rm p}$ \\
\hline
Radio-core $\delta{\rm var}$ &
3.70 & 23.12 & 2.36 &
0.39 & 0.72 &
$1.1\times10^{-5}$ &
48.53 &
21.88 &
$1.50\times10^{-5}$ &
$7.03\times10^{-3}$ \\

Minimum-$\Gamma$ (critical angle) &
3.45 & 16.86 & 3.45 &
0.54 & 1.12 &
$2.2\times10^{-10}$ &
26.93 &
8.60 &
$5.26\times10^{-5}$ &
$5.35\times10^{-2}$ \\

Illustrative $\delta=10$ &
5.59 & 3.44 & 10.00 &
0.82 & 0.96 &
$3.1\times10^{-9}$ &
4.85 &
0.80 &
$6.58\times10^{-4}$ &
8.45 \\

High-$\delta$ test, $\beta_{\rm app} =3.3$ &
13.72 & 0.51 & 27.01 &
0.90 & 1.06 &
$5.26\times10^{-2}$ &
0.33 &
0.10 &
0.40 &
$4.76\times10^{2}$ \\
\hline
\end{tabular}
\end{table*}

Pure SSC gives the lowest descriptive residual in each of the four calculations constrained by $\beta_{\rm app}=3.3$. In the four $\beta_{\rm app}$-constrained calculations, the EC-BLR contribution is negligible in the three cases with $\delta\leq10$ and reaches at most 5.3 per cent in the high-$\delta$ case. These results show that, for the adopted combinations of jet parameters, the sparse two-state SEDs can be represented by one-zone SSC solutions, while the relative EC-BLR contribution remains minor and dependent on the assumed jet parameters in the selected solutions.

These fits are descriptive and non-unique. They do not identify the transient $\gamma$-ray-emitting zone with the stationary VLBI radio core, nor do they establish SSC as the unique high-energy emission mechanism. The comparison shows only that a substantial BLR-seeded inverse-Compton contribution is not demanded by the specific sparse
SEDs and the models examined here. This conclusion
does not imply that the BLR radiation field is absent or that an external-Compton contribution can be excluded.

Acceptable pure-SSC solutions were obtained for all four $\beta_{\rm app}$-constrained cases. For each of these four cases, the model flux at every fitted main-peak data point differs from the corresponding observed flux by no more than a factor of 1.35. The five points comprise one optical $R$-band measurement, one 5--10 keV~X-ray measurement, and three LAT spectral measurements below 3 GeV. For the three fitted quiet-state detections, the maximum model-to-data discrepancy across the four cases is approximately a factor of 1.9. All reliable quiet-state LAT upper limits are satisfied.

The SED therefore does not uniquely determine the jet parameters. Table~\ref{tab:sed_parameters} presents representative parameters for the critical-angle solution. We use this solution for the illustrative SED curves in Fig.~\ref{fig:sed_two_state} because it corresponds to the minimum bulk Lorentz factor compatible with the adopted apparent speed $\beta_{\rm app}=3.3$. The choice is not based on the SED residual score: the radio-core $\delta_{\rm var}=2.36$ case has a lower main-peak SSC score ($\mathcal{S}_{\rm SSC}=0.39$ versus $0.54$ for the critical-angle case, Table~\ref{tab:sed_geometry}).

\begin{table}
\centering
\caption{Representative one-zone pure-SSC parameters for the quiet and main-peak states obtained for the critical-angle case. Within this parameter set, $R$, $\Gamma_{\rm j}$, $\theta$, $\delta$, and $p$ were fixed in both states. A dagger indicates that $\gamma_{\min}=10$ was fixed in the main-peak calculations; $\gamma_{\min}$ was allowed to vary in the quiet-state calculations. The listed parameters are representative solutions, not unique measurements or confidence intervals.}
\label{tab:sed_parameters}
\setlength{\tabcolsep}{4.0pt}
\begin{tabular}{lcc}
\hline
Parameter &
Quiet state &
Main peak \\
\hline
$R$ (cm) &
$3.0\times10^{16}$ &
$3.0\times10^{16}$ \\

$\Gamma_{\rm j}$ &
3.45 &
3.45 \\

$\theta$ (deg) &
16.86 &
16.86 \\

$\delta$ &
3.45 &
3.45 \\

$B$ (G) &
26.93 &
8.60 \\

$N$ (cm$^{-3}$) &
13.64 &
$2.77\times10^{3}$ \\

$\gamma_{\min}$ &
17.34 &
$10^{\dagger}$ \\

$\gamma_{\rm br}$ &
$3.62\times10^{4}$ &
$5.61\times10^{3}$ \\

$\gamma_{\max}$ &
$7.94\times10^{5}$ &
$3.94\times10^{6}$ \\

$p$ &
2.0 &
2.0 \\

$p_1$ &
7.04 &
3.62 \\
\hline
\end{tabular}
\end{table}

In the critical-angle case, the maximum model-to-data discrepancies are factors of 1.22 and 1.73 for the main-peak and quiet-state SEDs, respectively. Multiple acceptable solutions were found for this case and the other tested configurations. The parameters in Table~\ref{tab:sed_parameters} should therefore not be interpreted as formal measurements or confidence intervals.

In the critical-angle realization shown in
Fig.~\ref{fig:sed_two_state}, the representative main-peak curve has lower synchrotron and SSC peak frequencies than the selected
quiet-state curve. These apparent shifts are not established
observationally: the quiet-state Compton component is constrained by a single 3--10~GeV detection and upper limits, while the synchrotron component is sparsely sampled between the optical and X-ray bands. The inferred peak locations are therefore model dependent.

The inferred energy partition differs between the two fitted states and depends strongly on the assumed jet parameters. 

For the critical-angle case, \textsc{JetSeT} gives
\[
(U_B,U_{\rm e})_{\rm q}
=
(2.89\times10^{1},\,1.52\times10^{-3})
\ {\rm erg\,cm^{-3}},
\]
and
\[
(U_B,U_{\rm e})_{\rm p}
=
(2.94,\,1.57\times10^{-1})
\ {\rm erg\,cm^{-3}}.
\]
In this case, $U_{\rm e}/U_B$ increases from $5.26\times10^{-5}$ in the quiet state to $5.35\times10^{-2}$ at the main peak, by a factor of approximately $10^{3}$, although both representative states remain magnetically dominated.

Across the four selected pairs of pure-SSC solutions (Table~\ref{tab:sed_geometry}), the fitted magnetic field decreases from the quiet state to the main peak, with $B_{\rm p}/B_{\rm q}=0.17$--$0.45$. The corresponding magnetic energy density decreases by factors of approximately 5--37, while the electron energy density increases by factors of approximately 95--350. Consequently, $U_{\rm e}/U_B$ increases by factors of $4.7\times10^2$--$1.3\times10^4$. However, its main-peak value ranges from $7.0\times10^{-3}$ to $4.76\times10^2$, indicating a strong dependence of the inferred energy partition on the assumed jet parameters.

The consistent decrease in the fitted magnetic field, accompanied by an increase in the electron energy density, is compatible with magnetic energy dissipation, including possible magnetic reconnection. However, the quiet and main-peak SEDs were modelled independently, and these stationary solutions do not establish a direct conversion of magnetic energy into particle energy or uniquely identify the dissipation mechanism.

\section{Discussion}
\label{sec:discussion}

The multiwavelength evolution of the \sou non-thermal activity during 2024--2026 indicates that the current episode is not a single isolated flare but a sustained reconfiguration of the inner jet. Such prolonged and complex active phases can be modulated by the intermittent behaviour of the AGN central engine, where episodic accretion instabilities or magnetic reconnection events drive repetitive energy injections into the jet base \citep[e.g.][]{2025ApJ...987L..26W, 2026ApJ..1004L...5A}. The strongest evidence comes from the combination of rapid $\gamma$-ray and optical activity, an elevated X-ray state, frequency-dependent radio brightening, pronounced hardening of the high-frequency radio spectrum, and recent VLBI/VLA/ALMA evidence that the brightening is dominated by a compact core region rather than by a kiloparsec-scale jet. 

The radio behaviour is naturally interpreted in terms of opacity-driven evolution in a stratified synchrotron jet. The long-term radio brightening steepens markedly after 2022, with the strongest post-2022 rise observed at 22 GHz and progressively weaker and slower response towards lower frequencies. At the same time, the 11--22 GHz spectral index evolved from a steep state to flat or inverted values during the active phase, consistent with the emergence of a compact synchrotron component that becomes visible first at high frequencies, while the lower-frequency emission remains diluted by more extended and slowly varying regions. This interpretation is independently supported by recent VLBI observations obtained during the same active phase, which reveal a newly energized compact component in the core region, although no downstream knot has yet separated from the core \citep{2026A&A...710A..63L}. The single-dish radio evolution can therefore be understood as the opacity-dependent manifestation of this compact component as the disturbance develops within the stratified jet \citep[e.g.][]{1979ApJ...232...34B,1985ApJ...298..301H,1992A&A...254...71V,1998A&A...330...79L,2005AJ....130.1418J}.

The X-ray observations provide an important independent diagnostic of the broadband active state. The available X-ray measurements indicate that \sou entered an elevated X-ray state in 2025--2026, with the 5--10 keV flux reaching a level several times higher than the pre-flare measurements. The spectra are adequately described by a power law modified by Galactic absorption, with photon indices varying within $\Gamma_{\rm X}\simeq 0.9$--1.6. The sampling is too sparse for an independent X-ray flare decomposition or for a reliable X-ray lag measurement.

We also compared the X-ray photon index with the radio spectral indices using quasi-simultaneous measurements matched within 30 d. A tentative trend is seen in the sense that flatter or inverted high-frequency radio spectra tend to be accompanied by harder X-ray spectra, or equivalently by larger values of the X-ray energy index $\alpha_{\rm X}=1-\Gamma_{\rm X}$.

However, the trend is not statistically significant for the independent X-ray epochs ($N=10$ for $\alpha_{11-22}$, Pearson $r\simeq -0.46$, $p\simeq0.18$ for $\Gamma_{\rm X}$). Therefore, we regard this behaviour as suggestive of a possible connection between the X-ray emission and the compact high-frequency radio component, but not as evidence for a firm correlation.

The flare decomposition provides a phenomenological description of the high-energy and optical variability. The $\gamma$-ray light curve is represented by five flares, while the optical activity is described by a smoother sequence of partially blended subflares superimposed on a broader brightening episode. The fitted time-scales overlap for some events, although the $\gamma$-ray flares span a broader range, particularly in their rise times. At the same time, the optical activity forms a smoother and more strongly blended sequence of subflares, indicating differences in both temporal structure and morphology. A statistically significant flux correlation (Spearman $\rho\simeq0.66$,  $p\sim10^{-7}$) further supports a physical connection between the optical and $\gamma$-ray activity, although it is partly driven by a small number of strong and unevenly sampled events. The overlapping optical structure prevents a unique one-to-one association with individual $\gamma$-ray flares, and we therefore do not interpret the apparent offsets between individual subflares as robust inter-band lags. The two bands may instead respond differently to the same energy-injection episodes within an inhomogeneous compact emission region. 

The highest observed fluxes during the analysed interval in the $\gamma$-ray, X-ray, and optical bands occur within a $\simeq12$-d interval (Table~\ref{tab:maximum}). Although the sparse X-ray sampling prevents a precise lag measurement, this near-coincidence is consistent with the X-ray high state being associated with the same inner-jet activity episode as the $\gamma$-ray and optical flaring.  In contrast, the radio data do not reveal a uniquely identifiable counterpart flare. The radio emission instead exhibits a gradual, frequency-dependent brightening that began before the main high-energy event and remains ongoing. We therefore do not interpret the epochs of the largest observed radio flux densities as inter-band lags; the radio response is characterised by its long-term growth rates, spectral hardening, and increasing excess-component contribution towards higher frequencies.

The increase of fractional variability from low- to high-frequency radio bands, followed by substantially larger optical and $\gamma$-ray variability, is consistent with two variability regimes: a rapidly varying compact region dominating the high-energy and optical emission, and a slower radio response increasingly affected by opacity, extended emission, and propagation through the parsec-scale jet. This picture links the temporal evolution of the broadband outburst to the spatially resolved structural changes revealed by VLBI \citep{2026A&A...710A..63L}. Because the X-ray sampling is sparse and heterogeneous, its fractional variability is not compared directly with the other bands.

The compact-zone SED modelling is consistent with this two-regime interpretation, but the fits are descriptive and non-unique. For the four parameter combinations adopting the published VLBI apparent speed, the sparse quiet-state and main-peak SEDs admit one-zone SSC solutions, while the corresponding SSC+EC-BLR calculations yield only a minor EC-BLR contribution that depends on the assumed jet parameters.

These results neither establish SSC as the unique high-energy emission mechanism nor identify the transient $\gamma$-ray-emitting zone with the stationary VLBI radio core; the SED-derived energy partition should therefore not be attributed directly to the VLBI radio core.

Across all four representative pure-SSC solution pairs, $U_{\rm e}/U_B$ increases strongly from the quiet state to the main peak  (Table~\ref{tab:sed_geometry}), accompanied by a decrease in the fitted magnetic field with $B_{\rm p}/B_{\rm q}=0.17$--$0.45$. The inferred energy partition nevertheless depends strongly on the assumed jet parameters. This trend is compatible with magnetic dissipation, including possible reconnection, but the independently fitted stationary SEDs do not establish the dissipation mechanism. A similar one-zone SSC interpretation was proposed for the flaring CSS source 3C~216 by \citet{2025ApJ...986...99G}. In that source, denser quasi-simultaneous UV-to-$\gamma$-ray  observations allowed the authors to trace changes in the electron distribution over several days. Such a time-resolved analysis is not possible with the sparser two-state SEDs of \sou. In both sources, one-zone SSC provides a viable, although non-unique, description of the high-energy emission.

For \sou, a possible interpretation is that the 2024--2026 event was preceded by a gradual restructuring and energization of the compact jet. The long-term radio/mm brightening and spectral hardening indicate the development of a new compact synchrotron component rather than a single instantaneous flare. In this picture, the $\gamma$-ray flares occur when the disturbance energizes compact regions where particle acceleration and Doppler boosting are favourable for efficient inverse-Compton emission. These regions may correspond either to a single compact inner-jet zone or to several quasi-stationary sites, such as recollimation shocks. The optical subflares can then be interpreted as synchrotron responses from different parts of the same inhomogeneous compact region, while the smoother and slower radio/mm response traces the subsequent opacity-driven evolution of the synchrotron component as it becomes visible at centimetre and millimetre wavelengths.

\section{Summary}
\label{sec:summary}

We have reported and quantified the broadband activity of the CSS quasar \sou during 2024--2026 using \textit{Fermi}-LAT light curves, X-ray measurements with \textit{Swift}/XRT and \textit{SRG}/ART-XC (as well as two historical measurements by \textit{Chandra} and \textit{SRG}/eROSITA), $R$ band data  from the Zeiss-1000, AS-500/2 and ZTF optical monitoring, dense multifrequency radio observations with RATAN-600 and RT-32, and millimetre-band data from the public ALMA Calibrator Source Catalogue.

(i) The radio light curves show a clear transition after 2022, with post-2022 brightening rates significantly steeper than in the pre-2022 epoch, indicating the onset of a new long-term activity phase. The variability amplitude increases systematically with frequency in the radio band. The optical band shows substantially stronger variability, while the $\gamma$-ray emission exhibits the largest amplitude, indicating that the highest-energy emission originates in the most compact and rapidly varying regions.

(ii) The radio spectral indices exhibit pronounced frequency-dependent hardening. The high-frequency spectral index (11--22~GHz) evolves most strongly and reaches flat or inverted values during the active phase, consistent with the emergence of a compact synchrotron component becoming optically thin first at high frequencies.

(iii) The X-ray observations reveal an elevated state during the active phase, when the flux increased by a factor of more than three with respect to pre-flare measurements and the spectrum hardened from a photon index of $\Gamma_{\rm X}\simeq 1.6$ to $\Gamma_{\rm X}\simeq 0.9$ and softened back after the peak.

(iv) The flare decomposition of $\gamma$-ray and optical light curves (epochs 2024.75--2026.25) reveals five distinct $\gamma$-ray flares and a sequence of broader, partially blended optical subflares. The $\gamma$-ray and optical time-scales are broadly comparable for several events, while the radio brightening develops over much longer time-scales. The clearest time-scale contrast is therefore between the high-energy/optical activity and the slow opacity-affected radio response. 

(v) The highest observed fluxes during the analysed interval in the $\gamma$-ray, X-ray, and optical bands occur within a $\simeq 12$-day interval, supporting a common compact activity episode in the inner jet. In contrast, the radio emission evolves more gradually and reaches its strongest observed levels progressively later towards lower frequencies, consistent with opacity-driven propagation of a disturbance through a stratified jet.

(vi) For the four parameter combinations adopting the published VLBI apparent speed, the sparse quiet-state and main-peak SEDs admit descriptive, non-unique one-zone SSC solutions, while the corresponding SSC+EC-BLR calculations yield a minor EC-BLR contribution that depends on the assumed jet parameters. These fits neither establish SSC as the unique high-energy emission mechanism nor identify the transient $\gamma$-ray-emitting zone with the stationary VLBI radio core. In all four representative pure-SSC solution pairs, $U_{\rm e}/U_B$ increases strongly from the quiet state to the main peak,   accompanied by a decrease in the fitted magnetic field. The inferred energy partition and other fitted parameters remain dependent on the assumed jet parameters. The critical-angle case shown in Fig.~\ref{fig:sed_two_state} corresponds to the minimum Lorentz factor for the adopted $\beta_{\rm app}$ and is included for illustration, rather than as a unique determination of the jet parameters.

Overall, the observed timing behaviour, X-ray high-state evidence, spectral evolution, variability amplitudes, compact-zone SED modelling, and frequency-dependent radio brightening support a scenario in which the 2024--2026 activity was driven by a disturbance evolving through a structurally complex and opacity-stratified jet. The rapid high-energy and optical activity traces a compact inner region, while the smoother radio/mm brightening reflects the slower emergence of a compact synchrotron component as it becomes visible at progressively more transparent regions of the jet. This picture is consistent with recent VLBI/VLA/ALMA evidence for core-dominated brightening, although no separated new knot has yet been resolved.

While this paper was under preparation, \sou exhibited another exceptionally bright $\gamma$-ray flare in May 2026 \citep{2026ATel17808....1C}, indicating that the active phase is still ongoing. Because contemporaneous optical coverage is unavailable and any associated radio response may develop over subsequent months, multiwavelength analysis of this event is deferred to future work. 

\section*{Acknowledgements} 

This work is based on the data obtained with the RATAN-600 radio telescope, 6-metre BTA, Zeiss-1000, and AS-500/2 optical reflectors at the Special Astrophysical Observatory of the Russian Academy of Sciences (SAO RAS) within the government assignment work. The renovation of telescope equipment is currently provided within the national project “Science and Universities”.

This work is supported by the Tianshan Talent Training Program (grant No. 2023TSYCCX0099), the Xinjiang Tianchi Talent Program, and the Urumqi Nanshan Astronomy and Deep Space Exploration Observation and Research Station of Xinjiang (XJYWZ2303). 

The observations at 5.05 and 8.63 GHz were performed with the Svetloe, Badary, and Zelenchukskaya RT-32 radio telescopes operated by the Center of Shared Research Facility for the Quasar VLBI Network of IAA RAS (\url{https://iaaras.ru/cu-center/}).  

This work made use of observations with the Mikhail Pavlinsky ART-XC telescope, the hard X-ray instrument on board the SRG observatory. The SRG observatory was created by Roskosmos in the interests of the Russian Academy of Sciences represented by its Space Research Institute (IKI) in the framework of the Russian Federal Space Program, with the participation of Germany. The ART-XC team thanks Lavochkin Association (NPOL) with partners for the creation and operation of the SRG spacecraft (Navigator). This work also made use of data supplied by the UK Swift Science Data Centre at the University of Leicester.

YYK was supported by the MuSES project, which has received funding from the European Union (ERC grant agreement No 101142396). Views and opinions expressed are however those of the author(s) only and do not necessarily reflect those of the European Union or ERCEA. Neither the European Union nor the granting authority can be held responsible for them.

This paper makes use of the following ALMA data: ADS/JAO.ALMA\#2011.0.00001.CAL. ALMA is a partnership of ESO (representing its member states), NSF (USA) and NINS (Japan), together with NRC (Canada), NSTC and ASIAA (Taiwan), and KASI (Republic of Korea), in cooperation with the Republic of Chile. The Joint ALMA Observatory is operated by ESO, AUI/NRAO and NAOJ. This work made use of publicly available data from the Fermi Gamma-ray Space Telescope, obtained through the Fermi Science Support Center (FSSC). We acknowledge the Fermi-LAT Collaboration and the support of the agencies and institutes that have contributed to the development, operation, and scientific analysis of the LAT. This research also made use of data from the Catalina Sky Surveys, LPL, University of Arizona, and the Catalina Real-Time Transient Survey, Caltech, with value-added products served by IUCAA.

We thank Prof.~Eduardo Ros from the Max Planck Institute for Radio Astronomy for his valuable comments on the manuscript. The authors are grateful to the anonymous referee for constructive comments and suggestions, which helped to improve the presentation of the manuscript.



\section*{Data Availability}
The data underlying this article are available in the article and in its online supplementary material. The RATAN-600 and RT-32 radio flux-density measurements, together with the optical $R$ band measurements obtained with the AS-500/2 and Zeiss-1000 telescopes, will be deposited in VizieR. The X-ray fluxes and spectral parameters are listed in Table~\ref{tab:xray}. The remaining data were obtained from publicly available archives and catalogues, as described in Section~2.

\textit{Facilities}: RATAN-600, RT-32, ALMA, BTA, Zeiss-1000, AS-500/2, CRTS, ZTF, \textit{Fermi}-LAT, \textit{Swift}/XRT, \textit{SRG}/ART-XC, \textit{Chandra}, \textit{SRG}/eROSITA.

\bibliographystyle{mnras}
\bibliography{mufakharov}

@ARTICLE{2011ApJ...737..103S,
       author = {{Schlafly}, Edward F. and {Finkbeiner}, Douglas P.},
        title = "{Measuring Reddening with Sloan Digital Sky Survey Stellar Spectra and Recalibrating SFD}",
      journal = {\apj},
         year = 2011,
        month = aug,
       volume = {737},
       number = {2},
          eid = {103},
        pages = {103},
          doi = {10.1088/0004-637X/737/2/103},
archivePrefix = {arXiv},
       eprint = {1012.4804},
 primaryClass = {astro-ph.GA},
       adsurl = {https://ui.adsabs.harvard.edu/abs/2011ApJ...737..103S}
}

@ARTICLE{2015MNRAS.452.4274P,
       author = {{Pushkarev}, A.~B. and {Kovalev}, Y.~Y.},
        title = "{Milky Way scattering properties and intrinsic sizes of active galactic nuclei cores probed by very long baseline interferometry surveys of compact extragalactic radio sources}",
      journal = {\mnras},
         year = 2015,
        month = oct,
       volume = {452},
       number = {4},
        pages = {4274-4282},
          doi = {10.1093/mnras/stv1539},
archivePrefix = {arXiv},
       eprint = {1507.02459},
 primaryClass = {astro-ph.HE},
       adsurl = {https://ui.adsabs.harvard.edu/abs/2015MNRAS.452.4274P}
}

@ARTICLE{2022MNRAS.515.1736K,
       author = {{Koryukova}, T.~A. and {Pushkarev}, A.~B. and {Plavin}, A.~V. and {Kovalev}, Y.~Y.},
        title = "{Tracing Milky Way scattering by compact extragalactic radio sources}",
      journal = {\mnras},
         year = 2022,
        month = sep,
       volume = {515},
       number = {2},
        pages = {1736-1750},
          doi = {10.1093/mnras/stac1898},
archivePrefix = {arXiv},
       eprint = {2201.04359},
 primaryClass = {astro-ph.GA},
       adsurl = {https://ui.adsabs.harvard.edu/abs/2022MNRAS.515.1736K}
}

@ARTICLE{2026ApJ..1004L...5A,
       author = {{An}, Tao},
        title = "{Time-domain Radio-loudness of Active Galactic Nuclei: Intermittency, Memory, and Jet Escape}",
      journal = {\apjl},
         year = 2026,
        month = jun,
       volume = {1004},
       number = {1},
          eid = {L5},
        pages = {L5},
          doi = {10.3847/2041-8213/ae6cde},
archivePrefix = {arXiv},
       eprint = {2603.21119},
 primaryClass = {astro-ph.GA},
       adsurl = {https://ui.adsabs.harvard.edu/abs/2026ApJ..1004L...5A}
}

@ARTICLE{2026ApJS..283...69C,
       author = {{Cui}, Lang and {Mohana A}, Krishna and {Wang}, Xin and {Chang}, Ning and {Tan}, Guiping and {Liu}, Xiang},
        title = "{SMMAN: Quasi-simultaneous Multiwavelength Monitoring of Gamma-Ray-loud Active Galactic Nuclei with the Nanshan 26 m Radio Telescope}",
      journal = {\apjs},
         year = 2026,
        month = apr,
       volume = {283},
       number = {2},
          eid = {69},
        pages = {69},
          doi = {10.3847/1538-4365/ae4904},
archivePrefix = {arXiv},
       eprint = {2602.19524},
 primaryClass = {astro-ph.GA},
       adsurl = {https://ui.adsabs.harvard.edu/abs/2026ApJS..283...69C}
}

@ARTICLE{2016A&A...596A..45F,
       author = {{Fuhrmann}, L. and {Angelakis}, E. and {Zensus}, J.~A. and {Nestoras}, I. and {Marchili}, N. and {Pavlidou}, V. and {Karamanavis}, V. and {Ungerechts}, H. and {Krichbaum}, T.~P. and {Larsson}, S. and {Lee}, S.~S. and {Max-Moerbeck}, W. and {Myserlis}, I. and {Pearson}, T.~J. and {Readhead}, A.~C.~S. and {Richards}, J.~L. and {Sievers}, A. and {Sohn}, B.~W.},
        title = "{The F-GAMMA programme: multi-frequency study of active galactic nuclei in the Fermi era. Programme description and the first 2.5 years of monitoring}",
      journal = {\aap},
         year = 2016,
        month = nov,
       volume = {596},
          eid = {A45},
        pages = {A45},
          doi = {10.1051/0004-6361/201528034},
archivePrefix = {arXiv},
       eprint = {1608.02580},
 primaryClass = {astro-ph.HE},
       adsurl = {https://ui.adsabs.harvard.edu/abs/2016A&A...596A..45F}
}

@ARTICLE{2021MNRAS.507.4564P,
       author = {{Principe}, G. and {Di Venere}, L. and {Orienti}, M. and {Migliori}, G. and {D'Ammando}, F. and {Mazziotta}, M.~N. and {Giroletti}, M.},
        title = "{Gamma-ray emission from young radio galaxies and quasars}",
      journal = {\mnras},
         year = 2021,
        month = nov,
       volume = {507},
       number = {3},
        pages = {4564-4583},
          doi = {10.1093/mnras/stab2357},
archivePrefix = {arXiv},
       eprint = {2107.12963},
 primaryClass = {astro-ph.HE},
       adsurl = {https://ui.adsabs.harvard.edu/abs/2021MNRAS.507.4564P}
}

@ARTICLE{2022MNRAS.511.214P,
       author = {{Nascimento}, R.S. and {Rodriguez-Ardila}, A. and {Dahmer-Hahn}, L. and {Fonseca-Faria}, M.A. and {Riffel}, R. and {Marinello}, M. and {Beuhert}, T. and {Callingham}, J.R.},
        title = "{Optical properties of Peaked Spectrum rario sources}",
      journal = {\mnras},
         year = 2022,
        month = jan,
       volume = {511},
       number = {1},
        pages = {214-230},
          doi = {10.1093/mnras/stab3791},
archivePrefix = {arXiv},
 primaryClass = {astro-ph.HE},
       adsurl = {https://ui.adsabs.harvard.edu/abs/2022MNRAS.511.214P}
}

@ARTICLE{2011BaltA..20..363A,
       author = {{Afanasiev}, V.~L. and {Moiseev}, A.~V.},
        title = "{Scorpio on the 6 m Telescope: Current State and Perspectives for Spectroscopy of Galactic and Extragalactic Objects}",
      journal = {Baltic Astronomy},
         year = 2011,
        month = aug,
       volume = {20},
        pages = {363-370},
          doi = {10.1515/astro-2017-0305},
archivePrefix = {arXiv},
       eprint = {1106.2020},
 primaryClass = {astro-ph.IM},
       adsurl = {https://ui.adsabs.harvard.edu/abs/2011BaltA..20..363A}
}

@ARTICLE{2022A&A...661A..38P,
       author = {{Pavlinsky}, M. and {Sazonov}, S. and {Burenin}, R. and {Filippova}, E. and {Krivonos}, R. and {Arefiev}, V. and {Buntov}, M. and {Chen}, C.-T. and {Ehlert}, S. and {Lapshov}, I. and {Levin}, V. and {Lutovinov}, A. and {Lyapin}, A. and {Mereminskiy}, I. and {Molkov}, S. and {Ramsey}, B.~D. and {Semena}, A. and {Semena}, N. and {Shtykovsky}, A. and {Sunyaev}, R. and {Tkachenko}, A. and {Swartz}, D.~A. and {Vikhlinin}, A.},
        title = "{SRG/ART-XC all-sky X-ray survey: Catalog of sources detected during the first year}",
      journal = {\aap},
         year = 2022,
        month = may,
       volume = {661},
          eid = {A38},
        pages = {A38},
          doi = {10.1051/0004-6361/202141770},
archivePrefix = {arXiv},
       eprint = {2107.05879},
 primaryClass = {astro-ph.HE},
       adsurl = {https://ui.adsabs.harvard.edu/abs/2022A&A...661A..38P}
}

@ARTICLE{2001A&A...370...65S,
       author = {{Shen}, Z.-Q. and {Jiang}, D.~R. and {Kameno}, S. and {Chen}, Y.~J.},
        title = "{Superluminal motion in a compact steep spectrum radio source 3C 138}",
      journal = {\aap},
         year = 2001,
        month = apr,
       volume = {370},
        pages = {65-69},
          doi = {10.1051/0004-6361:20010193},
archivePrefix = {arXiv},
       eprint = {astro-ph/0102083},
 primaryClass = {astro-ph},
       adsurl = {https://ui.adsabs.harvard.edu/abs/2001A&A...370...65S}
}

@ARTICLE{2023A&A...671...A32,
       author = {{Capetti}, A. and {B. Balmaverde}, B. and {Baldi}, R.D. and {Baum}, S. and {Chiaberge}, M. and {Grandi}, P and {Marconi}, A. and {O’Dea}, C. and3, and {Venturi}, G.},
        title = "{The MURALES survey.VII. Optical spectral properties of the nuclei of 3C radio sources at 0.3 < z < 0.82}",
      journal = {\aap},
         year = 2023,
        month = nov,
       volume = {671},
        pages = {A32-A47},
          doi = {10.1051/0004-6361:202244606},
       adsurl = {https://ui.adsabs.harvard.edu/abs/2023A&A...671...A32}
}

@ARTICLE{2026ATel17808....1C,
       author = {{Casaburo}, F. and {La Mura}, G. and {Cheung}, C.~C. and {Ciprini}, S.},
        title = "{Fermi LAT detection of historical maximum gamma-ray flux from 3C 138}",
      journal = {The Astronomer's Telegram},
         year = 2026,
        month = may,
       volume = {17808},
        pages = {1},
       adsurl = {https://ui.adsabs.harvard.edu/abs/2026ATel17808....1C}
}

@ARTICLE{2003A&A...406...43C,
       author = {{Cotton}, W.~D. and {Dallacasa}, D. and {Fanti}, C. and {Fanti}, R. and {Foley}, A.~R. and {Schilizzi}, R.~T. and {Spencer}, R.~E.},
        title = "{The Faraday screen near the nucleus of the CSS quasar 3C 138}",
      journal = {\aap},
         year = 2003,
        month = jul,
       volume = {406},
        pages = {43-50},
          doi = {10.1051/0004-6361:20030523},
       adsurl = {https://ui.adsabs.harvard.edu/abs/2003A&A...406...43C}
}

@ARTICLE{2026MNRAS.547ag333S,
       author = {{Sotnikova}, Yu V. and {Mufakharov}, T.~V. and {Volvach}, A.~E. and {Vlasyuk}, V.~V. and {Khabibullina}, M.~L. and {Mikhailov}, A.~G. and {An}, T. and {Kudryavtsev}, D.~O. and {Kovalev}, Yu A. and {Kovalev}, Yu Yu and {Popkov}, A.~V. and {Savchenko}, S.~S. and {Erkenov}, A.~K. and {Morozova}, D.~A. and {Semenova}, T.~A. and {Spiridonova}, O.~I. and {Kharinov}, M.~A. and {Rakhimov}, I.~A. and {Andreeva}, T.~S. and {Cui}, L. and {Wang}, X. and {Chang}, N. and {Udovitskiy}, R. Yu and {Zhekanis}, P.~G. and {Borman}, G.~A. and {Grishina}, T.~S. and {Kopatskaya}, E.~N. and {Larionova}, E.~G. and {Troitskiy}, I.~S. and {Troitskaya}, Yu V. and {Vasilyev}, A.~A. and {Zhovtan}, A.~V. and {Kratov}, D.~V. and {Volvach}, L.~N. and {Shishkina}, E.~V. and {Dmytrotsa}, A.~I. and {Zharov}, V.~I.},
        title = "{Multiwavelength quasi-periodic variability of the blazar Ton 599}",
      journal = {\mnras},
         year = 2026,
        month = apr,
       volume = {547},
       number = {2},
          eid = {stag333},
        pages = {stag333},
          doi = {10.1093/mnras/stag333},
archivePrefix = {arXiv},
       eprint = {2603.05894},
 primaryClass = {astro-ph.HE},
       adsurl = {https://ui.adsabs.harvard.edu/abs/2026MNRAS.547ag333S}
}

@ARTICLE{2021A&A...647A...1P,
       author = {{Predehl}, P. and {Andritschke}, R. and {Arefiev}, V. and {Babyshkin}, V. and {Batanov}, O. and {Becker}, W. and {B{\"o}hringer}, H. and {Bogomolov}, A. and {Boller}, T. and {Borm}, K. and {Bornemann}, W. and {Br{\"a}uninger}, H. and {Br{\"u}ggen}, M. and {Brunner}, H. and {Brusa}, M. and {Bulbul}, E. and {Buntov}, M. and {Burwitz}, V. and {Burkert}, W. and {Clerc}, N. and {Churazov}, E. and {Coutinho}, D. and {Dauser}, T. and {Dennerl}, K. and {Doroshenko}, V. and {Eder}, J. and {Emberger}, V. and {Eraerds}, T. and {Finoguenov}, A. and {Freyberg}, M. and {Friedrich}, P. and {Friedrich}, S. and {F{\"u}rmetz}, M. and {Georgakakis}, A. and {Gilfanov}, M. and {Granato}, S. and {Grossberger}, C. and {Gueguen}, A. and {Gureev}, P. and {Haberl}, F. and {H{\"a}lker}, O. and {Hartner}, G. and {Hasinger}, G. and {Huber}, H. and {Ji}, L. and {Kienlin}, A. v. and {Kink}, W. and {Korotkov}, F. and {Kreykenbohm}, I. and {Lamer}, G. and {Lomakin}, I. and {Lapshov}, I. and {Liu}, T. and {Maitra}, C. and {Meidinger}, N. and {Menz}, B. and {Merloni}, A. and {Mernik}, T. and {Mican}, B. and {Mohr}, J. and {M{\"u}ller}, S. and {Nandra}, K. and {Nazarov}, V. and {Pacaud}, F. and {Pavlinsky}, M. and {Perinati}, E. and {Pfeffermann}, E. and {Pietschner}, D. and {Ramos-Ceja}, M.~E. and {Rau}, A. and {Reiffers}, J. and {Reiprich}, T.~H. and {Robrade}, J. and {Salvato}, M. and {Sanders}, J. and {Santangelo}, A. and {Sasaki}, M. and {Scheuerle}, H. and {Schmid}, C. and {Schmitt}, J. and {Schwope}, A. and {Shirshakov}, A. and {Steinmetz}, M. and {Stewart}, I. and {Str{\"u}der}, L. and {Sunyaev}, R. and {Tenzer}, C. and {Tiedemann}, L. and {Tr{\"u}mper}, J. and {Voron}, V. and {Weber}, P. and {Wilms}, J. and {Yaroshenko}, V.},
        title = "{The eROSITA X-ray telescope on SRG}",
      journal = {\aap},
         year = 2021,
        month = mar,
       volume = {647},
          eid = {A1},
        pages = {A1},
          doi = {10.1051/0004-6361/202039313},
archivePrefix = {arXiv},
       eprint = {2010.03477},
 primaryClass = {astro-ph.HE},
       adsurl = {https://ui.adsabs.harvard.edu/abs/2021A&A...647A...1P}
}

@ARTICLE{4FGL-DR4,
       author = {{Ballet}, J. and {Bruel}, P. and {Burnett}, T.~H. and {Lott}, B. and {The Fermi-LAT collaboration}},
        title = "{Fermi Large Area Telescope Fourth Source Catalog Data Release 4 (4FGL-DR4)}",
      journal = {arXiv e-prints},
         year = 2023,
        month = jul,
          eid = {arXiv:2307.12546},
        pages = {arXiv:2307.12546},
          doi = {10.48550/arXiv.2307.12546},
archivePrefix = {arXiv},
       eprint = {2307.12546},
 primaryClass = {astro-ph.HE},
       adsurl = {https://ui.adsabs.harvard.edu/abs/2023arXiv230712546B}
}

@ARTICLE{RFC,
       author = {{Petrov}, L.~Y. and {Kovalev}, Y.~Y.},
        title = "{The Radio Fundamental Catalog. I. Astrometry}",
      journal = {\apjs},
         year = 2025,
        month = feb,
       volume = {276},
       number = {2},
          eid = {38},
        pages = {38},
          doi = {10.3847/1538-4365/ad8c36},
archivePrefix = {arXiv},
       eprint = {2410.11794},
 primaryClass = {astro-ph.IM},
       adsurl = {https://ui.adsabs.harvard.edu/abs/2025ApJS..276...38P}
}

@ARTICLE{Lott12,
       author = {{Lott}, B. and {Escande}, L. and {Larsson}, S. and {Ballet}, J.},
        title = "{An adaptive-binning method for generating constant-uncertainty/constant-significance light curves with Fermi-LAT data}",
      journal = {\aap},
         year = 2012,
        month = aug,
       volume = {544},
          eid = {A6},
        pages = {A6},
          doi = {10.1051/0004-6361/201218873},
archivePrefix = {arXiv},
       eprint = {1201.4851},
 primaryClass = {astro-ph.HE},
       adsurl = {https://ui.adsabs.harvard.edu/abs/2012A&A...544A...6L}
}

@ARTICLE{2025MNRAS.540.3170U,
       author = {{Uskov}, G.~S. and {Sazonov}, S. and {Lapshov}, I. and {Mikhailov}, A.~G. and {Filippova}, E. and {Lutovinov}, A. and {Mereminskiy}, I.~A. and {Mochalina}, M. and {Semena}, A. and {Tkachenko}, A.},
        title = "{SRGA J2306 + 1556: an extremely X-ray luminous, heavily obscured, radio-loud quasar at z = 0.44 discovered by SRG/ART-XC}",
      journal = {\mnras},
         year = 2025,
        month = jul,
       volume = {540},
       number = {4},
        pages = {3170-3185},
          doi = {10.1093/mnras/staf924},
archivePrefix = {arXiv},
       eprint = {2504.13658},
 primaryClass = {astro-ph.HE},
       adsurl = {https://ui.adsabs.harvard.edu/abs/2025MNRAS.540.3170U}
}

@ARTICLE{2021A&A...656A.132S,
       author = {{Sunyaev}, R. and {Arefiev}, V. and {Babyshkin}, V. and {Bogomolov}, A. and {Borisov}, K. and {Buntov}, M. and {Brunner}, H. and {Burenin}, R. and {Churazov}, E. and {Coutinho}, D. and {Eder}, J. and {Eismont}, N. and {Freyberg}, M. and {Gilfanov}, M. and {Gureyev}, P. and {Hasinger}, G. and {Khabibullin}, I. and {Kolmykov}, V. and {Komovkin}, S. and {Krivonos}, R. and {Lapshov}, I. and {Levin}, V. and {Lomakin}, I. and {Lutovinov}, A. and {Medvedev}, P. and {Merloni}, A. and {Mernik}, T. and {Mikhailov}, E. and {Molodtsov}, V. and {Mzhelsky}, P. and {M{\"u}ller}, S. and {Nandra}, K. and {Nazarov}, V. and {Pavlinsky}, M. and {Poghodin}, A. and {Predehl}, P. and {Robrade}, J. and {Sazonov}, S. and {Scheuerle}, H. and {Shirshakov}, A. and {Tkachenko}, A. and {Voron}, V.},
        title = "{SRG X-ray orbital observatory. Its telescopes and first scientific results}",
      journal = {\aap},
         year = 2021,
        month = dec,
       volume = {656},
          eid = {A132},
        pages = {A132},
          doi = {10.1051/0004-6361/202141179},
archivePrefix = {arXiv},
       eprint = {2104.13267},
 primaryClass = {astro-ph.HE},
       adsurl = {https://ui.adsabs.harvard.edu/abs/2021A&A...656A.132S}
}

@ARTICLE{2021A&A...650A..42P,
       author = {{Pavlinsky}, M. and {Tkachenko}, A. and {Levin}, V. and {Alexandrovich}, N. and {Arefiev}, V. and {Babyshkin}, V. and {Batanov}, O. and {Bodnar}, Yu. and {Bogomolov}, A. and {Bubnov}, A. and {Buntov}, M. and {Burenin}, R. and {Chelovekov}, I. and {Chen}, C.-T. and {Drozdova}, T. and {Ehlert}, S. and {Filippova}, E. and {Frolov}, S. and {Gamkov}, D. and {Garanin}, S. and {Garin}, M. and {Glushenko}, A. and {Gorelov}, A. and {Grebenev}, S. and {Grigorovich}, S. and {Gureev}, P. and {Gurova}, E. and {Ilkaev}, R. and {Katasonov}, I. and {Krivchenko}, A. and {Krivonos}, R. and {Korotkov}, F. and {Kudelin}, M. and {Kuznetsova}, M. and {Lazarchuk}, V. and {Lomakin}, I. and {Lapshov}, I. and {Lipilin}, V. and {Lutovinov}, A. and {Mereminskiy}, I. and {Molkov}, S. and {Nazarov}, V. and {Oleinikov}, V. and {Pikalov}, E. and {Ramsey}, B.~D. and {Roiz}, I. and {Rotin}, A. and {Ryadov}, A. and {Sankin}, E. and {Sazonov}, S. and {Sedov}, D. and {Semena}, A. and {Semena}, N. and {Serbinov}, D. and {Shirshakov}, A. and {Shtykovsky}, A. and {Shvetsov}, A. and {Sunyaev}, R. and {Swartz}, D.~A. and {Tambov}, V. and {Voron}, V. and {Yaskovich}, A.},
        title = "{The ART-XC telescope on board the SRG observatory}",
      journal = {\aap},
         year = 2021,
        month = jun,
       volume = {650},
          eid = {A42},
        pages = {A42},
          doi = {10.1051/0004-6361/202040265},
archivePrefix = {arXiv},
       eprint = {2103.12479},
 primaryClass = {astro-ph.HE},
       adsurl = {https://ui.adsabs.harvard.edu/abs/2021A&A...650A..42P}
}

@ARTICLE{2004ApJ...611.1005G,
       author = {{Gehrels}, N. and {Chincarini}, G. and {Giommi}, P. and {Mason}, K.~O. and {Nousek}, J.~A. and {Wells}, A.~A. and {White}, N.~E. and {Barthelmy}, S.~D. and {Burrows}, D.~N. and {Cominsky}, L.~R. and {Hurley}, K.~C. and {Marshall}, F.~E. and {M{\'e}sz{\'a}ros}, P. and {Roming}, P.~W.~A. and {Angelini}, L. and {Barbier}, L.~M. and {Belloni}, T. and {Campana}, S. and {Caraveo}, P.~A. and {Chester}, M.~M. and {Citterio}, O. and {Cline}, T.~L. and {Cropper}, M.~S. and {Cummings}, J.~R. and {Dean}, A.~J. and {Feigelson}, E.~D. and {Fenimore}, E.~E. and {Frail}, D.~A. and {Fruchter}, A.~S. and {Garmire}, G.~P. and {Gendreau}, K. and {Ghisellini}, G. and {Greiner}, J. and {Hill}, J.~E. and {Hunsberger}, S.~D. and {Krimm}, H.~A. and {Kulkarni}, S.~R. and {Kumar}, P. and {Lebrun}, F. and {Lloyd-Ronning}, N.~M. and {Markwardt}, C.~B. and {Mattson}, B.~J. and {Mushotzky}, R.~F. and {Norris}, J.~P. and {Osborne}, J. and {Paczynski}, B. and {Palmer}, D.~M. and {Park}, H.-S. and {Parsons}, A.~M. and {Paul}, J. and {Rees}, M.~J. and {Reynolds}, C.~S. and {Rhoads}, J.~E. and {Sasseen}, T.~P. and {Schaefer}, B.~E. and {Short}, A.~T. and {Smale}, A.~P. and {Smith}, I.~A. and {Stella}, L. and {Tagliaferri}, G. and {Takahashi}, T. and {Tashiro}, M. and {Townsley}, L.~K. and {Tueller}, J. and {Turner}, M.~J.~L. and {Vietri}, M. and {Voges}, W. and {Ward}, M.~J. and {Willingale}, R. and {Zerbi}, F.~M. and {Zhang}, W.~W.},
        title = "{The Swift Gamma-Ray Burst Mission}",
      journal = {\apj},
         year = 2004,
        month = aug,
       volume = {611},
       number = {2},
        pages = {1005-1020},
          doi = {10.1086/422091},
archivePrefix = {arXiv},
       eprint = {astro-ph/0405233},
 primaryClass = {astro-ph},
       adsurl = {https://ui.adsabs.harvard.edu/abs/2004ApJ...611.1005G}
}

@ARTICLE{2025ATel17540....1V,
       author = {{Vlasyuk}, V.~V. and {Spiridonova}, O.~I.},
        title = "{The CSS 3C 138: secondary R band maximum after 40 days}",
      journal = {The Astronomer's Telegram},
         year = 2025,
        month = dec,
       volume = {17540},
        pages = {1},
       adsurl = {https://ui.adsabs.harvard.edu/abs/2025ATel17540....1V}
}

@ARTICLE{2009ApJ...696..870D,
       author = {{Drake}, A.~J. and {Djorgovski}, S.~G. and {Mahabal}, A. and {Beshore}, E. and {Larson}, S. and {Graham}, M.~J. and {Williams}, R. and {Christensen}, E. and {Catelan}, M. and {Boattini}, A. and {Gibbs}, A. and {Hill}, R. and {Kowalski}, R.},
        title = "{First Results from the Catalina Real-Time Transient Survey}",
      journal = {\apj},
         year = 2009,
        month = may,
       volume = {696},
       number = {1},
        pages = {870-884},
          doi = {10.1088/0004-637X/696/1/870},
archivePrefix = {arXiv},
       eprint = {0809.1394},
 primaryClass = {astro-ph},
       adsurl = {https://ui.adsabs.harvard.edu/abs/2009ApJ...696..870D}
}

@ARTICLE{2024MNRAS.535.2775V,
       author = {{Vlasyuk}, V.~V. and {Sotnikova}, Y.~V. and {Volvach}, A.~E. and {Mufakharov}, T.~V. and {Kovalev}, Y.~A. and {Spiridonova}, O.~I. and {Khabibullina}, M.~L. and {Kovalev}, Y.~Y. and {Mikhailov}, A.~G. and {Stolyarov}, V.~A. and {Kudryavtsev}, D.~O. and {Mingaliev}, M.~G. and {Razzaque}, S. and {Semenova}, T.~A. and {Kudryashova}, A.~K. and {Bursov}, N.~N. and {Trushkin}, S.~A. and {Popkov}, A.~V. and {Erkenov}, A.~K. and {Rakhimov}, I.~A. and {Kharinov}, M.~A. and {Gurwell}, M.~A. and {Tsybulev}, P.~G. and {Moskvitin}, A.~S. and {Fatkhullin}, T.~A. and {Emelianov}, E.~V. and {Arshinova}, A. and {Iuzhanina}, K.~V. and {Andreeva}, T.~S. and {Volvach}, L.~N. and {Ghosh}, A.},
        title = "{Multiwavelength variability of the blazar AO 0235+164}",
      journal = {\mnras},
         year = 2024,
        month = dec,
       volume = {535},
       number = {3},
        pages = {2775-2799},
          doi = {10.1093/mnras/stae2491},
archivePrefix = {arXiv},
       eprint = {2411.01497},
 primaryClass = {astro-ph.HE},
       adsurl = {https://ui.adsabs.harvard.edu/abs/2024MNRAS.535.2775V}
}

@ARTICLE{2005SSRv..120..165B,
       author = {{Burrows}, David N. and {Hill}, J.~E. and {Nousek}, J.~A. and {Kennea}, J.~A. and {Wells}, A. and {Osborne}, J.~P. and {Abbey}, A.~F. and {Beardmore}, A. and {Mukerjee}, K. and {Short}, A.~D.~T. and {Chincarini}, G. and {Campana}, S. and {Citterio}, O. and {Moretti}, A. and {Pagani}, C. and {Tagliaferri}, G. and {Giommi}, P. and {Capalbi}, M. and {Tamburelli}, F. and {Angelini}, L. and {Cusumano}, G. and {Br{\"a}uninger}, H.~W. and {Burkert}, W. and {Hartner}, G.~D.},
        title = "{The Swift X-Ray Telescope}",
      journal = {\ssr},
         year = 2005,
        month = oct,
       volume = {120},
       number = {3-4},
        pages = {165-195},
          doi = {10.1007/s11214-005-5097-2},
archivePrefix = {arXiv},
       eprint = {astro-ph/0508071},
 primaryClass = {astro-ph},
       adsurl = {https://ui.adsabs.harvard.edu/abs/2005SSRv..120..165B}
}

@ARTICLE{2026ATel17645....1U,
       author = {{Uskov}, G. and {Sazonov}, S. and {Lapshov}, I. and {Tkachenko}, A. and {Lutovinov}, A. and {Mikhailov}, A.~G. and {Sotnikova}, Yu. V. and {Mufakharov}, T.~V.},
        title = "{SRG/ART-XC captures 3C 138 in a bright X-ray state}",
      journal = {The Astronomer's Telegram},
         year = 2026,
        month = feb,
       volume = {17645},
        pages = {1},
       adsurl = {https://ui.adsabs.harvard.edu/abs/2026ATel17645....1U}
}

@ARTICLE{2025ApJ...987L..26W,
       author = {{Wang}, Ailing and {An}, Tao and {Kellermann}, Kenneth I. and {Feng}, Hua and {Bempong-Manful}, Emmanuel K. and {Timmerman}, Roland and {Guo}, Shaoguang},
        title = "{A Relativistic Jet in the Radio-quiet Active Galactic Nucleus Mrk 110}",
      journal = {\apjl},
         year = 2025,
        month = jul,
       volume = {987},
       number = {2},
          eid = {L26},
        pages = {L26},
          doi = {10.3847/2041-8213/ade14a},
archivePrefix = {arXiv},
       eprint = {2506.03970},
 primaryClass = {astro-ph.GA},
       adsurl = {https://ui.adsabs.harvard.edu/abs/2025ApJ...987L..26W}
}

@ARTICLE{1980ARA&A..18..165M,
       author = {{Miley}, G.},
        title = "{The structure of extended extragalactic radio sources}",
      journal = {\araa},
         year = 1980,
        month = jan,
       volume = {18},
        pages = {165-218},
          doi = {10.1146/annurev.aa.18.090180.001121},
       adsurl = {https://ui.adsabs.harvard.edu/abs/1980ARA&A..18..165M}
}

@ARTICLE{1999A&A...349...45T,
       author = {{T{\"u}rler}, M. and {Courvoisier}, T.~J. -L. and {Paltani}, S.},
        title = "{Modelling the submillimetre-to-radio flaring behaviour of 3C 273}",
      journal = {\aap},
         year = 1999,
        month = sep,
       volume = {349},
        pages = {45-54},
          doi = {10.48550/arXiv.astro-ph/9906274},
archivePrefix = {arXiv},
       eprint = {astro-ph/9906274},
 primaryClass = {astro-ph},
       adsurl = {https://ui.adsabs.harvard.edu/abs/1999A&A...349...45T}
}

@INPROCEEDINGS{2020gbar.conf...32S,
       author = {{Sotnikova}, Yu. V.},
        title = "{RATAN-600 Radio Telescope: Observing Programs and Outlook}",
    booktitle = {Ground-Based Astronomy in Russia. 21st Century},
         year = 2020,
       editor = {{Romanyuk}, I.~I. and {Yakunin}, I.~A. and
                 {Valeev}, A.~F. and {Kudryavtsev}, D.~O.},
    publisher = {Special Astrophysical Observatory of the Russian Academy of Sciences},
      address = {Nizhny Arkhyz, Russia},
        month = dec,
        pages = {32--40},
          doi = {10.26119/978-5-6045062-0-2_2020_32},
         isbn = {978-5-6045062-0-2},
       adsurl = {https://ui.adsabs.harvard.edu/abs/2020gbar.conf...32S}
}

@ARTICLE{2009A&A...494..527H,
       author = {{Hovatta}, T. and {Valtaoja}, E. and {Tornikoski}, M. and {L{\"a}hteenm{\"a}ki}, A.},
        title = "{Doppler factors, Lorentz factors and viewing angles for quasars, BL Lacertae objects and radio galaxies}",
      journal = {\aap},
         year = 2009,
        month = feb,
       volume = {494},
       number = {2},
        pages = {527-537},
          doi = {10.1051/0004-6361:200811150},
archivePrefix = {arXiv},
       eprint = {0811.4278},
 primaryClass = {astro-ph},
       adsurl = {https://ui.adsabs.harvard.edu/abs/2009A&A...494..527H}
}

@ARTICLE{2018AstBu..73..494T,
       author = {{Tsybulev}, P.~G. and {Nizhelskii}, N.~A. and {Dugin}, M.~V. and {Borisov}, A.~N. and {Kratov}, D.~V. and {Udovitskii}, R. Yu.},
        title = "{C-Band Radiometer for Continuum Observations at RATAN-600 Radio Telescope}",
      journal = {Astrophysical Bulletin},
         year = 2018,
        month = oct,
       volume = {73},
       number = {4},
        pages = {494-500},
          doi = {10.1134/S1990341318040132},
       adsurl = {https://ui.adsabs.harvard.edu/abs/2018AstBu..73..494T}
}

@ARTICLE{1990A&A...231..333F,
       author = {{Fanti}, R. and {Fanti}, C. and {Schilizzi}, R.~T. and {Spencer}, R.~E. and {Nan Rendong} and {Parma}, P. and {van Breugel}, W.~J.~M. and {Venturi}, T.},
        title = "{On the nature of compact steep spectrum radio sources.}",
      journal = {\aap},
         year = 1990,
        month = may,
       volume = {231},
        pages = {333-346},
       adsurl = {https://ui.adsabs.harvard.edu/abs/1990A&A...231..333F}
}

@ARTICLE{2019MNRAS.482.2336P,
       author = {{Pushkarev}, A.~B. and {Butuzova}, M.~S. and {Kovalev}, Y.~Y. and {Hovatta}, T.},
        title = "{Multifrequency study of the gamma-ray flaring BL Lacertae object PKS 2233-148 in 2009-2012}",
      journal = {\mnras},
         year = 2019,
        month = jan,
       volume = {482},
       number = {2},
        pages = {2336-2353},
          doi = {10.1093/mnras/sty2724},
archivePrefix = {arXiv},
       eprint = {1808.06138},
 primaryClass = {astro-ph.HE},
       adsurl = {https://ui.adsabs.harvard.edu/abs/2019MNRAS.482.2336P}
}

@ARTICLE{2026A&A...710A..63L,
       author = {{Li}, Shan and {Lee}, Sang-Sung and {Cheong}, Whee Yeon and {An}, Tao and {Kameno}, Seiji and {Kneissl}, Ruediger},
        title = "{Constraining the magnetic field strength of a flaring radio core in the compact steep spectrum source 3C 138}",
      journal = {\aap},
         year = 2026,
        month = jun,
       volume = {710},
          eid = {A63},
        pages = {A63},
          doi = {10.1051/0004-6361/202659759},
archivePrefix = {arXiv},
       eprint = {2605.18283},
 primaryClass = {astro-ph.GA},
       adsurl = {https://ui.adsabs.harvard.edu/abs/2026A&A...710A..63L}
}

@ARTICLE{2005ApJ...622..797K,
       author = {{Kataoka}, Jun and {Stawarz}, {\L}ukasz},
        title = "{X-Ray Emission Properties of Large-Scale Jets, Hot Spots, and Lobes in Active Galactic Nuclei}",
      journal = {\apj},
         year = 2005,
        month = apr,
       volume = {622},
       number = {2},
        pages = {797-810},
          doi = {10.1086/428083},
archivePrefix = {arXiv},
       eprint = {astro-ph/0411042},
 primaryClass = {astro-ph},
       adsurl = {https://ui.adsabs.harvard.edu/abs/2005ApJ...622..797K}
}

@ARTICLE{2014Msngr.155...19F,
       author = {{Fomalont}, E. and {van Kempen}, T. and {Kneissl}, R. and {Marcelino}, N. and {Barkats}, D. and {Corder}, S. and {Cortes}, P. and {Hills}, R. and {Lucas}, R. and {Manning}, A. and {Peck}, A.},
        title = "{The Calibration of ALMA using Radio Sources}",
      journal = {The Messenger},
         year = 2014,
        month = mar,
       volume = {155},
        pages = {19-22},
       adsurl = {https://ui.adsabs.harvard.edu/abs/2014Msngr.155...19F}
}

@ARTICLE{2003MNRAS.345.1271V,
       author = {{Vaughan}, S. and {Edelson}, R. and {Warwick}, R.~S. and {Uttley}, P.},
        title = "{On characterizing the variability properties of X-ray light curves from active galaxies}",
      journal = {\mnras},
         year = 2003,
        month = nov,
       volume = {345},
       number = {4},
        pages = {1271-1284},
          doi = {10.1046/j.1365-2966.2003.07042.x},
archivePrefix = {arXiv},
       eprint = {astro-ph/0307420},
 primaryClass = {astro-ph},
       adsurl = {https://ui.adsabs.harvard.edu/abs/2003MNRAS.345.1271V}
}

@ARTICLE{1995A&A...302..317F,
       author = {{Fanti}, C. and {Fanti}, R. and {Dallacasa}, D. and {Schilizzi}, R.~T. and {Spencer}, R.~E. and {Stanghellini}, C.},
        title = "{Are compact steep-spectrum sources young?}",
      journal = {\aap},
         year = 1995,
        month = oct,
       volume = {302},
        pages = {317},
       adsurl = {https://ui.adsabs.harvard.edu/abs/1995A&A...302..317F}
}

@ARTICLE{2009DatSJ...8...34V,
       author = {{Verkhodanov}, O.~V. and {Trushkin}, S.~A. and {Andernach}, H. and {Chernenkov}, V.~N.},
        title = "{The CATS Service: An Astrophysical Research Tool}",
      journal = {Data Science Journal},
         year = 2009,
        month = mar,
       volume = {8},
        pages = {34-40},
          doi = {10.2481/dsj.8.34},
archivePrefix = {arXiv},
       eprint = {0901.3118},
 primaryClass = {astro-ph.IM},
       adsurl = {https://ui.adsabs.harvard.edu/abs/2009DatSJ...8...34V}
}

@STRING(pasp="PASP")

@ARTICLE{1983ApJ...264..296M,
   author = {{Marscher}, A.~P.},
    title = "{Accurate formula for the self-Compton X-ray flux density from a uniform, spherical, compact radio source}",
  journal = {\apj},
     year = 1983,
    month = jan,
   volume = 264,
    pages = {296-+},
      doi = {10.1086/160597},
   adsurl = {http://adsabs.harvard.edu/abs/1983ApJ...264..296M}
}

@ARTICLE{1979ApJ...232...34B,
       author = {{Blandford}, R.~D. and {K{\"o}nigl}, A.},
        title = "{Relativistic jets as compact radio sources.}",
      journal = {\apj},
         year = 1979,
        month = aug,
       volume = {232},
        pages = {34-48},
          doi = {10.1086/157262},
       adsurl = {https://ui.adsabs.harvard.edu/abs/1979ApJ...232...34B}
}

@ARTICLE{1985ApJ...298..301H,
       author = {{Hughes}, P.~A. and {Aller}, H.~D. and {Aller}, M.~F.},
        title = "{Polarized radio outbursts in BL Lacertae. II. The flux and polarization of a piston-driven shock.}",
      journal = {\apj},
         year = 1985,
        month = nov,
       volume = {298},
        pages = {301-315},
          doi = {10.1086/163611},
       adsurl = {https://ui.adsabs.harvard.edu/abs/1985ApJ...298..301H}
}

@ARTICLE{1992A&A...254...71V,
       author = {{Valtaoja}, E. and {Terasranta}, H. and {Urpo}, S. and {Nesterov}, N.~S. and {Lainela}, M. and {Valtonen}, M.},
        title = "{Five years monitoring of extragalactic radio sources. III. Generalized shock models and the dependence of variability on frequency.}",
      journal = {\aap},
         year = 1992,
        month = feb,
       volume = {254},
        pages = {71-79},
       adsurl = {https://ui.adsabs.harvard.edu/abs/1992A&A...254...71V}
}

@ARTICLE{1998A&A...330...79L,
       author = {{Lobanov}, A.~P.},
        title = "{Ultracompact jets in active galactic nuclei}",
      journal = {\aap},
         year = 1998,
        month = feb,
       volume = {330},
        pages = {79-89},
          doi = {10.48550/arXiv.astro-ph/9712132},
archivePrefix = {arXiv},
       eprint = {astro-ph/9712132},
 primaryClass = {astro-ph},
       adsurl = {https://ui.adsabs.harvard.edu/abs/1998A&A...330...79L}
}

@ARTICLE{2005AJ....130.1418J,
       author = {{Jorstad}, Svetlana G. and {Marscher}, Alan P. and {Lister}, Matthew L. and {Stirling}, Alastair M. and {Cawthorne}, Timothy V. and {Gear}, Walter K. and {G{\'o}mez}, Jos{\'e} L. and {Stevens}, Jason A. and {Smith}, Paul S. and {Forster}, James R. and {Robson}, E. Ian},
        title = "{Polarimetric Observations of 15 Active Galactic Nuclei at High Frequencies: Jet Kinematics from Bimonthly Monitoring with the Very Long Baseline Array}",
      journal = {\aj},
         year = 2005,
        month = oct,
       volume = {130},
       number = {4},
        pages = {1418-1465},
          doi = {10.1086/444593},
archivePrefix = {arXiv},
       eprint = {astro-ph/0502501},
 primaryClass = {astro-ph},
       adsurl = {https://ui.adsabs.harvard.edu/abs/2005AJ....130.1418J}
}

@ARTICLE{1994A&A...284..331O,
   author = {{Ott}, M. and {Witzel}, A. and {Quirrenbach}, A. and {Krichbaum}, T.~P. and
	{Standke}, K.~J. and {Schalinski}, C.~J. and {Hummel}, C.~A.},
    title = "{An updated list of radio flux density calibrators}",
  journal = {\aap},
     year = 1994,
    month = apr,
   volume = 284,
    pages = {331-339},
   adsurl = {http://adsabs.harvard.edu/abs/1994A%26A...284..331O}
}

@ARTICLE{1980A&AS...39..379T,
   author = {{Tabara}, H. and {Inoue}, M.},
    title = "{A catalogue of linear polarization of radio sources}",
  journal = {\aaps},
     year = 1980,
    month = mar,
   volume = 39,
    pages = {379-393},
   adsurl = {http://adsabs.harvard.edu/abs/980A%26AS...39..379T}
}

@ARTICLE{1997ASPC..125...46V,
   author = {{Verkhodanov}, O.~V.},
    title = "{Multiwave Continuum Data Reduction at RATAN-600}",
  journal = {Astronomical Data Analysis Software and Systems VI, A.S.P. Conference Series},
     year = 1997,
    month = mar,
   volume = 125,
    pages = {46-49},
   adsurl = {http://adsabs.harvard.edu/abs/1997ASPC..125...46V}
}

@ARTICLE{1995PASP..107..803U,
   author = {{Urry}, C.~M. and {Padovani}, P.},
    title = "{Unified Schemes for Radio-Loud Active Galactic Nuclei}",
  journal = {\pasp},
     year = 1995,
    month = sep,
   volume = 107,
    pages = {803-845},
      doi = {10.1086/133630},
   adsurl = {http://adsabs.harvard.edu/abs/1995PASP..107..803U}
}

@ARTICLE{1993IAPM...35....7P,
   author = {{Parijskij}, Y.~N.},
    title = "{RATAN-600 - The world's biggest reflector at the 'cross roads'}",
  journal = {IEEE Antennas and Propagation Magazine},
     year = 1993,
    month = aug,
   volume = 35,
    pages = {7-12},
      doi = {10.1109/74.229840},
   adsurl = {http://adsabs.harvard.edu/abs/1993IAPM...35....7P}
}

@article{2000AJ....120.1579P,
doi = {10.1086/301513},
url = {https://doi.org/10.1086/301513},
year = {2000},
month = {sep},
publisher = {},
volume = {120},
number = {3},
pages = {1579},
author = {York, Donald G. and Adelman, J. and Anderson, Jr., John E. and Anderson, Scott F. and Annis, James and Bahcall, Neta A. and Bakken, J. A. and Barkhouser, Robert and Bastian, Steven and Berman, Eileen and Boroski, William N. and Bracker, Steve and Briegel, Charlie and Briggs, John W. and Brinkmann, J. and Brunner, Robert and Burles, Scott and Carey, Larry and Carr, Michael A. and Castander, Francisco J. and Chen, Bing and Colestock, Patrick L. and Connolly, A. J. and Crocker, J. H. and Csabai, István and Czarapata, Paul C. and Davis, John Eric and Doi, Mamoru and Dombeck, Tom and Eisenstein, Daniel and Ellman, Nancy and Elms, Brian R. and Evans, Michael L. and Fan, Xiaohui and Federwitz, Glenn R. and Fiscelli, Larry and Friedman, Scott and Frieman, Joshua A. and Fukugita, Masataka and Gillespie, Bruce and Gunn, James E. and Gurbani, Vijay K. and de Haas, Ernst and Haldeman, Merle and Harris, Frederick H. and Hayes, J. and Heckman, Timothy M. and Hennessy, G. S. and Hindsley, Robert B. and Holm, Scott and Holmgren, Donald J. and Huang, Chi-hao and Hull, Charles and Husby, Don and Ichikawa, Shin-Ichi and Ichikawa, Takashi and Ivezić, Željko and Kent, Stephen and Kim, Rita S. J. and Kinney, E. and Klaene, Mark and Kleinman, A. N. and Kleinman, S. and Knapp, G. R. and Korienek, John and Kron, Richard G. and Kunszt, Peter Z. and Lamb, D. Q. and Lee, B. and Leger, R. French and Limmongkol, Siriluk and Lindenmeyer, Carl and Long, Daniel C. and Loomis, Craig and Loveday, Jon and Lucinio, Rich and Lupton, Robert H. and MacKinnon, Bryan and Mannery, Edward J. and Mantsch, P. M. and Margon, Bruce and McGehee, Peregrine and McKay, Timothy A. and Meiksin, Avery and Merelli, Aronne and Monet, David G. and Munn, Jeffrey A. and Narayanan, Vijay K. and Nash, Thomas and Neilsen, Eric and Neswold, Rich and Newberg, Heidi Jo and Nichol, R. C. and Nicinski, Tom and Nonino, Mario and Okada, Norio and Okamura, Sadanori and Ostriker, Jeremiah P. and Owen, Russell and Pauls, A. George and Peoples, John and Peterson, R. L. and Petravick, Donald and Pier, Jeffrey R. and Pope, Adrian and Pordes, Ruth and Prosapio, Angela and Rechenmacher, Ron and Quinn, Thomas R. and Richards, Gordon T. and Richmond, Michael W. and Rivetta, Claudio H. and Rockosi, Constance M. and Ruthmansdorfer, Kurt and Sandford, Dale and Schlegel, David J. and Schneider, Donald P. and Sekiguchi, Maki and Sergey, Gary and Shimasaku, Kazuhiro and Siegmund, Walter A. and Smee, Stephen and Smith, J. Allyn and Snedden, S. and Stone, R. and Stoughton, Chris and Strauss, Michael A. and Stubbs, Christopher and SubbaRao, Mark and Szalay, Alexander S. and Szapudi, Istvan and Szokoly, Gyula P. and Thakar, Anirudda R. and Tremonti, Christy and Tucker, Douglas L. and Uomoto, Alan and Vanden Berk, Dan and Vogeley, Michael S. and Waddell, Patrick and Wang, Shu-i and Watanabe, Masaru and Weinberg, David H. and Yanny, Brian and Yasuda, Naoki},
title = {The Sloan Digital Sky Survey: Technical Summary},
journal = {The Astronomical Journal}
}

@ARTICLE{2011AstBu..66..109T,
   author = {{Tsybulev}, P.~G.},
    title = "{New-generation data acquisition and control system for continuum radio-astronomic observations with RATAN-600 radio telescope: Development, observations, and measurements}",
  journal = {Astrophysical Bulletin},
     year = 2011,
    month = jan,
   volume = 66,
    pages = {109-122},
      doi = {10.1134/S199034131101010X},
   adsurl = {http://adsabs.harvard.edu/abs/2011AstBu..66..109T}
}

@ARTICLE{1991ApJ...374..431S,
   author = {{Stickel}, M. and {Padovani}, P. and {Urry}, C.~M. and {Fried}, J.~W. and 
	{Kuehr}, H.},
    title = "{The complete sample of 1 Jansky BL Lacertae objects. I - Summary properties}",
  journal = {\apj},
     year = 1991,
    month = jun,
   volume = 374,
    pages = {431-439},
      doi = {10.1086/170133},
   adsurl = {http://adsabs.harvard.edu/abs/1991ApJ...374..431S}
}

@ARTICLE{1979S&T....57..324K,
   author = {{Korolkov}, D.~V. and {Pariiskii}, I.~N.},
    title = "{The Soviet RATAN-600 radio telescope}",
  journal = {\skytel},
     year = 1979,
    month = apr,
   volume = 57,
    pages = {324-329},
   adsurl = {http://adsabs.harvard.edu/abs/1979S%26T....57..324K}
}

@ARTICLE{1977A&A....61...99B,
   author = {{Baars}, J.~W.~M. and {Genzel}, R. and {Pauliny-Toth}, I.~I.~K. and
    {Witzel}, A.},
    title = "{The absolute spectrum of CAS A - an accurate flux density scale and a set of secondary calibrators}",
  journal = {\aap},
     year = 1977,
    month = oct,
   volume = 61,
    pages = {99-106},
   adsurl = {http://adsabs.harvard.edu/abs/1977A%26A....61...99B}
}

@ARTICLE{2014A&A...572A..59M,
author = {{Mingaliev}, M.~G. and {Sotnikova}, Y.~V. and {Udovitskiy}, R.~Y. and
    {Mufakharov}, T.~V. and {Nieppola}, E. and {Erkenov}, A.~K.},
    title = "{RATAN-600 multi-frequency data for the BL Lacertae objects}",
  journal = {\aap},
archivePrefix = "arXiv",
   eprint = {1410.2835},
     year = 2014,
    month = dec,
   volume = 572,
      eid = {A59},
    pages = {A59},
      doi = {10.1051/0004-6361/201424437},
   adsurl = {http://adsabs.harvard.edu/abs/2014A%26A...572A..59M}
}

@ARTICLE{2026ATel17681....1R,
       author = {{Rani}, S. and {Lewis}, T. and {La Mura}, G.},
        title = "{Fermi-LAT detection of renewed gamma-ray activity from the CSS 3C 138}",
      journal = {The Astronomer's Telegram},
         year = 2026,
        month = feb,
       volume = {17681},
        pages = {1},
       adsurl = {https://ui.adsabs.harvard.edu/abs/2026ATel17681....1R}
}

@ARTICLE{2011ApJS..194...29R,
   author = {{Richards}, J.~L. and {Max-Moerbeck}, W. and {Pavlidou}, V. and 
	{King}, O.~G. and {Pearson}, T.~J. and {Readhead}, A.~C.~S. and 
	{Reeves}, R. and {Shepherd}, M.~C. and {Stevenson}, M.~A. and 
	{Weintraub}, L.~C. and {Fuhrmann}, L. and {Angelakis}, E. and 
	{Zensus}, J.~A. and {Healey}, S.~E. and {Romani}, R.~W. and 
	{Shaw}, M.~S. and {Grainge}, K. and {Birkinshaw}, M. and {Lancaster}, K. and 
	{Worrall}, D.~M. and {Taylor}, G.~B. and {Cotter}, G. and {Bustos}, R.
	},
    title = "{Blazars in the Fermi Era: The OVRO 40 m Telescope Monitoring Program}",
  journal = {\apjs},
archivePrefix = "arXiv",
   eprint = {1011.3111},
 primaryClass = "astro-ph.CO",
     year = 2011,
    month = jun,
   volume = 194,
      eid = {29},
    pages = {29},
      doi = {10.1088/0067-0049/194/2/29},
   adsurl = {http://adsabs.harvard.edu/abs/2011ApJS..194...29R}
}

@ARTICLE{2016AstBu..71..496U,
       author = {{Udovitskiy}, R. Yu. and {Sotnikova}, Yu. V. and {Mingaliev}, M.~G. and
         {Tsybulev}, P.~G. and {Zhekanis}, G.~V. and {Nizhelskij}, N.~A.},
        title = "{Automated system for reduction of observational data on RATAN-600 radio telescope}",
      journal = {Astrophysical Bulletin},
         year = 2016,
        month = oct,
       volume = {71},
       number = {4},
        pages = {496-505},
          doi = {10.1134/S1990341316040131},
       adsurl = {https://ui.adsabs.harvard.edu/abs/2016AstBu..71..496U}
}

@ARTICLE{2025ATel17104....1S,
       author = {{Sotnikova}, Yu. V. and {Mufakharov}, T.~V. and {Erkenov}, A.~K. and {Udovitskiy}, R. Yu. and {Semenova}, T.~A. and {Trushkin}, S.~A. and {Kovalev}, Yu. A. and {Kovalev}, Y.~Y. and {Popkov}, A.~V. and {An}, Tao},
        title = "{Radio flare in CSS quasar 3C 138}",
      journal = {The Astronomer's Telegram},
         year = 2025,
        month = mar,
       volume = {17104},
        pages = {1},
       adsurl = {https://ui.adsabs.harvard.edu/abs/2025ATel17104....1S}
}

@ARTICLE{2025ATel17496....1H,
       author = {{Hu}, Ruixiang and {Yang}, Shuaikang and {Fan}, Xiao and {He}, Han and {Xu}, Saien and {You}, Bei and {Zhu}, Zonghong and {Li}, Xiaoyan and {Zhang}, Kai and {Li}, Zhengyang and {Zhou}, Tong and {Cong}, Jianan and {Zheng}, Kaiwen and {Yang}, Yiqiao and {Chen}, Chao and {Wu}, Jiajia and {Liang}, Yuan},
        title = "{Photometric Follow-up of the Quasar 3C 138}",
      journal = {The Astronomer's Telegram},
         year = 2025,
        month = nov,
       volume = {17496},
        pages = {1},
       adsurl = {https://ui.adsabs.harvard.edu/abs/2025ATel17496....1H}
}

@ARTICLE{2025ATel17107....1M,
       author = {{Monti-Guarnieri}, Pietro and {La Mura}, Giovanni},
        title = "{Fermi-LAT detection of renewed gamma-ray activity from the CSS Quasar 3C 138}",
      journal = {The Astronomer's Telegram},
         year = 2025,
        month = mar,
       volume = {17107},
        pages = {1},
       adsurl = {https://ui.adsabs.harvard.edu/abs/2025ATel17107....1M}
}

@ARTICLE{2025ATel17461....1L,
       author = {{Lopez-Perez}, S. and {La Mura}, G.},
        title = "{Fermi-LAT detection of renewed gamma-ray activity from the CSS 3C 138}",
      journal = {The Astronomer's Telegram},
         year = 2025,
        month = oct,
       volume = {17461},
        pages = {1},
       adsurl = {https://ui.adsabs.harvard.edu/abs/2025ATel17461....1L}
}

@ARTICLE{2025ATel17142....1G,
       author = {{Giacchino}, Federica and {Monti-Guarnieri}, Pietro and {La Mura}, Giovanni},
        title = "{Swift XRT follow-up observations of 3C138}",
      journal = {The Astronomer's Telegram},
         year = 2025,
        month = apr,
       volume = {17142},
        pages = {1},
       adsurl = {https://ui.adsabs.harvard.edu/abs/2025ATel17142....1G}
}

@ARTICLE{2025ATel17180....1W,
       author = {{Wagner}, S.},
        title = "{Fermi-LAT detection of renewed gamma-ray activity from the CSS 3C 138}",
      journal = {The Astronomer's Telegram},
         year = 2025,
        month = may,
       volume = {17180},
        pages = {1},
       adsurl = {https://ui.adsabs.harvard.edu/abs/2025ATel17180....1W}
}

@ARTICLE{2025ATel17077....1L,
       author = {{Li}, Shan and {Lee}, Sang-Sung and {Cheong}, Whee Yeon},
        title = "{Significant Radio Brightening of 3C 138 Revealed by KVN Multi-Frequency Observations at 22-129 GHz}",
      journal = {The Astronomer's Telegram},
         year = 2025,
        month = mar,
       volume = {17077},
        pages = {1},
       adsurl = {https://ui.adsabs.harvard.edu/abs/2025ATel17077....1L}
}

@ARTICLE{2025ATel17193....1K,
       author = {{Kameno}, Seiji and {Artur}, Elizabeth and {Asaki}, Yoshiharu and {Cerrigone}, Luciano and {Cortes}, Paulo and {Guzman}, Andres and {Harrington}, Kevin and {Kneissl}, Ruediger and {Koumpia}, Evgenia and {Lopez}, Cristian and {Marinello}, Gabriel and {Martin}, Sergio and {Messias}, Hugo and {Morgado}, Jorge and {Nowajewski}, Priscilla and {Perez}, Andres and {Plarre}, Kurt and {Radiszcz}, Matias and {Sawada}, Tsuyoshi and {Song}, Yiqing and {Toledo}, Ignacio and {Verdugo}, Celia and {Vilaro}, Baltasar Vila and {Cataldi}, Gianni and {Nagai}, Hiroshi and {Nakanishi}, Kouichiro and {Fomalont}, Edward and {Paradino}, Rosita},
        title = "{3C 138: Millimeter/submillimeter polarized flare}",
      journal = {The Astronomer's Telegram},
         year = 2025,
        month = may,
       volume = {17193},
        pages = {1},
       adsurl = {https://ui.adsabs.harvard.edu/abs/2025ATel17193....1K}
}

@ARTICLE{1966ApJ...144.1244L,
       author = {{Lynds}, C.~R. and {Hill}, S.~J. and {Heere}, Karen and {Stockton}, A.~N.},
        title = "{New Spectroscopic Observations of Fourteen Quasi-Stellar Sources}",
      journal = {\apj},
         year = 1966,
        month = jun,
       volume = {144},
        pages = {1244},
          doi = {10.1086/148730},
       adsurl = {https://ui.adsabs.harvard.edu/abs/1966ApJ...144.1244L}
}

@ARTICLE{2024A&A...682A..34M,
       author = {{Merloni}, A. and {Lamer}, G. and {Liu}, T. and {Ramos-Ceja}, M.~E. and {Brunner}, H. and {Bulbul}, E. and {Dennerl}, K. and {Doroshenko}, V. and {Freyberg}, M.~J. and {Friedrich}, S. and {Gatuzz}, E. and {Georgakakis}, A. and {Haberl}, F. and {Igo}, Z. and {Kreykenbohm}, I. and {Liu}, A. and {Maitra}, C. and {Malyali}, A. and {Mayer}, M.~G.~F. and {Nandra}, K. and {Predehl}, P. and {Robrade}, J. and {Salvato}, M. and {Sanders}, J.~S. and {Stewart}, I. and {Tub{\'\i}n-Arenas}, D. and {Weber}, P. and {Wilms}, J. and {Arcodia}, R. and {Artis}, E. and {Aschersleben}, J. and {Avakyan}, A. and {Aydar}, C. and {Bahar}, Y.~E. and {Balzer}, F. and {Becker}, W. and {Berger}, K. and {Boller}, T. and {Bornemann}, W. and {Br{\"u}ggen}, M. and {Brusa}, M. and {Buchner}, J. and {Burwitz}, V. and {Camilloni}, F. and {Clerc}, N. and {Comparat}, J. and {Coutinho}, D. and {Czesla}, S. and {Dannhauer}, S.~M. and {Dauner}, L. and {Dauser}, T. and {Dietl}, J. and {Dolag}, K. and {Dwelly}, T. and {Egg}, K. and {Ehl}, E. and {Freund}, S. and {Friedrich}, P. and {Gaida}, R. and {Garrel}, C. and {Ghirardini}, V. and {Gokus}, A. and {Gr{\"u}nwald}, G. and {Grandis}, S. and {Grotova}, I. and {Gruen}, D. and {Gueguen}, A. and {H{\"a}mmerich}, S. and {Hamaus}, N. and {Hasinger}, G. and {Haubner}, K. and {Homan}, D. and {Ider Chitham}, J. and {Joseph}, W.~M. and {Joyce}, A. and {K{\"o}nig}, O. and {Kaltenbrunner}, D.~M. and {Khokhriakova}, A. and {Kink}, W. and {Kirsch}, C. and {Kluge}, M. and {Knies}, J. and {Krippendorf}, S. and {Krumpe}, M. and {Kurpas}, J. and {Li}, P. and {Liu}, Z. and {Locatelli}, N. and {Lorenz}, M. and {M{\"u}ller}, S. and {Magaudda}, E. and {Mannes}, C. and {McCall}, H. and {Meidinger}, N. and {Michailidis}, M. and {Migkas}, K. and {Mu{\~n}oz-Giraldo}, D. and {Musiimenta}, B. and {Nguyen-Dang}, N.~T. and {Ni}, Q. and {Olechowska}, A. and {Ota}, N. and {Pacaud}, F. and {Pasini}, T. and {Perinati}, E. and {Pires}, A.~M. and {Pommranz}, C. and {Ponti}, G. and {Poppenhaeger}, K. and {P{\"u}hlhofer}, G. and {Rau}, A. and {Reh}, M. and {Reiprich}, T.~H. and {Roster}, W. and {Saeedi}, S. and {Santangelo}, A. and {Sasaki}, M. and {Schmitt}, J. and {Schneider}, P.~C. and {Schrabback}, T. and {Schuster}, N. and {Schwope}, A. and {Seppi}, R. and {Serim}, M.~M. and {Shreeram}, S. and {Sokolova-Lapa}, E. and {Starck}, H. and {Stelzer}, B. and {Stierhof}, J. and {Suleimanov}, V. and {Tenzer}, C. and {Traulsen}, I. and {Tr{\"u}mper}, J. and {Tsuge}, K. and {Urrutia}, T. and {Veronica}, A. and {Waddell}, S.~G.~H. and {Willer}, R. and {Wolf}, J. and {Yeung}, M.~C.~H. and {Zainab}, A. and {Zangrandi}, F. and {Zhang}, X. and {Zhang}, Y. and {Zheng}, X.},
        title = "{The SRG/eROSITA all-sky survey. First X-ray catalogues and data release of the western Galactic hemisphere}",
      journal = {\aap},
         year = 2024,
        month = feb,
       volume = {682},
          eid = {A34},
        pages = {A34},
          doi = {10.1051/0004-6361/202347165},
archivePrefix = {arXiv},
       eprint = {2401.17274},
 primaryClass = {astro-ph.HE},
       adsurl = {https://ui.adsabs.harvard.edu/abs/2024A&A...682A..34M}
}

@ARTICLE{2024A&A...687A.183S,
       author = {{Sazonov}, S. and {Burenin}, R. and {Filippova}, E. and {Krivonos}, R. and {Arefiev}, V. and {Borisov}, K. and {Buntov}, M. and {Chen}, C.-T. and {Ehlert}, S. and {Garanin}, S. and {Garin}, M. and {Grigorovich}, S. and {Lapshov}, I. and {Levin}, V. and {Lutovinov}, A. and {Mereminskiy}, I. and {Molkov}, S. and {Pavlinsky}, M. and {Ramsey}, B.~D. and {Semena}, A. and {Semena}, N. and {Shtykovsky}, A. and {Sunyaev}, R. and {Tkachenko}, A. and {Swartz}, D.~A. and {Uskov}, G. and {Vikhlinin}, A. and {Voron}, V. and {Zakharov}, E. and {Zaznobin}, I.},
        title = "{SRG/ART-XC all-sky X-ray survey: Catalog of sources detected during the first five surveys}",
      journal = {\aap},
         year = 2024,
        month = jul,
       volume = {687},
          eid = {A183},
        pages = {A183},
          doi = {10.1051/0004-6361/202348950},
archivePrefix = {arXiv},
       eprint = {2405.09184},
 primaryClass = {astro-ph.HE},
       adsurl = {https://ui.adsabs.harvard.edu/abs/2024A&A...687A.183S}
}

@ARTICLE{2016ApJS..224...40W,
       author = {{Wang}, Song and {Liu}, Jifeng and {Qiu}, Yanli and {Bai}, Yu and {Yang}, Huiqin and {Guo}, Jincheng and {Zhang}, Peng},
        title = "{CHANDRA ACIS Survey of X-Ray Point Sources: The Source Catalog}",
      journal = {\apjs},
         year = 2016,
        month = jun,
       volume = {224},
       number = {2},
          eid = {40},
        pages = {40},
          doi = {10.3847/0067-0049/224/2/40},
archivePrefix = {arXiv},
       eprint = {1603.08353},
 primaryClass = {astro-ph.SR},
       adsurl = {https://ui.adsabs.harvard.edu/abs/2016ApJS..224...40W}
}

@ARTICLE{2024ATel16845....1B,
       author = {{Bronzini}, Ettore and {Cheung}, C.~C. and {La Mura}, G.},
        title = "{Fermi-LAT detection of enhanced gamma-ray activity from the CSS Quasar 3C 138}",
      journal = {The Astronomer's Telegram},
         year = 2024,
        month = oct,
       volume = {16845},
        pages = {1},
       adsurl = {https://ui.adsabs.harvard.edu/abs/2024ATel16845....1B}
}

@ARTICLE{2017ApJS..230....7P,
       author = {{Perley}, R.~A. and {Butler}, B.~J.},
        title = "{An Accurate Flux Density Scale from 50 MHz to 50 GHz}",
      journal = {\apjs},
         year = 2017,
        month = may,
       volume = {230},
       number = {1},
          eid = {7},
        pages = {7},
          doi = {10.3847/1538-4365/aa6df9},
archivePrefix = {arXiv},
       eprint = {1609.05940},
 primaryClass = {astro-ph.IM},
       adsurl = {https://ui.adsabs.harvard.edu/abs/2017ApJS..230....7P}
}

@ARTICLE{2013ApJS..204...19P,
       author = {{Perley}, R.~A. and {Butler}, B.~J.},
        title = "{An Accurate Flux Density Scale from 1 to 50 GHz}",
      journal = {\apjs},
         year = 2013,
        month = feb,
       volume = {204},
       number = {2},
          eid = {19},
        pages = {19},
          doi = {10.1088/0067-0049/204/2/19},
archivePrefix = {arXiv},
       eprint = {1211.1300},
 primaryClass = {astro-ph.IM},
       adsurl = {https://ui.adsabs.harvard.edu/abs/2013ApJS..204...19P}
}

@ARTICLE{1985ApJ...298..114M,
       author = {{Marscher}, A.~P. and {Gear}, W.~K.},
        title = "{Models for high-frequency radio outbursts in extragalactic sources, with application to the early 1983 millimeter-to-infrared flare of 3C 273.}",
      journal = {\apj},
         year = 1985,
        month = nov,
       volume = {298},
        pages = {114-127},
          doi = {10.1086/163592},
       adsurl = {https://ui.adsabs.harvard.edu/abs/1985ApJ...298..114M}
}

@ARTICLE{2016MNRAS.462.3325P,
       author = {{Petropoulou}, Maria and {Giannios}, Dimitrios and {Sironi}, Lorenzo},
        title = "{Blazar flares powered by plasmoids in relativistic reconnection}",
      journal = {\mnras},
         year = 2016,
        month = nov,
       volume = {462},
       number = {3},
        pages = {3325-3343},
          doi = {10.1093/mnras/stw1832},
archivePrefix = {arXiv},
       eprint = {1606.07447},
 primaryClass = {astro-ph.HE},
       adsurl = {https://ui.adsabs.harvard.edu/abs/2016MNRAS.462.3325P}
}

@article{kharinov2012,
  author = {M.~A. Kharinov and A.~E. Yablokova},
  issue = {24},
  journal = {Tr. IPA RAN},
  note = {russian},
  pages = {342--347},
  title = {Class Visual: modernization of the package for processing the single dish observations data},
  url = {http://iaaras.ru/library/paper/877/},
  year = {2012}
}

@ARTICLE{1999A&AS..139..545K,
       author = {{Kovalev}, Y.~Y. and {Nizhelsky}, N.~A. and {Kovalev}, Yu. A. and {Berlin}, A.~B. and {Zhekanis}, G.~V. and {Mingaliev}, M.~G. and {Bogdantsov}, A.~V.},
        title = "{Survey of instantaneous 1-22 GHz spectra of 550 compact extragalactic objects with declinations from -30$^{deg}$ to +43$^{deg}$}",
      journal = {\aaps},
         year = 1999,
        month = nov,
       volume = {139},
        pages = {545-554},
          doi = {10.1051/aas:1999406},
archivePrefix = {arXiv},
       eprint = {astro-ph/0408264},
 primaryClass = {astro-ph},
       adsurl = {https://ui.adsabs.harvard.edu/abs/1999A&AS..139..545K}
}

@ARTICLE{1993BSAO...36..107V, 
       author = {{Vlasyuk}, V.~V.},
        title = "{Software for reduction of spectral data obtained with panoramic detectors of the 6 m telescope}",
      journal = {Bulletin of Special Astrophysical Observatory},
         year = 1993,
        month = dec,
       volume = {36},
       number = {3},
        pages = {107-117},
       adsurl = {https://ui.adsabs.harvard.edu/abs/1993BSAO...36..107V}
}

@ARTICLE{1990A&AS...83..183M,
       author = {{Mead}, A.~R.~G. and {Ballard}, K.~R. and {Brand}, P.~W.~J.~L. and {Hough}, J.~H. and {Brindle}, C. and {Bailey}, J.~A.},
        title = "{Optical and infrared polarimetry and photometry of blazars.}",
      journal = {\aaps},
         year = 1990,
        month = apr,
       volume = {83},
        pages = {183-204},
       adsurl = {https://ui.adsabs.harvard.edu/abs/1990A&AS...83..183M}
}

@ARTICLE{2009ApJ...697.1071A,
       author = {{Atwood}, W.~B. and {Abdo}, A.~A. and {Ackermann}, M. and {Althouse}, W. and {Anderson}, B. and {Axelsson}, M. and {Baldini}, L. and {Ballet}, J. and {Band}, D.~L. and {Barbiellini}, G. and {Bartelt}, J. and {Bastieri}, D. and {Baughman}, B.~M. and {Bechtol}, K. and {B{\'e}d{\'e}r{\`e}de}, D. and {Bellardi}, F. and {Bellazzini}, R. and {Berenji}, B. and {Bignami}, G.~F. and {Bisello}, D. and {Bissaldi}, E. and {Blandford}, R.~D. and {Bloom}, E.~D. and {Bogart}, J.~R. and {Bonamente}, E. and {Bonnell}, J. and {Borgland}, A.~W. and {Bouvier}, A. and {Bregeon}, J. and {Brez}, A. and {Brigida}, M. and {Bruel}, P. and {Burnett}, T.~H. and {Busetto}, G. and {Caliandro}, G.~A. and {Cameron}, R.~A. and {Caraveo}, P.~A. and {Carius}, S. and {Carlson}, P. and {Casandjian}, J.~M. and {Cavazzuti}, E. and {Ceccanti}, M. and {Cecchi}, C. and {Charles}, E. and {Chekhtman}, A. and {Cheung}, C.~C. and {Chiang}, J. and {Chipaux}, R. and {Cillis}, A.~N. and {Ciprini}, S. and {Claus}, R. and {Cohen-Tanugi}, J. and {Condamoor}, S. and {Conrad}, J. and {Corbet}, R. and {Corucci}, L. and {Costamante}, L. and {Cutini}, S. and {Davis}, D.~S. and {Decotigny}, D. and {DeKlotz}, M. and {Dermer}, C.~D. and {de Angelis}, A. and {Digel}, S.~W. and {do Couto e Silva}, E. and {Drell}, P.~S. and {Dubois}, R. and {Dumora}, D. and {Edmonds}, Y. and {Fabiani}, D. and {Farnier}, C. and {Favuzzi}, C. and {Flath}, D.~L. and {Fleury}, P. and {Focke}, W.~B. and {Funk}, S. and {Fusco}, P. and {Gargano}, F. and {Gasparrini}, D. and {Gehrels}, N. and {Gentit}, F. -X. and {Germani}, S. and {Giebels}, B. and {Giglietto}, N. and {Giommi}, P. and {Giordano}, F. and {Glanzman}, T. and {Godfrey}, G. and {Grenier}, I.~A. and {Grondin}, M. -H. and {Grove}, J.~E. and {Guillemot}, L. and {Guiriec}, S. and {Haller}, G. and {Harding}, A.~K. and {Hart}, P.~A. and {Hays}, E. and {Healey}, S.~E. and {Hirayama}, M. and {Hjalmarsdotter}, L. and {Horn}, R. and {Hughes}, R.~E. and {J{\'o}hannesson}, G. and {Johansson}, G. and {Johnson}, A.~S. and {Johnson}, R.~P. and {Johnson}, T.~J. and {Johnson}, W.~N. and {Kamae}, T. and {Katagiri}, H. and {Kataoka}, J. and {Kavelaars}, A. and {Kawai}, N. and {Kelly}, H. and {Kerr}, M. and {Klamra}, W. and {Kn{\"o}dlseder}, J. and {Kocian}, M.~L. and {Komin}, N. and {Kuehn}, F. and {Kuss}, M. and {Landriu}, D. and {Latronico}, L. and {Lee}, B. and {Lee}, S. -H. and {Lemoine-Goumard}, M. and {Lionetto}, A.~M. and {Longo}, F. and {Loparco}, F. and {Lott}, B. and {Lovellette}, M.~N. and {Lubrano}, P. and {Madejski}, G.~M. and {Makeev}, A. and {Marangelli}, B. and {Massai}, M.~M. and {Mazziotta}, M.~N. and {McEnery}, J.~E. and {Menon}, N. and {Meurer}, C. and {Michelson}, P.~F. and {Minuti}, M. and {Mirizzi}, N. and {Mitthumsiri}, W. and {Mizuno}, T. and {Moiseev}, A.~A. and {Monte}, C. and {Monzani}, M.~E. and {Moretti}, E. and {Morselli}, A. and {Moskalenko}, I.~V. and {Murgia}, S. and {Nakamori}, T. and {Nishino}, S. and {Nolan}, P.~L. and {Norris}, J.~P. and {Nuss}, E. and {Ohno}, M. and {Ohsugi}, T. and {Omodei}, N. and {Orlando}, E. and {Ormes}, J.~F. and {Paccagnella}, A. and {Paneque}, D. and {Panetta}, J.~H. and {Parent}, D. and {Pearce}, M. and {Pepe}, M. and {Perazzo}, A. and {Pesce-Rollins}, M. and {Picozza}, P. and {Pieri}, L. and {Pinchera}, M. and {Piron}, F. and {Porter}, T.~A. and {Poupard}, L. and {Rain{\`o}}, S. and {Rando}, R. and {Rapposelli}, E. and {Razzano}, M. and {Reimer}, A. and {Reimer}, O. and {Reposeur}, T. and {Reyes}, L.~C. and {Ritz}, S. and {Rochester}, L.~S. and {Rodriguez}, A.~Y. and {Romani}, R.~W. and {Roth}, M. and {Russell}, J.~J. and {Ryde}, F. and {Sabatini}, S. and {Sadrozinski}, H.~F. -W. and {Sanchez}, D. and {Sander}, A. and {Sapozhnikov}, L. and {Parkinson}, P.~M. Saz and {Scargle}, J.~D. and {Schalk}, T.~L. and {Scolieri}, G. and {Sgr{\`o}}, C. and {Share}, G.~H. and {Shaw}, M. and {Shimokawabe}, T. and {Shrader}, C. and {Sierpowska-Bartosik}, A. and {Siskind}, E.~J. and {Smith}, D.~A. and {Smith}, P.~D. and {Spandre}, G. and {Spinelli}, P. and {Starck}, J. -L. and {Stephens}, T.~E. and {Strickman}, M.~S. and {Strong}, A.~W. and {Suson}, D.~J. and {Tajima}, H. and {Takahashi}, H. and {Takahashi}, T. and {Tanaka}, T. and {Tenze}, A. and {Tether}, S. and {Thayer}, J.~B. and {Thayer}, J.~G. and {Thompson}, D.~J. and {Tibaldo}, L. and {Tibolla}, O. and {Torres}, D.~F. and {Tosti}, G. and {Tramacere}, A. and {Turri}, M. and {Usher}, T.~L. and {Vilchez}, N. and {Vitale}, V. and {Wang}, P. and {Watters}, K. and {Winer}, B.~L. and {Wood}, K.~S. and {Ylinen}, T. and {Ziegler}, M.},
        title = "{The Large Area Telescope on the Fermi Gamma-Ray Space Telescope Mission}",
      journal = {\apj},
         year = 2009,
        month = jun,
       volume = {697},
       number = {2},
        pages = {1071-1102},
          doi = {10.1088/0004-637X/697/2/1071},
archivePrefix = {arXiv},
       eprint = {0902.1089},
 primaryClass = {astro-ph.IM},
       adsurl = {https://ui.adsabs.harvard.edu/abs/2009ApJ...697.1071A}
}

@article{2019..VLBI..Quasar,
author = {Shuygina, N. and Ivanov, Dmitrii and Ipatov, A. and Gayazov, I. and Marshalov, D. and Melnikov, Alexey and Kurdubov, Sergei and Vasilyev, Mikhail and Ilin, G. and Skurikhina, E. and Surkis, I. and Mardyshkin, V. and Mikhailov, A. and Salnikov, A. and Vytnov, A. and Rakhimov, I. and Dyakov, A. and Olifirov, V.},
year = {2019},
month = {3},
pages = {150-156},
title = {Russian VLBI network “Quasar”: Current status and outlook},
volume = {10},
journal = {Geodesy and Geodynamics},
doi = {10.1016/j.geog.2018.09.008}
}

@ARTICLE{2019PASP..131a8002B,
       author = {{Bellm}, Eric C. and {Kulkarni}, Shrinivas R. and {Graham}, Matthew J. and {Dekany}, Richard and {Smith}, Roger M. and {Riddle}, Reed and {Masci}, Frank J. and {Helou}, George and {Prince}, Thomas A. and {Adams}, Scott M. and {Barbarino}, C. and {Barlow}, Tom and {Bauer}, James and {Beck}, Ron and {Belicki}, Justin and {Biswas}, Rahul and {Blagorodnova}, Nadejda and {Bodewits}, Dennis and {Bolin}, Bryce and {Brinnel}, Valery and {Brooke}, Tim and {Bue}, Brian and {Bulla}, Mattia and {Burruss}, Rick and {Cenko}, S. Bradley and {Chang}, Chan-Kao and {Connolly}, Andrew and {Coughlin}, Michael and {Cromer}, John and {Cunningham}, Virginia and {De}, Kishalay and {Delacroix}, Alex and {Desai}, Vandana and {Duev}, Dmitry A. and {Eadie}, Gwendolyn and {Farnham}, Tony L. and {Feeney}, Michael and {Feindt}, Ulrich and {Flynn}, David and {Franckowiak}, Anna and {Frederick}, S. and {Fremling}, C. and {Gal-Yam}, Avishay and {Gezari}, Suvi and {Giomi}, Matteo and {Goldstein}, Daniel A. and {Golkhou}, V. Zach and {Goobar}, Ariel and {Groom}, Steven and {Hacopians}, Eugean and {Hale}, David and {Henning}, John and {Ho}, Anna Y.~Q. and {Hover}, David and {Howell}, Justin and {Hung}, Tiara and {Huppenkothen}, Daniela and {Imel}, David and {Ip}, Wing-Huen and {Ivezi{\'c}}, {\v{Z}}eljko and {Jackson}, Edward and {Jones}, Lynne and {Juric}, Mario and {Kasliwal}, Mansi M. and {Kaspi}, S. and {Kaye}, Stephen and {Kelley}, Michael S.~P. and {Kowalski}, Marek and {Kramer}, Emily and {Kupfer}, Thomas and {Landry}, Walter and {Laher}, Russ R. and {Lee}, Chien-De and {Lin}, Hsing Wen and {Lin}, Zhong-Yi and {Lunnan}, Ragnhild and {Giomi}, Matteo and {Mahabal}, Ashish and {Mao}, Peter and {Miller}, Adam A. and {Monkewitz}, Serge and {Murphy}, Patrick and {Ngeow}, Chow-Choong and {Nordin}, Jakob and {Nugent}, Peter and {Ofek}, Eran and {Patterson}, Maria T. and {Penprase}, Bryan and {Porter}, Michael and {Rauch}, Ludwig and {Rebbapragada}, Umaa and {Reiley}, Dan and {Rigault}, Mickael and {Rodriguez}, Hector and {van Roestel}, Jan and {Rusholme}, Ben and {van Santen}, Jakob and {Schulze}, S. and {Shupe}, David L. and {Singer}, Leo P. and {Soumagnac}, Maayane T. and {Stein}, Robert and {Surace}, Jason and {Sollerman}, Jesper and {Szkody}, Paula and {Taddia}, F. and {Terek}, Scott and {Van Sistine}, Angela and {van Velzen}, Sjoert and {Vestrand}, W. Thomas and {Walters}, Richard and {Ward}, Charlotte and {Ye}, Quan-Zhi and {Yu}, Po-Chieh and {Yan}, Lin and {Zolkower}, Jeffry},
        title = "{The Zwicky Transient Facility: System Overview, Performance, and First Results}",
      journal = {\pasp},
         year = 2019,
        month = jan,
       volume = {131},
       number = {995},
        pages = {018002},
          doi = {10.1088/1538-3873/aaecbe},
archivePrefix = {arXiv},
       eprint = {1902.01932},
 primaryClass = {astro-ph.IM},
       adsurl = {https://ui.adsabs.harvard.edu/abs/2019PASP..131a8002B}
}

@ARTICLE{2020ApJS..247...33A,
       author = {{Abdollahi}, S. and {Acero}, F. and {Ackermann}, M. and {Ajello}, M. and {Atwood}, W.~B. and {Axelsson}, M. and {Baldini}, L. and {Ballet}, J. and {Barbiellini}, G. and {Bastieri}, D. and {Becerra Gonzalez}, J. and {Bellazzini}, R. and {Berretta}, A. and {Bissaldi}, E. and {Blandford}, R.~D. and {Bloom}, E.~D. and {Bonino}, R. and {Bottacini}, E. and {Brandt}, T.~J. and {Bregeon}, J. and {Bruel}, P. and {Buehler}, R. and {Burnett}, T.~H. and {Buson}, S. and {Cameron}, R.~A. and {Caputo}, R. and {Caraveo}, P.~A. and {Casandjian}, J.~M. and {Castro}, D. and {Cavazzuti}, E. and {Charles}, E. and {Chaty}, S. and {Chen}, S. and {Cheung}, C.~C. and {Chiaro}, G. and {Ciprini}, S. and {Cohen-Tanugi}, J. and {Cominsky}, L.~R. and {Coronado-Bl{\'a}zquez}, J. and {Costantin}, D. and {Cuoco}, A. and {Cutini}, S. and {D'Ammando}, F. and {DeKlotz}, M. and {de la Torre Luque}, P. and {de Palma}, F. and {Desai}, A. and {Digel}, S.~W. and {Di Lalla}, N. and {Di Mauro}, M. and {Di Venere}, L. and {Dom{\'\i}nguez}, A. and {Dumora}, D. and {Fana Dirirsa}, F. and {Fegan}, S.~J. and {Ferrara}, E.~C. and {Franckowiak}, A. and {Fukazawa}, Y. and {Funk}, S. and {Fusco}, P. and {Gargano}, F. and {Gasparrini}, D. and {Giglietto}, N. and {Giommi}, P. and {Giordano}, F. and {Giroletti}, M. and {Glanzman}, T. and {Green}, D. and {Grenier}, I.~A. and {Griffin}, S. and {Grondin}, M. -H. and {Grove}, J.~E. and {Guiriec}, S. and {Harding}, A.~K. and {Hayashi}, K. and {Hays}, E. and {Hewitt}, J.~W. and {Horan}, D. and {J{\'o}hannesson}, G. and {Johnson}, T.~J. and {Kamae}, T. and {Kerr}, M. and {Kocevski}, D. and {Kovac'evic'}, M. and {Kuss}, M. and {Landriu}, D. and {Larsson}, S. and {Latronico}, L. and {Lemoine-Goumard}, M. and {Li}, J. and {Liodakis}, I. and {Longo}, F. and {Loparco}, F. and {Lott}, B. and {Lovellette}, M.~N. and {Lubrano}, P. and {Madejski}, G.~M. and {Maldera}, S. and {Malyshev}, D. and {Manfreda}, A. and {Marchesini}, E.~J. and {Marcotulli}, L. and {Mart{\'\i}-Devesa}, G. and {Martin}, P. and {Massaro}, F. and {Mazziotta}, M.~N. and {McEnery}, J.~E. and {Mereu}, I. and {Meyer}, M. and {Michelson}, P.~F. and {Mirabal}, N. and {Mizuno}, T. and {Monzani}, M.~E. and {Morselli}, A. and {Moskalenko}, I.~V. and {Negro}, M. and {Nuss}, E. and {Ojha}, R. and {Omodei}, N. and {Orienti}, M. and {Orlando}, E. and {Ormes}, J.~F. and {Palatiello}, M. and {Paliya}, V.~S. and {Paneque}, D. and {Pei}, Z. and {Pe{\~n}a-Herazo}, H. and {Perkins}, J.~S. and {Persic}, M. and {Pesce-Rollins}, M. and {Petrosian}, V. and {Petrov}, L. and {Piron}, F. and {Poon}, H. and {Porter}, T.~A. and {Principe}, G. and {Rain{\`o}}, S. and {Rando}, R. and {Razzano}, M. and {Razzaque}, S. and {Reimer}, A. and {Reimer}, O. and {Remy}, Q. and {Reposeur}, T. and {Romani}, R.~W. and {Saz Parkinson}, P.~M. and {Schinzel}, F.~K. and {Serini}, D. and {Sgr{\`o}}, C. and {Siskind}, E.~J. and {Smith}, D.~A. and {Spandre}, G. and {Spinelli}, P. and {Strong}, A.~W. and {Suson}, D.~J. and {Tajima}, H. and {Takahashi}, M.~N. and {Tak}, D. and {Thayer}, J.~B. and {Thompson}, D.~J. and {Tibaldo}, L. and {Torres}, D.~F. and {Torresi}, E. and {Valverde}, J. and {Van Klaveren}, B. and {van Zyl}, P. and {Wood}, K. and {Yassine}, M. and {Zaharijas}, G.},
        title = "{Fermi Large Area Telescope Fourth Source Catalog}",
      journal = {\apjs},
         year = 2020,
        month = mar,
       volume = {247},
       number = {1},
          eid = {33},
        pages = {33},
          doi = {10.3847/1538-4365/ab6bcb},
archivePrefix = {arXiv},
       eprint = {1902.10045},
 primaryClass = {astro-ph.HE},
       adsurl = {https://ui.adsabs.harvard.edu/abs/2020ApJS..247...33A}
}

@ARTICLE{1988ApJ...328..315M,
       author = {{Massey}, Philip and {Strobel}, Kevin and {Barnes}, Jeannette V. and {Anderson}, Edwin},
        title = "{Spectrophotometric Standards}",
      journal = {\apj},
         year = 1988,
        month = may,
       volume = {328},
        pages = {315},
          doi = {10.1086/166294},
       adsurl = {https://ui.adsabs.harvard.edu/abs/1988ApJ...328..315M}
}

@ARTICLE{2021JOSS....6.3001B,
       author = {{Buchner}, Johannes},
        title = "{UltraNest - a robust, general purpose Bayesian inference engine}",
      journal = {The Journal of Open Source Software},
         year = 2021,
        month = apr,
       volume = {6},
       number = {60},
          eid = {3001},
        pages = {3001},
          doi = {10.21105/joss.03001},
archivePrefix = {arXiv},
       eprint = {2101.09604},
 primaryClass = {stat.CO},
       adsurl = {https://ui.adsabs.harvard.edu/abs/2021JOSS....6.3001B}
}

@INPROCEEDINGS{1996ASPC..101...17A,
       author = {{Arnaud}, K.~A.},
        title = "{XSPEC: The First Ten Years}",
    booktitle = {Astronomical Data Analysis Software and Systems V},
         year = 1996,
       editor = {{Jacoby}, George H. and {Barnes}, Jeannette},
       series = {Astronomical Society of the Pacific Conference Series},
       volume = {101},
        month = jan,
        pages = {17},
       adsurl = {https://ui.adsabs.harvard.edu/abs/1996ASPC..101...17A}
}

@ARTICLE{2000ApJ...542..914W,
       author = {{Wilms}, J. and {Allen}, A. and {McCray}, R.},
        title = "{On the Absorption of X-Rays in the Interstellar Medium}",
      journal = {\apj},
         year = 2000,
        month = oct,
       volume = {542},
       number = {2},
        pages = {914-924},
          doi = {10.1086/317016},
archivePrefix = {arXiv},
       eprint = {astro-ph/0008425},
 primaryClass = {astro-ph},
       adsurl = {https://ui.adsabs.harvard.edu/abs/2000ApJ...542..914W}
}

@ARTICLE{2016A&A...594A.116H,
       author = {{HI4PI Collaboration} and {Ben Bekhti}, N. and {Fl{\"o}er}, L. and {Keller}, R. and {Kerp}, J. and {Lenz}, D. and {Winkel}, B. and {Bailin}, J. and {Calabretta}, M.~R. and {Dedes}, L. and {Ford}, H.~A. and {Gibson}, B.~K. and {Haud}, U. and {Janowiecki}, S. and {Kalberla}, P.~M.~W. and {Lockman}, F.~J. and {McClure-Griffiths}, N.~M. and {Murphy}, T. and {Nakanishi}, H. and {Pisano}, D.~J. and {Staveley-Smith}, L.},
        title = "{HI4PI: A full-sky H I survey based on EBHIS and GASS}",
      journal = {\aap},
         year = 2016,
        month = oct,
       volume = {594},
          eid = {A116},
        pages = {A116},
          doi = {10.1051/0004-6361/201629178},
archivePrefix = {arXiv},
       eprint = {1610.06175},
 primaryClass = {astro-ph.GA},
       adsurl = {https://ui.adsabs.harvard.edu/abs/2016A&A...594A.116H}
}

@ARTICLE{1989GeCoA..53..197A,
       author = {{Anders}, E. and {Grevesse}, N.},
        title = "{Abundances of the elements: Meteoritic and solar}",
      journal = {\gca},
         year = 1989,
        month = jan,
       volume = {53},
       number = {1},
        pages = {197-214},
          doi = {10.1016/0016-7037(89)90286-X},
       adsurl = {https://ui.adsabs.harvard.edu/abs/1989GeCoA..53..197A}
}

@ARTICLE{2009A&A...501..879T,
       author = {{Tramacere}, A. and {Giommi}, P. and {Perri}, M. and {Verrecchia}, F. and {Tosti}, G.},
        title = "{Swift observations of the very intense flaring activity of Mrk 421 during 2006. I. Phenomenological picture of electron acceleration and predictions for MeV/GeV emission}",
      journal = {\aap},
         year = 2009,
        month = jul,
       volume = {501},
       number = {3},
        pages = {879-898},
          doi = {10.1051/0004-6361/200810865},
archivePrefix = {arXiv},
       eprint = {0901.4124},
 primaryClass = {astro-ph.HE},
       adsurl = {https://ui.adsabs.harvard.edu/abs/2009A&A...501..879T}
}

@software{2020ascl.soft09001T,
       author = {{Tramacere}, Andrea},
        title = "{JetSeT: Numerical modeling and SED fitting tool for relativistic jets}",
 howpublished = {Astrophysics Source Code Library, record ascl:2009.001},
         year = 2020,
        month = sep,
          eid = {ascl:2009.001},
archivePrefix = {ascl},
       eprint = {2009.001},
       adsurl = {https://ui.adsabs.harvard.edu/abs/2020ascl.soft09001T}
}

@ARTICLE{2011ApJ...739...66T,
       author = {{Tramacere}, A. and {Massaro}, E. and {Taylor}, A.~M.},
        title = "{Stochastic Acceleration and the Evolution of Spectral Distributions in Synchro-Self-Compton Sources: A Self-consistent Modeling of Blazars' Flares}",
      journal = {\apj},
         year = 2011,
        month = oct,
       volume = {739},
       number = {2},
          eid = {66},
        pages = {66},
          doi = {10.1088/0004-637X/739/2/66},
archivePrefix = {arXiv},
       eprint = {1107.1879},
 primaryClass = {astro-ph.HE},
       adsurl = {https://ui.adsabs.harvard.edu/abs/2011ApJ...739...66T}
}

@ARTICLE{2025ApJ...986...99G,
       author = {{Giacchino}, Federica and {La Mura}, Giovanni and {Ciprini}, Stefano and {Gasparrini}, Dario and {Giroletti}, Marcello and {Laurenti}, Marco},
        title = "{Fermi-LAT Discovery of a Gamma-Ray Outburst from the Peculiar Compact Steep Spectrum Radiogalaxy 3C 216}",
      journal = {\apj},
         year = 2025,
        month = jun,
       volume = {986},
       number = {1},
          eid = {99},
        pages = {99},
          doi = {10.3847/1538-4357/add150},
archivePrefix = {arXiv},
       eprint = {2501.00582},
 primaryClass = {astro-ph.HE},
       adsurl = {https://ui.adsabs.harvard.edu/abs/2025ApJ...986...99G}
}

@ARTICLE{2021AJ....161...88P,
       author = {{Popkov}, A.~V. and {Kovalev}, Y.~Y. and {Petrov}, L.~Y. and {Kovalev}, Yu. A.},
        title = "{Parsec-scale Properties of Steep- and Flat-spectrum Extragalactic Radio Sources from a VLBA Survey of a Complete North Polar Cap Sample}",
      journal = {\aj},
         year = 2021,
        month = feb,
       volume = {161},
       number = {2},
          eid = {88},
        pages = {88},
          doi = {10.3847/1538-3881/abd18c},
archivePrefix = {arXiv},
       eprint = {2008.06803},
 primaryClass = {astro-ph.GA},
       adsurl = {https://ui.adsabs.harvard.edu/abs/2021AJ....161...88P}
}

\section*{Supplementary material}
Supplementary material is available at MNRAS online.

\appendix

\section{Light-curve diagnostics}
\label{app:1GHz}

To assess whether the apparent low-frequency response is driven only by uneven sampling, we performed a separate diagnostic analysis of the 1 GHz light curve. Repeated observations obtained on the same date were first combined using daily medians. The pre-2022 data show only a mild long-term decline, with an unweighted linear slope of
\(-0.17\pm0.06~{\rm Jy~yr^{-1}}\), consistent with
\(-0.17\pm0.10~{\rm Jy~yr^{-1}}\) from a weighted fit. The source then remained in a relatively low and variable state during 2023--2025.70, with a median flux density of 8.25 Jy (Table~\ref{tab:1GHz_diagnostic}). In contrast, the early-2026 measurements have a median flux density of 10.16 Jy, about 1.9 Jy above the 2023--2025.70 level. The positive 2023--2026 trend remains present when the fit is repeated without weights, with the quoted measurement errors, with an additional 0.5 Jy error floor, or using monthly median points. Thus, the recent 1 GHz increase is unlikely to be purely an artefact of uneven sampling. The measurements themselves were checked and the elevated flux level persists outside the dense campaign. However, the unusually strong 1 GHz response relative to the neighbouring radio bands and the sparse pre-2025 sampling prevent us from associating it unambiguously with the same opacity-driven synchrotron component seen at higher frequencies. We therefore do not use the 1 GHz behaviour to derive an inter-band delay or a physical variability time-scale.

\begin{table}
\centering
\caption{Diagnostic statistics of the 1 GHz light curve. Repeated measurements obtained on the same date were combined using daily medians. The weighted mean uses the quoted flux-density uncertainties.}
\label{tab:1GHz_diagnostic}
\begin{tabular}{lccc}
\hline
Interval & $N$ & Median $S_{\rm 1\,GHz}$ & Weighted mean \\
         &     & (Jy) & (Jy) \\
\hline
pre-2022        & 37 & 8.78 & $8.74\pm0.15$ \\
2023--2025.70   & 10 & 8.25 & $8.32\pm0.27$ \\
2026.0--2026.21 & 13 & 10.16 & $10.16\pm0.06$ \\
2025.83--2026.21 & 42 & 10.17 & $10.18\pm0.03$ \\
\hline
\end{tabular}
\end{table}

\begin{figure}
\centerline{\includegraphics[width=0.49\textwidth]{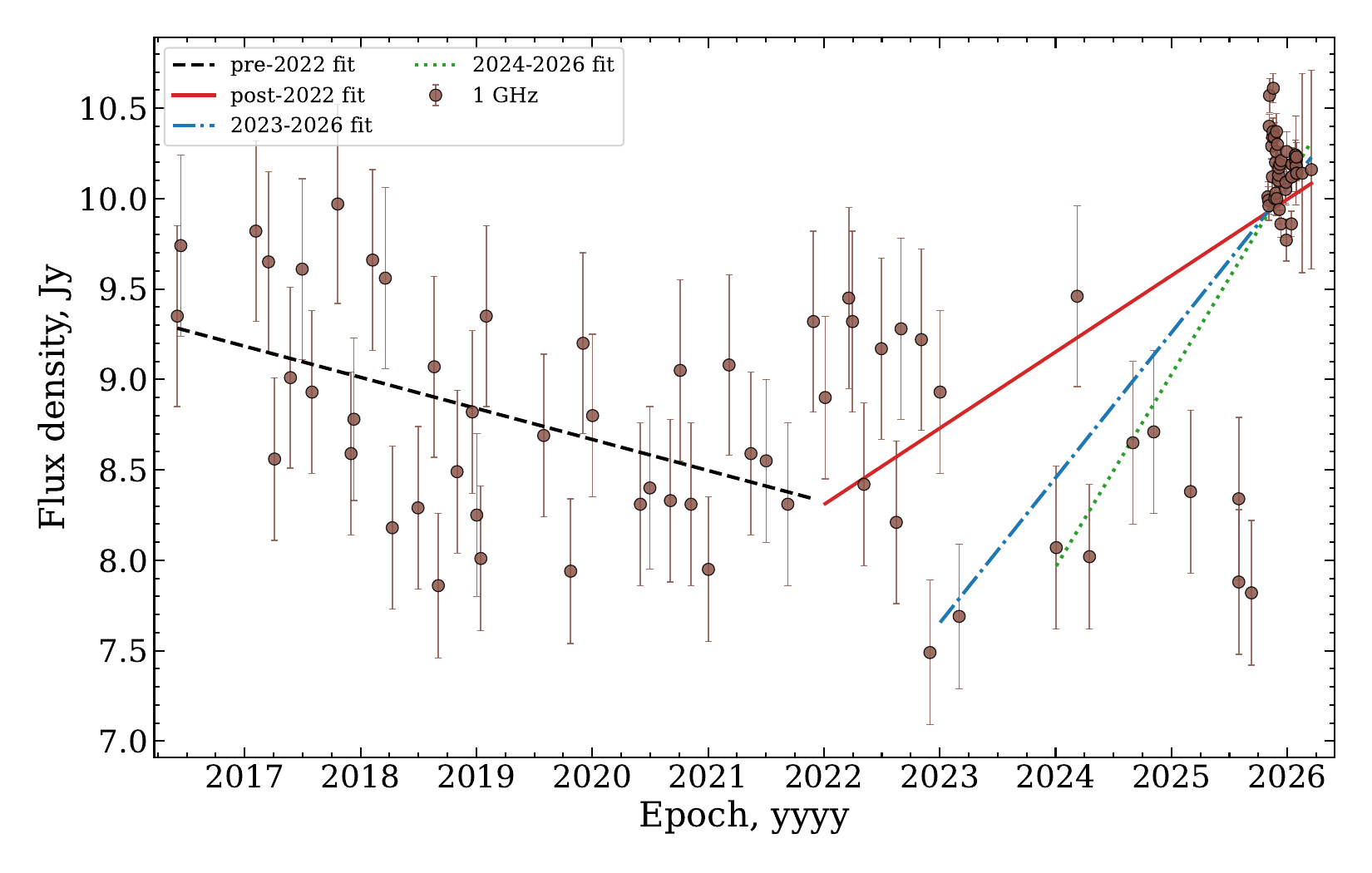}}
\caption{Radio light curve of \sou at 1 GHz with linear fits computed over different time intervals (pre-2022, post-2022, 2023--2026, and 2024--2026). The steep slope obtained for the post-2022 interval is primarily driven by the combination of sparse sampling at earlier epochs and a dense cluster of high-flux measurements in 2025--2026. This illustrates that the fitted OLS trend at 1 GHz is sensitive to the time window and sampling pattern, and therefore does not robustly trace the intrinsic long-term evolution at this frequency.}
\label{fig:1GHz}
\end{figure}

\clearpage
\bsp
\label{lastpage}

\end{document}